\documentclass[12pt,english]{article}

\usepackage[T1]{fontenc}
\usepackage[latin9]{inputenc}
\usepackage{geometry}
\usepackage{color}
\usepackage{babel}
\usepackage{textcomp}
\usepackage{url}
\usepackage{amsmath}
\usepackage{amssymb}
\usepackage{stmaryrd}
\usepackage{graphicx}
\usepackage{rotfloat}
\usepackage{setspace}
\usepackage{esint}
\usepackage[authoryear]{natbib}
\PassOptionsToPackage{normalem}{ulem}
\usepackage{ulem}
\usepackage[unicode=true]
 {hyperref}

\makeatletter

\providecommand{\tabularnewline}{\\}

\@ifundefined{date}{}{\date{}}
\makeatother

\begin{document}
\title{Climate, Agriculture and Food\\
{\normalsize{}Submitted as a chapter to the }\textit{\normalsize{}Handbook
of Agricultural Economics}}
\author{Ariel Ortiz-Bobea\thanks{Associate Professor, Charles H. Dyson School of Applied Economics
and Management, Cornell University. Email: ao332@cornell.edu. }\thanks{I am thankful for useful comments provided by the editors Christopher
Barrett and David Just as well as by Thomas Hertel and Christophe
Gouel. Code and data necessary to reproduce the figures and analysis
discussed in the chapter are available in a permanent repository at
the Cornell Institute for Social and Economic Research (CISER): \protect\url{https://doi.org/10.6077/fb1a-c376}.}}
\maketitle
\begin{center}
May 2021
\par\end{center}
\begin{abstract}
Agriculture is arguably the most climate-sensitive sector of the economy.
Growing concerns about anthropogenic climate change have increased
research interest in assessing its potential impact on the sector
and in identifying policies and adaptation strategies to help the
sector cope with a changing climate. This chapter provides an overview
of recent advancements in the analysis of climate change impacts and
adaptation in agriculture with an emphasis on methods. The chapter
provides an overview of recent research efforts addressing key conceptual
and empirical challenges. The chapter also discusses practical matters
about conducting research in this area and provides reproducible R
code to perform common tasks of data preparation and model estimation
in this literature. The chapter provides a hands-on introduction to
new researchers in this area.
\end{abstract}
\textbf{Approximate length:} 31,200 words.

\newpage{}

\tableofcontents{}

\newpage{}

\section{Introduction }

Climate has always been critical to the development of agriculture.
For instance, changes in climate are believed to have played an imporant
role in the origin of agriculture \citep{gupta_origin_2004,matranga_ant_2017}.
More recently, the story surrounding the territorial expansion of
agriculture over the past few centuries was also one about adapting
farming practices and existing crops to new climates (e.g. \citealp{olmstead_adapting_2011}). 

But climate is now changing at an unprecedented rate overwhelmingly
due to human causes \citep{pachauri_climate_2014}. And even as the
extent of global agricultural land stabilizes, and agricultural productivity
continues to rise, climatic shocks continue to play a central role
in explaining fluctuations in agricultural production \citep{lesk_influence_2016}.
In fact, recent climate change appears to have already substantially
slowed down global agricultural productivity growth \citep{ortiz-bobea_anthropogenic_2021}.
In this context, research is needed not only to understand potential
future impacts of anthropogenic climate change on agriculture, but
also to identify efficient strategies to enhance adaptation to a changing
climate. This includes identifying market failures and possible barriers
to adaptation.

Unlike mitigation of climate change, which requires a coordinated
international effort to reduce greenhouse emissions, adaption is generally
framed as a local matter, something that only local private agents
have to deal with. However, farmers rely directly or indirectly on
public infrastructure, and buy technologies and sell products in markets
with important government presence and regulation. Moreover, the increasing
globalization of agricultural markets and technologies challenge this
view. So agricultural adaptation to climate change goes beyond the
boundaries of the farm.

This chapter is primarily aimed to introduce new researchers to the
analysis of climate change impacts on the agricultural sector. Most
of the content should be highly accessible but familiarity with matrix
algebra is necessary to take fully advantage of the content. Importantly,
the chapter provides code and data to reproduce common tasks while
conducting research in this area, including data preparation and cleaning
as well as the estimation of semi-parametric models. All of the figures
illustrating these techniques are fully reproducible in R. R is a
an increasingly popular open source statistical software that will
ensure a wide access to this material to all researchers.

The chapter is organized in 3 main sections. Section \ref{sec:Basic-concepts}
covers important basic concepts and terminology regarding climate
change and agriculture and discusses various aspects regarding common
datasets used in economic analysis in the field. Subsection \ref{subsec:Weather-and-climate}
defines weather as a random variable and climate as describing the
moments of the underlying distribution of that weather variable. Weather
change and climate change mean different things and this clarification
will allow a more precise discussion surrounding these concepts throughout
the chapter. Subsection \ref{subsec:Adaptation} defines what we mean
by adaptation to climate change following the formal definition adopted
by the Intergovernmental Panel of Climate Change (IPCC). 

Conducting research in this area also require familiarity with different
types of weather data. Subsection \ref{subsec:Weather-data} discusses
important features about historical weather datasets including their
format (e.g. gridded, weather stations). I also provide the names
of common datasets used in the literature. Subsection \ref{subsec:Climate-models}
also discusses the basics of General Circulation Models (GCM) and
how the climate science community improves these models over time
within global inter-comparison projects that feed into the IPCC reports.
Subsection \ref{subsec:Degree-days-and-agriculture} covers some basic
concepts regarding the use and interpretation of degree days. As it
will become apparent, temperature has been found to be a critical
driver of agricultural production. The use of degree days is not new
in agriculture science, and I provide a historical perspective on
the concept and how its use has evolved in the more recent economic
literature. 

Section \ref{sec:Climate-change-impacts} dives into specific areas
of research or methodologies to assess the economic impacts of extreme
weather or climate change on agriculture. An important emphasis in
some of the techniques and approaches presented in this section deal
with the extent to which farmer adaptations are captured and represented.
Subsection \ref{subsec:Biophysical-approaches} provides an overview
of process-based biophysical approach of modeling crop yields and
how these are integrated into economic models to simulate climate
change impacts to the agricultural sector. That presentation provides
an overview of the early literature as well as recent trends toward
the adoption of multi-model ensembles and inter-comparison projects. 

Subsection \ref{subsec:The-Ricardian-approach} transitions to discuss
the cross-sectional Ricardian approach, one of the first econometric
approaches introduced to evaluate the impact of climate change on
agriculture. Here I discuss some the advantages and pitfalls of this
approach including recent advances. The following subsection \ref{subsec:Panel-profit-and}
deals with models estimating the effects of weather fluctuations on
agricultural profits or aggregate measures of productivity based on
panel data. Because these models include location fixed effects, they
provide a more credible identification of the effect of weather than
correctional approaches. I also provide an overview of their limitations. 

Subsection \ref{subsec:Statistical-crop-yield} discusses the rise
of statistical crop yield models in agricultural economics and related
fields. These models are also based on panel data but their focus
on specific crop yields allows researchers to engage in more detailed
crop-specific treatment of weather variables. I try to provide a historical
perspective of the origin of these models before the renewed interest
in the context of climate change. I also briefly discuss new creative
ways to combine statistical and biophysical approaches to modeling
crop yields in subsection \ref{subsec:Mixed-statistical-and}. These
new efforts provide exciting new frontiers of collaboration with natural
scientists. 

In subsection \ref{subsec:Joint-estimation-of} I discuss an emerging
area of research proposing new methods to overcome certain perceived
limitations of both cross-sectional and panel approaches. I describe
these recent advances and their limitations. I highlight new work
that clarifies the theoretical interpretation of panel estimates.
I also provide a brief overview to retrospective climate change studies
in subsection \ref{subsec:Retrospective-studies}. Most of the research
emphasis has focused on assessing future potential impacts of climate
change on the sector. However, anthropogenic forces have already changed
climate which is about 1°C warmer than in pre-industrial times. As
climate continues to change, such retrospective studies are likely
to increase in popularity. I provide a brief overview of early and
more recent studies. The rest of section \ref{sec:Climate-change-impacts}
is less about methods, and more about various aspects of capturing
climate change impacts and farmer and market responses.

There is a vast literature assessing impacts on crop yields, but there
is much less work focusing on the impact on crop quality. Subsection
\ref{subsec:Statistical-crop-quality} presents recent work on this
topic and lays out some challenges for future work. Subsection \ref{subsec:Modeling-planting-and}
discusses the analysis of planting and harvesting decisions, ranging
from the timing of planting, to the decision to increase cropping
frequency (e.g. double cropping). Naturally, irrigation is central
to agriculture and is often perceived as an important mechanism for
farmers to adapt a water scarcity in a changing climate. I cover this
topic in subsection \ref{subsec:Irrigation-and-other}, where I discuss
early studies but also provide a look to more recent work based on
new data sources collected from satellites or providing high-frequency
information about water use at the farm level. I also discuss the
role of trade in the analysis of climate change impacts on agriculture
in subsection \ref{subsec:Trade-and-general}.

I conclude section \ref{sec:Climate-change-impacts} addressing various
questions that still seem unsettled and where more research is likely
needed (subsection \ref{subsec:Some-unsettled-questions}). For instance,
there is insufficient work on the economic of agricultural innovation
in the context of a changing climate. For instance, it is unclear
whether current levels and the current nature of agricultural R\&D
is adequate in a rapidly changing climate. In addition, there are
many lingering uncertainties regarding the important of changing pest
pressure as well as the rise of soil salinity in a warming world.
There is also little emphasis on climate justice.

Section \ref{sec:Coding-and-other} provides more practical and hands-on
guidance regarding common empirical tasks in this area of research.
The chapter is accompanied with reproducible R code that illustrates
how to perform many of these tasks in a systematic way. Providing
the accompanying code seems important for new researchers entering
this field because many of the data management and estimation techniques
are yet not standard in agricultural economic curricula. The code
provided can be easily adapted to new projects and accelerate the
learning curve of new entrants. For instance, the section provides
an overview on how to efficiently aggregate point and gridded weather
datasets (subsections \ref{subsec:Aggregation-of-point} and \ref{subsec:Aggregation-of-gridded}).
The techniques that I introduce are based on matrix algebra and sparse
matrices to speed up aggregation relative to standard ``canned''
functions in R packages.

An important empirical consideration in the analysis of the effects
of extreme weather and climate change on agriculture is the presence
of nonlinearities and thresholds. Climate change will lead to more
frequent extreme weather and thus capturing such nonlinearities is
critical. Importantly, temporal and spatial averaging of weather conditions
can conceal exposure to extreme weather and thus more advanced techniques
are required to overcome such obstacles. In subsection \ref{subsec:Estimating-non-linear-effects}
I describe how to estimate non-linear effects of weather variables
semi-parametrically based on bins representing the entire distribution
of time exposure to varying environmental conditions (e.g. temperature).
I also describe how to construct these datasets. The R code provided
illustrates not only how to construct these data but also how to estimate
these models. I also clarify certain misconceptions and confusion
regarding the estimation of these models. In subsection \ref{subsec:Estimating-within-season-varying}
I also present a two-dimensional generalization of the semi-parametric
model above that allows simultaneously for non-linear and within-season
varying effects. This is particularly valuable when trying to capture
varying sensitivities to environmental conditions within the growing
season.

A common feature of agricultural and climate data is spatial dependence,
which generally translates into spatial dependence in a regression
setting. Subsection \ref{subsec:Spatial-dependence} introduces a
few approaches to deal with this by either correcting for spatial
dependence or by harnessing spatial dependence to obtain a more efficient
estimator. Finally, subsection \ref{subsec:Common-robustness-and}
discusses common robustness checks in the literature as well as strategies
to present these in a concise manner in a ``specification chart''.
I also provide reproducible code to conduct these sensitivity checks.

I finally conclude in section \ref{sec:Conclusion} where I provide
some final thoughts about the potential for new collaborations and
for enhancing the impact of economic research in this field.

\section{Basic concepts and data\label{sec:Basic-concepts}}

Here I discuss basic concepts and terminology regarding climate change
and how farmers respond to it. This preliminary step is critical in
helping new researchers understand the common language used in this
field.

\subsection{Weather and climate\label{subsec:Weather-and-climate}}

Throughout this chapter I refer to ``weather'' and ``climate''
which are related but distinct concepts. A useful way to characterize
their relationship is to think about weather as a random variable
representing the state of the atmosphere. That random variable could
represent, say the \textquotedblleft average temperature during the
month July in Ithaca, NY\textquotedblright . The value taken by this
variable on a given year represents weather conditions. 

In contrast, ``climate'' refers to the moments of the probability
distribution of that random variable. Thus, quantities such as the
\textquotedblleft long term average\textquotedblright{} or the \textquotedblleft inter-annual
variability\textquotedblright{} of this random variable relate to
climate. It is often the case that climatologists (and economists
by extension) refer to \textquotedblleft climatology\textquotedblright{}
as the 30-year average of a weather variable, and as \textquotedblleft climate
variability\textquotedblright{} as the inter-annual variance of a
weather variable. This concept is different from the term \textquotedblleft intra-annual
weather variability\textquotedblright{} which refers to the variability
of weather conditions between contiguous time periods within a season
or the year. 

As a result, the term \textquotedblleft climate change\textquotedblright{}
refers to the change in the long term distribution of weather conditions
at a given location. By definition, it is a long term process because
climatologies are defined over several decades. It follows that climate
change eventually results in the rising frequency of weather events
that were previously considered unusual or extreme. That being said,
because of natural variability in the climate system, a sequence of
extreme weather events is not evidence of climate change per se. Indeed,
assessing changes in climate requires a relatively long time series
of weather observations. 

In addition, the term ``climate change'' should also not be conflated
with \textquotedblleft weather change\textquotedblright{} which may
refer to either inter-annual or intra-annual fluctuations in weather
conditions depending on the context. Moreover, recent weather trends
should not be conflated with climate change. There are well known
cyclical components in the climate system such as El Niño-Southern
Oscillation (ENSO) which are linked to cyclical changes in weather
patterns in various parts of the world. These multi-year weather patterns
do not constitute climate change, though climate change may alter
their intensity. 

Climate change can have both natural or human causes. An entire field
in climate science called \textquotedblleft detection and attribution\textquotedblright{}
focuses on determining whether observed historical changes in climate
can be attributed to specific causes. The term \textquotedblleft anthropogenic
climate change\textquotedblright{} thus refers to changes in climate
that have been shown to originate from human activities, including
through the emissions of greenhouse gases. 

Making the clear distinction between weather and climate is important.
Not only does it help avoid ambiguous terminology and helps convey
ideas more clearly, but it also has economic implications. The reason
is that while economic agents cope directly with weather conditions,
they actually form expectations about climate.

\subsection{Adaptation\label{subsec:Adaptation}}

According to the \citet{ipcc_climate_2014}, adaptation to climate
change in human systems is \textit{``the process of adjustment to
actual or expected climate and its effects, in order to moderate harm
or exploit beneficial opportunities''}. It should be clear why assessing
the potential impacts of climate change on agriculture should require
considering the degree to which farmers would adapt a new climate.
Not accounting for adaptation would naturally overstate potential
damages and under appreciate potential opportunities.

Agricultural economists have mostly focused on adaptation undertaken
by farmers. If you consider weather as a stochastic essential input,
then the adjustments of traditional inputs under the control of the
farmer to maximize profit or utility in response to weather fluctuations
are forms of farmer adaptation. Input decisions are typically sequential
in nature \citep{antle_sequential_1983}, so certain decisions are
committed irreversibly early in the season (e.g. crop and parcel choice,
acreage, etc.) before the farmer gets to observe the weather realization.
As a result, some inputs remain fixed throughout the growing season,
constraining the farmer to a limited range of adaptations. In general,
these short-run adaptations to weather fluctuations understate the
range of adaptations undertaken by farmers when considering long run
adjustments in response to a changing climate. 

However, certain short run adjustments may not be available to the
farmer in the long run. One example may be irrigation. For instance,
a farmer with land equipped for irrigation may increase the amount
of water used in response to dryer weather conditions. However, if
the source of irrigation water is projected to be depleted or water
prices are expected to be much higher in the future, then adjustments
made in the short run may not be indicative of those available in
the long run. 

The empirical characterization of future farmer adaptations generally
relies on historical behavior. But past behavior relative to a change
in a weather shock (or small changes in climate) could mischaracterize
the degree to which farmers may adapt in the long run. This is somewhat
related to the Lucas critique applied to the climate change context
(see \citealp{kahn_climate_2014}).

Accounting for adaptation is critical to the estimation of potential
future climate change impacts on agriculture. This has been a major
emphasis in the literature. Not accounting for adaptation would naturally
overstate damages. Thus researchers typically seek to constrain or
characterize their findings depending on the degree to which farmers
can adapt in their modeling approach.

Capturing or measuring long run adaptations to climate change is challenging.
It ultimately requires capturing adjustments to the production process
in response to a long term change in the distribution of weather conditions.
Ideally, characterizing this process requires a long time series of
weather and production decisions. Long longitudinal datasets with
detailed information about production practices are very rare, making
detection of adaptation activities elusive. 

Explicitly accounting or modeling all possible farmer adaptation to
a changing climate is intractable. Farmers have numerous potential
adjustments to their production decisions. This could include changes
in input use, tilling practices, planting dates or crop mix for crop
production, or the change in management, feed, animal breeds, equipment
or infrastructure for livestock production. As a result, researchers
often rely on indirect evidence to identify or quantify adaptation
(or lack of).

The \citet{ipcc_climate_2014} also employs the term ``maladaptation''
which refers to \textit{``actions that may lead to increased risk
of adverse climate-related outcomes, increased vulnerability to climate
change, or diminished welfare, now or in the future''}. However,
this term is rarely used in mainstream economic academic discussions.
In the agricultural context, this would mean that agriculture is growing
more vulnerable to climate change, such as becoming increasingly sensitive
to higher temperature (e.g. \citealt{lobell_greater_2014,ortiz-bobea_growing_2018,ortiz-bobea_historical_2020}).
However, such changes in sensitivity to extreme temperature may result
from an optimal tradeoff so referring to such phenomena as maladaptation
which carries a undesirable connotation may be misleading.

Note that the notion of economic efficiency is absent from the IPCC
characterization of adaptation. Economists bring a unique perspective
to analyze the economic desirability of adaptive investments from
a welfare perspective. This is more often than not absent for the
analysis and discussions surrounding adaptation.

\subsection{Weather data\label{subsec:Weather-data}}

Weather data is a fundamental component of conducting empirical analysis
of climate change impacts and adaptation. Here I highlight some key
features of such data without being exhaustive. I encourage readers
to consult \citet{auffhammer_using_2013} for a complete guide on
how to use weather data and climate model output in economic analysis.

Basic weather variables like air temperature and precipitation are
commonly measured in weather stations. These are commonly (but not
always) government-run facilities with the necessary instrumentation
to record information about atmospheric conditions. In these facilities,
temperature has historically been measured directly via thermometers,
whereas precipitation is measured via rain gauges, which measure the
amount of precipitation falling within a time interval, typically
a day. In certain countries like the US and parts of Western Europe,
this instrumental record dates back to the 19th century. Air temperature
varies throughout the day, and it is sometimes possible to obtain
hourly data for certain regions in recent years. But data is more
commonly available at the daily, monthly or annual scales. In such
cases summary statistics are reported including maximum, average and
minimum temperature or total precipitation over the given time period.
Note that strict rules and protocols govern the recording and reporting
of official weather data. 

Temperature measurements prior to the use of thermometers are based
on proxy variables (e.g. tree rings) and are used to reconstruct past
weather conditions in paleoclimatology. Since the late 1970s, researchers
can also obtain remotely-sensed temperature from satellites which
are derived indirectly from microwave radiation. 

Because temperature and precipitation are commonly measured at specific
locations (weather stations) throughout the landscape, such type of
weather data is referred to as ``point data'' in Geographical Information
Systems (GIS). A point is associated with precise geographical coordinates
and is thus said to be geo-referenced. 

\begin{figure}
\includegraphics[scale=0.25]{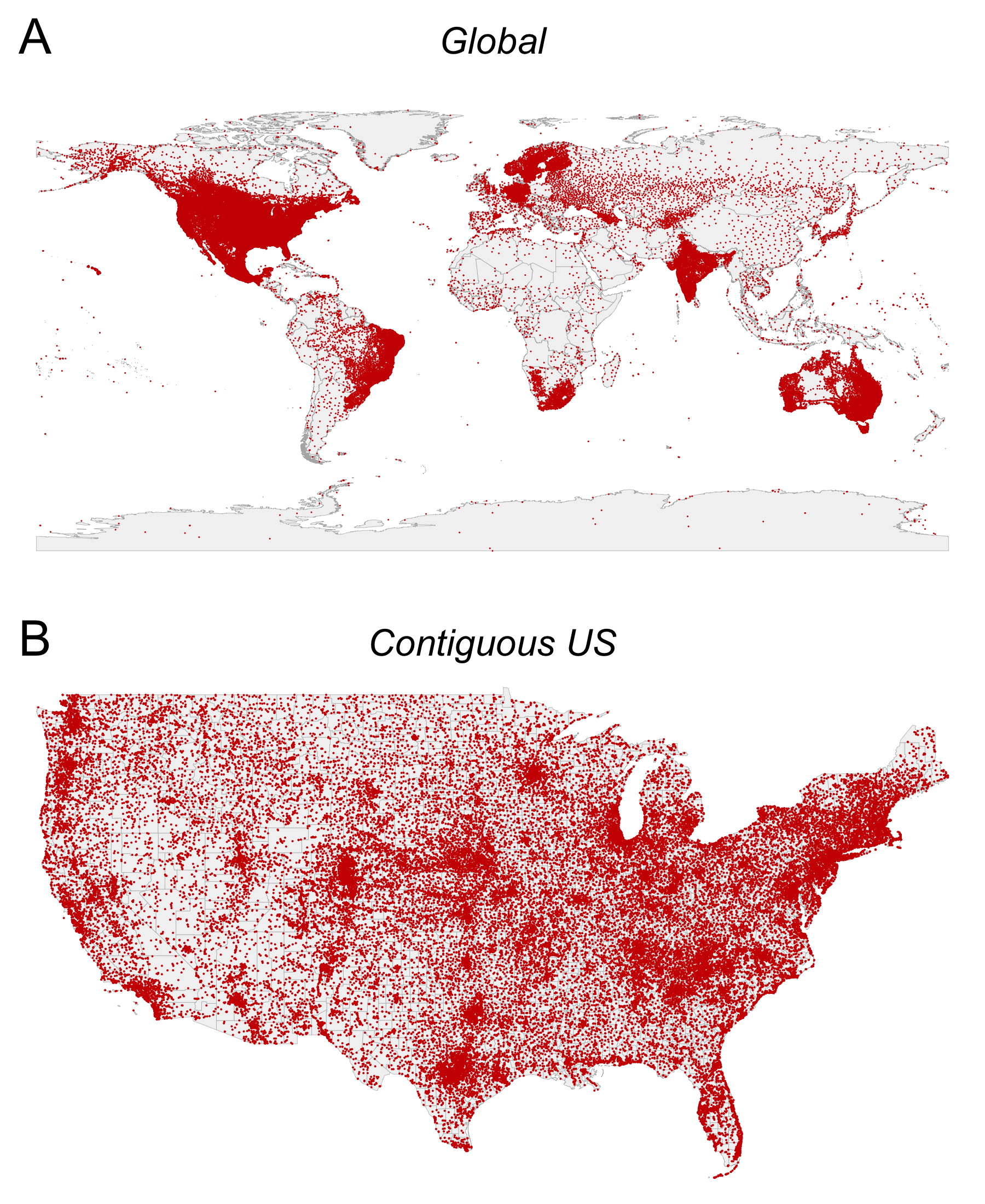}

\caption{Spatial distribution of weather stations in the Global Historical
Climatology Network in 2020.\label{fig:Spatial-distribution}}

\end{figure}

Figure \ref{fig:Spatial-distribution}A provides an overview of the
spatial distribution of the more than 100,000 weather stations reporting
daily information in the Global Historical Climatology Network (GHCN)
in 2020. The GHCN is the world's largest database of climate summaries
from land surface stations across the globe and it is managed by the
National Oceanic and Atmospheric Administration (NOAA). The distribution
of weather stations can be very sparse across the world, even within
countries like the US (Fig. \ref{fig:Spatial-distribution}B). This
spatial sparsity raises challenges for obtaining correct weather information
in areas located far from weather stations, particularly when the
landscape has pronounced orography. In section \ref{subsec:Aggregation-of-point}
I provide a brief and reproducible introduction on (very) basic weather
station data interpolation.

With the goal of providing more complete spatial coverage in a consistent
fashion, climatologists have developed ``gridded'' weather datasets
based on various interpolation techniques. These interpolation techniques
often rely on elevation and other physical factors that are known
to affect both temperature and precipitation. These procedures are
considerably more sophisticated than a simple spatial interpolation.
These geo-referenced datasets come on a regular grid and are referred
to as ``raster'' data in GIS. They are characterized by a spatial
resolution, often measured in degrees or distance. Raster data is
fundamentally structured as a matrix, where each entry corresponds
to a patch of of the Earth's surface. 

It is important to clarify that some of gridded weather datasets can
sometimes incorporate numerical weather models (similar to those used
for weather forecasts). In that case these gridded weather data are
referred to as ``reanalysis''. This is one example of ``modeled''
data that incorporates both observations (from weather stations) and
information from a mechanistic weather simulation model. The advantage
of such data is that they can provide a spatially and temporally consistent
field of weather information even when there are gaps in the underlying
weather station data. They can also provide output of variables that
are actually not being actually measured in any consistent way (e.g.
temperature or wind speed at high altitudes).

So far I have only discussed weather variables and data relating to
atmospheric conditions. For certain applications the researcher might
be interested in more direct measures of water content or temperature
in the soil. Direct measurement of soil water content and temperature
are very rare and only available over limited areas and time periods
in a handful of nations. Obtaining data on these variables at larger
scales generally requires relying on modeled data, although new satellite
sensors are increasingly able to indirectly measure some of these
soil moisture variables.

The evolution of soil water content and temperature is a complex process
that is typically modeled with a Land Surface Model (LSMs). An LSM
takes ``forcing'' or exogenous variables as inputs (e.g. surface
temperature, precipitation, wind, air humidity, etc.) to characterize
the evolution of soil conditions over time. Some of the key variables
of interest include soil water content but also soil temperature.
These models provide a modeled snapshot in time of these variables
at various depths in the soil, often down to a couple meters. The
use of certain highly detailed LSM datasets can be cumbersome as they
may require manipulating terabytes of data. For some applications,
the use of simpler drought indices, such as the Palmer Drought Severity
Index (PDSI) or the Standardized Precipitation Index (SPI), may suffice
(see \citealp{heim_review_2002}). These indices seek to approximate
water deficit conditions based on water supply (precipitation) and
demand (evapotranspiration and runoff) with relatively simple algorithms.

\begin{table}
\begin{centering}
\begin{tabular}{cccccc}
\hline 
 & \multicolumn{2}{c}{Spatial} & \multicolumn{2}{c}{Temporal} & \tabularnewline
Name & Coverage & Resolution & Coverage & Resolution & Source\tabularnewline
\hline 
\href{https://prism.oregonstate.edu/}{PRISM} & CONUS & 4 km & 1981 -- & daily & \citet{daly_prism_1997}\tabularnewline
\href{https://daymet.ornl.gov/}{Daymet} & North America & 1 km & 1980 -- & daily & \citet{thornton_daymet_2014}\tabularnewline
\href{https://www.ncdc.noaa.gov/data-access/model-data/model-datasets/north-american-regional-reanalysis-narr}{NARR} & North America & 0.3 deg & 1979 -- & 3-hourly & \citet{mesinger_north_2006}\tabularnewline
\href{https://ldas.gsfc.nasa.gov/nldas}{NLDAS} & North America & 0.125 deg & 1979 -- & hourly & \citet{xia_continental-scale_2012}\tabularnewline
\href{https://ldas.gsfc.nasa.gov/gldas}{GLDAS} & Global & 0.25 deg & 1948 -- & 3-hourly & \citet{rodell_global_2004}\tabularnewline
\href{https://hydrology.princeton.edu/data.pgf.php}{GMFD} & Global & 0.25 deg & 1948 -- 2016 & daily & \citet{sheffield_development_2006}\tabularnewline
\hline 
\end{tabular}
\par\end{centering}
\caption{Commonly used gridded weather datasets.\label{tab:gridded}}
\end{table}

In table \ref{tab:gridded} I include a list of commonly used gridded
weather and land surface datasets with at least a daily temporal resolution.
This list is by no means exhaustive but provides the reader with a
starting point in their analysis. Monthly datasets are easier to come
by and typically offer longer temporal coverage. For instance, the
widely used monthly version of Oregon State University's Parameter-elevation
Regressions on Independent Slopes Model (PRISM) dataset over the contiguous
United States (CONUS) is available since 1896.

An important point regarding empirical work in this literature is
the common mismatch in spatial resolution between agricultural and
weather data. Agricultural data is often available to researchers
after being aggregated to administrative units such as counties, states
or even countries. This aggregation is sometimes performed from micro-data
from surveys or census to preserve anonymity of individual farmers.
Unless a researcher is dealing with field or farm-level data, the
spatial resolution of agricultural data is typically coarser than
that of gridded weather data. That is, several grid cells fall within
the boundaries of the administrative unit. As a result, researchers
end up aggregating the gridded weather data to the administrative
unit level. 

This naturally raises the question of how should gridded weather data
be aggregated to administrative levels. Certain administrative units
(e.g. US states) can be fairly large and heterogeneous and contain
areas with little to no agricultural activity (e.g. like high mountains
or deserts). In other words, weather conditions in certain parts of
the administrative unit may be irrelevant for agricultural production
within that unit. A common practice is to rely on fine scale land
cover data (e.g. cropland, pastures, or a combination) to use as an
aggregation weight. Land cover data comes in raster format and with
spatial resolutions ranging anywhere from 30m to 1km depending on
the region of the world.

\begin{figure}
\includegraphics[scale=0.25]{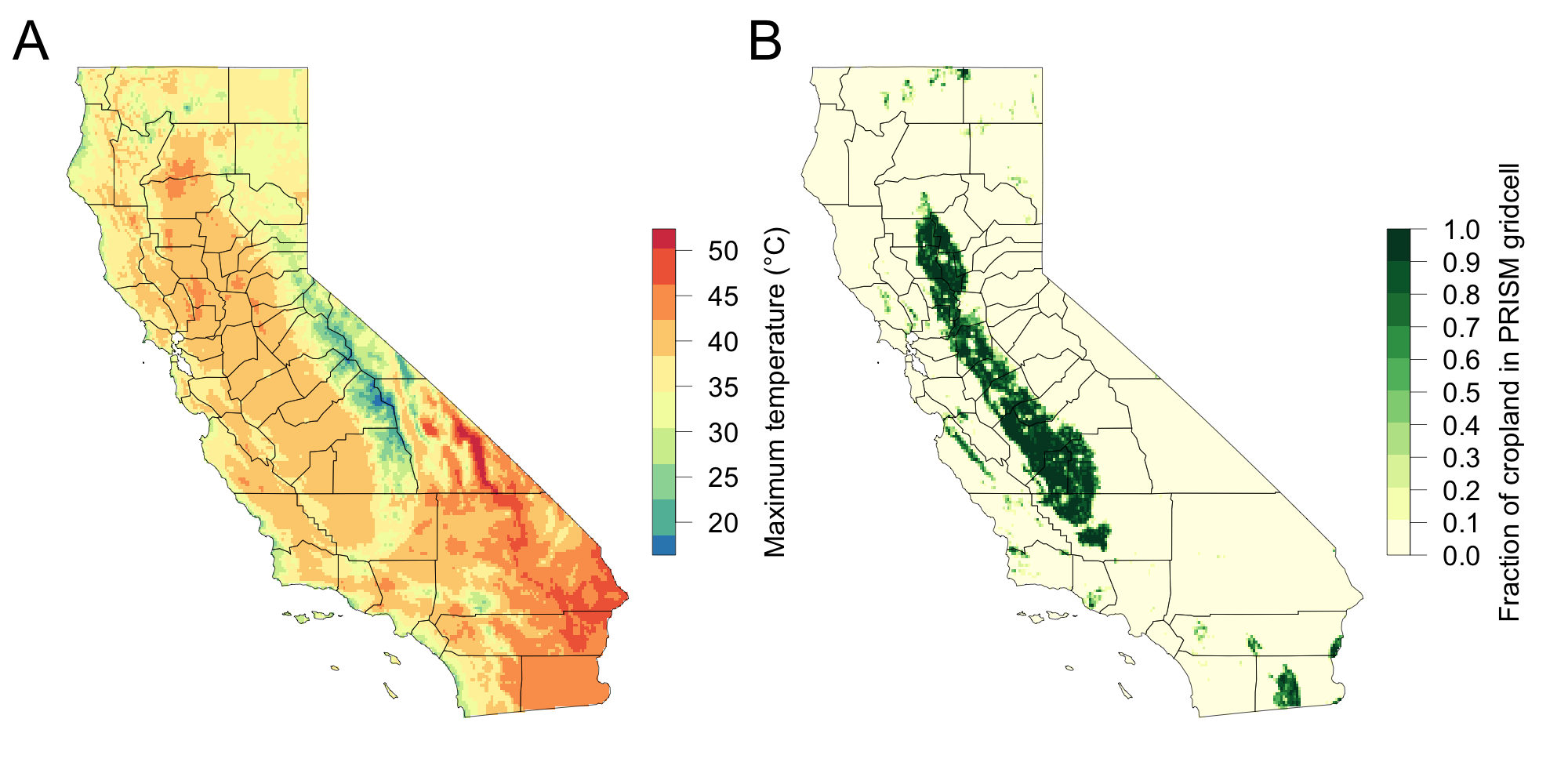}

\noindent\begin{minipage}[t]{1\columnwidth}%
\textit{\footnotesize{}Notes:}{\footnotesize{} Panel A shows maximum
temperature corresponding to August 16 of 2020 over California. This
is gridded daily data from PRISM. Panel B shows the share of cropland
contained in each of the PRISM grid cells. The grid cell share was
derived from finer scale 30m land cover data for 2016 from the National
Land Cover Database (NLCD).}%
\end{minipage}

\caption{Gridded data and cropland land cover in California. \label{fig:raster-aggregation}}
\end{figure}

To illustrate this point, Fig. \ref{fig:raster-aggregation}A shows
maximum temperature in California on August 16, 2020 when possibly
the highest temperature ever recorded on Earth (54.4 °C) was measured
in the Death Valley (darkest shade of red). This daily gridded data
is from PRISM and shows wildly varying weather conditions across the
state on the very same day. However, agriculture is mostly concentrated
in the Central Valley region. This can be seen in Fig. \ref{fig:raster-aggregation}B
showing the share of cropland within each PRISM grid cell. A common
practice is to aggregate the weather (panel A) variable within each
county (or within the state) based on weights proportional to the
cropland cover (panel B). The large climatic variations within California
illustrate how critical land cover information can be for representing
environmental conditions within administrative units.

Intuitively, spatial weighting procedures should make little difference
if we are located in a relative small or homogeneous administrative
units. However, potentially substantial differences could arise between
different weighting schemes in the presence of climatically diverse
units. Some have suggested that the weighting should be performed
by value (rather than land cover) although it seems unclear how livestock
would be accounted for in such circumstances. One way to think about
issues about spatial aggregation is to frame it in terms of measurement
error. It turns out that implications of these practices have not
been chracterized. Researchers often end up showing regression results
under alternative weighting schemes to assuage concerns during the
peer review process. This practice seems suboptimal and more systematic
analysis of the consequences of various strategies of spatial data
aggregation are needed.

There are also issues regarding temporal aggregation in weather data.
Aggregating weather data over time can also result in measurement
error if the weather conditions are non-additive and if non-linearities
in exposure to various levels of weather conditions are important,
which they likely are. 

\subsection{Climate models\label{subsec:Climate-models}}

Climate scientists have developed Global Circulation Models (GCMs)
to simulate the evolution of the climate system. These models are
fundamentally similarly to numerical weather models used in weather
forecasting, but incorporate a more complete representation of energy
exchanges between land, oceans, sea ice and the many layers of the
atmosphere. Central to these models are the Navier-Stokes equations,
partial differential equations that describe the movement of viscous
fluids. Solving these equations to describe moving air masses in three
dimensions requires considerable computational power, which is why
running these models requires super computers. Major countries have
research groups and labs with their own version of these models. 

In order to learn more about factors affecting our climate system,
modeling groups have joined a global inter-comparison project called
the Coupled Model Inter-comparison Project (CMIP). Four of such inter-comparisons
have been completed (CMIP Phases 1, 2, 3 and 5) and there is one under
way (CMIP Phase 6 or CMIP6). The key feature of CMIP is the parallel
implementation of identical climate experiments across a wide range
of GCMs. Some of these experiments are designed to learn about specific
aspects of these models, so that GCMs can be improved. However, some
of these climate experiments are much more policy relevant and seek
to understand how the global climate system is influenced by anthropogenic
influences.

Various experiments seem particularly policy relevant and are commonly
used by economists. Using CMIP6 terminology, these include the Shared
Socioeconomic Pathways (SSPs, see \citealp{riahi_shared_2017}), including
SSP1-2.6, SSP2-2.5, SSP3-7.0 and SSP5-8.5. The SSPs correspond to
different scenarios about the nature economic development and the
pathway of climate forcing (e.g. emissions) throughout the century.
The appended numbers to these scenarios (i.e. 2.6, 4.5, 7.0 and 8.5)
represent the additional radiative forcing on our climate system in
Watts/m$^{2}$ in the year 2100. The higher the number, the higher
the additional radiative forcing, and the higher global temperatures
are projected to rise. These scenarios are analogous to the Representative
Concentration Pathways (RCPs) scenarios used in CMIP5, and the Special
Report on Emissions Scenarios (SRES) used in CMIP3. Researchers can
relate these projected future states of the atmosphere under various
GCMs and SSPs with the ``historical'' experiment that seeks to replicate
the conditions of our historical climate system from the nineteen
century through 2015 in the case of CMIP6. The ``historical'' experiment
considers both historical levels of both natural (e.g. volcanic eruptions
from el El Chichón in 1982 and Pinatubo in 1991) and anthropogenic
forcing (e.g. greenhouse gas emissions since the industrial revolution).

Other relevant climate experiments that are relatively underused by
economists are the ``historicalNat'' in CMIP5 and ``hist-nat''
in CMIP6. These experiments run a historical simulation but only with
natural forcing. That is, the output of these experiments provide
a counterfactual sequence of modeled weather trajectories that exclude
human influence from the climate system. In climate science, the comparison
of these experiments and the ``historical'' experiment is a foundation
of attribution studies that seek to establish to what extent extreme
weather events (e.g. heat waves) are likely to arise because of human
influences, and not because of natural variability. Some studies have
used this approach to analyze the historical impact of anthropogenic
climate change on agriculture (See section \ref{subsec:Retrospective-studies}).

It should be noted that output from climate models is gridded in nature
(raster format) and can be manipulated in the same way than gridded
weather datasets previously discussed. 

Finally, it is worth noting that the CMIPs serve as the basis of the
Assessment Reports (AR) for the first working group (WGI) of the Intergovernmental
Panel of Climate Change (IPCC) charged with describing the physical
basis of the factors affecting our climate system. In fact, the name
of the CMIP and the AR are in phase, so that the lessons from CMIP6
feed into the the Sixth Assessment Report of AR6. The ARs serve as
an input for international negotiations regarding adaptation and mitigation
of anthropogenic climate change within the United Nations Framework
Convention on Climate Change (UNFCCC).

\subsection{Degree-days and agriculture\label{subsec:Degree-days-and-agriculture}}

\begin{onehalfspace}
A growing number of studies rely on variables representing ``degree
days'', ``growing degree days'' (GDD), ``damaging degree days''
(DDD), ``extreme degree days'' (EDD) or ``killing degree days (KDD)
for analyzing the effect of cumulative temperature exposure on agricultural
production. Although the growing popularity of degree-day measures
seems relatively recent in agricultural economic research, the concept
has roots that are centuries-old \citep{reaumur_observations_1735}.
Here I provide a brief background on the concept and how to compute
these variables.

A degree-day is one of the many units of measurement of thermal time.
Thermal time is a physical quantity measured in units of temperature
$\times$ time. This concept is very familiar to scientists studying
phenology, which is the study of how the periodicity of biological
cycles are influenced by their environment. The concept emerged as
an heuristic tool to predict the \emph{length} of the different phases
of plant life cycles \citep{reaumur_observations_1735}. Some measures
of thermal time are highly correlated with the timing of numerous
phenological events, or cyclical natural phenomena, such as insect
or plant development. In plants, such development phases are typically
signaled by the appearance of new and differentiated organs such as
the emergence of subsequent leafs and flowers or the formation of
fruit. French scientist René-Antoine Ferchault de Réaumur laid the
foundations of thermal time in the eighteenth century as recalled
by \citet{wang_critique_1960}: \emph{``He summed up the mean daily
air temperatures for 91 days during the months of April, May and June
in his locality and found the sum to be a nearly constant value for
the development of any plant from year to year.''} Thermal time subsequently
became a pivotal concept in phenology (\citealp{hudson_phenological_2009},
ch. 1).

Indeed, the timing of many biological cycles, particularly in plants
and insects, is closely correlated with thermal time accumulation.
For this reason, biologists coined a term that measures thermal time
accumulation, \emph{Growing Degree-Days} (GDD). This biological thermal
time corresponds to cumulative temperature during a period of time.
Intuitively, it is an amount of accumulated exposure to heat.\footnote{Although strictly speaking this is inexact because ``heat'', in
thermodynamics, is a form of energy which is measured in Joules, not
in temperature units. There is, however, a direct relationship between
temperature change of a body and the heat it receives.} This quantity is often defined mathematically. As shown in equation
\ref{eq:TT general}, thermal time $T$ is typically expressed as
a function of two temperature thresholds, $\underline{h}$ and $\overline{h}$
and two points in time, $t_{0}$ and $t_{1}$. 

\begin{equation}
T(\overline{h},\underline{h},t_{0},t_{1})=\intop_{t_{0}}^{t_{1}}H(t)dt\qquad\qquad\textrm{where }H(t)=\begin{cases}
\overline{h}-\underline{h} & \textrm{if }h(t)>\overline{h}\\
h(t)-\underline{h} & \textrm{if }h(t)\in\:]\underline{h};\overline{h}]\\
0 & \textrm{if }h(t)\leqslant\underline{h}
\end{cases}\label{eq:TT general}
\end{equation}

\end{onehalfspace}

This definition implies that any measurement of thermal time is a
function of the two temperature thresholds and a duration. In fact,
the two temperature thresholds, $\underline{h}$ and $\overline{h}$,
are experimentally determined to yield a nearly linear relationship
between thermal time accumulation and biological development.\footnote{\begin{onehalfspace}
For instance, for the early-maturing dwarf hybrid sunflower plant
(Pioneer 6150) with $\underline{h}$ defined as 0°C (undefined $\overline{h}$),
the first two true leaves appear after exposure to 249-313 GDD, flowering
begins at 935-1077 GDD, and maturity is achieved after 1780-1972 GDD.
\end{onehalfspace}
} 

The link between temperature and the timing of biological cycles can
be explained at the molecular level within cells. Air temperature,
at the most basic level, affects cellular function, and animals and
plants have developed strategies to take advantage of climatic conditions
conducive to their growth and largely avoid conditions that are harmful.
In the case of crops, when air temperature is below the lower threshold,
$\underline{h}$, also referred to as the crop-specific ``base temperature,''
or above the higher threshold, $\overline{h}$, crop development essentially
stops \citep{ritchie_temperature_1991,mcmaster_growing_1997}. Enzymes,
which are proteins that accelerate biochemical reactions within cells,
become too rigid at low temperatures and coagulate at very high temperatures,
leading to slow or entirely inhibited crop growth \citep{bonhomme_bases_2000}.
In other words, because plants cannot regulate much their own temperature,
their metabolism is subject to outside temperature, which affects
the speed of biochemical reactions.\footnote{Organisms unable to actively regulate inner temperature to favorable
levels are coined ``ectotherms''. This is not only the case of plants,
but also of ``cold blooded'' animals such as insects, reptiles,
etc.}

\begin{onehalfspace}
This relationship is not perfectly linear because the timing of these
development stages are also dependent on adequate light, water, and
nutrients in addition to appropriate temperature. However, temperature
remains the major factor explaining the timing of development stages.
This explains why agronomists and farmers use GDDs to estimate stages
of crop development and weed and pest life cycles. 
\end{onehalfspace}

Note that variables that measure the amount of time exposed to certain
temperature levels are closely related to the concept of degree days.
Degree day are measures of cumulative temperature exposure between
two temperature thresholds. The concept is very general and is used
outside of agronomic sciences. For instance, engineers and energy
analysts rely on Cooling Degree Days (CDD) and Heating Degree Days
(HDD) to predict energy consumption for cooling in the summer, and
for heating in the winter, respectively. 

Possibly the oldest use of degree days was to predict phenology, the
timing of life stages in plants and certain animals \citep{reaumur_observations_1735}.
The modern incarnation of this concept applied to plants is reflected
in the modern use of Growing Degree-Days (GDD) which are precisely
reserved to predict crop stages \citep{bonhomme_bases_2000}. Field
crops are characterized by their crop maturity rating which indicates
the amount of ``cumulative temperature'' measured in GDD to reach
maturity. Short-season cultivars require less heat of the growing
season to reach maturity. For this reason such cultivars are used
in colder climate in temperature countries like the US, where the
non-freezing period that is fatal to most crops is relatively short.
Thus, farmers choose a crop maturity rating based on their local climate. 

The implication is that once farmers plant a given cultivar, unexpectedly
warm or cold conditions can accelerate or delay the timing to reach
crop maturity and harvest. As a result, changes in GDD can affect
yield, because shortening the time the crop spends on the field shortens
the time the crop has to accumulate biomass. Similarly, lengthening
the timing the crop spends on the field can expose the crop to perilous
conditions at the end of the season (e.g. damaging Fall frost). However,
the concept of GDD is not intended to predict yield, but to predict
phenology. Many economic studies rely on ``GDD'' measures to predict
yield. This is incorrect. A more rigorous use of the term that does
not introduce confusion with its agronomic use is to simply describe
degree day variables in terms of the range of temperature used to
defined them, like ``degree days 8-30°C''. Note that degree days
capturing exposure to relatively high thresholds, say 30°C, are commonly
referred to as Extreme Degree Days (EDD), Killing Degree Days (KDD)
or Damaging Degrees Days (DDD). Similarly, describing degree days
by their threshold, such as ``degree days over 30°C'' seems generally
more appropriate.

\section{Climate change impacts and adaptation\label{sec:Climate-change-impacts}}

This section provides an overview of the main research questions and
the methods used to evaluate climate change impacts and adaptation
in the agricultural sector. While the discussion will cover developments
over the past 2 decades, I put some emphasis on the evolution of the
literature as well as recent contributions. I also spend some time
discussing unresolved or relatively unexplored research questions.
I also invite the reader to consult several overview articles on methods
of assessing climate change impacts on agriculture and other sectors,
including \citet{blanc_use_2017}, \citet{carter_identifying_2018}
and \citet{kolstad_estimating_2020}, to name just a few.

\subsection{Biophysical approaches\label{subsec:Biophysical-approaches}}

Most of the early work assessing climate change impacts on agriculture
is fundamentally based on biophysical models (e.g. \citealt{adams_global_1989,adams_global_1990,rosenzweig_potential_1994}).
These plant science models mechanistically characterize the effect
of environmental conditions (e.g. sunlight, water availability, air
and soil temperature, carbon dioxide concentrations, air humidity,
etc.) on the physiological processes that underly crop yield formation.
This type of approach typically assume farmers adopt a series of more
or less sophisticated management practices including choice of cultivar,
fertilization decisions, planting time, etc. These models are subsequently
coupled with climate models and supply-demand economic models to simulate
the effect of climate change on agricultural production and welfare.
I revisit the integration with economic models in subsection \ref{subsec:Trade-and-general}
on market equilibrium and trade. 

One of the main advantages of biophysical approaches is that they
can provide a transparent understanding of the exact channels through
which climate change impacts occur. For instance, these approaches
allow unpacking the role of CO$_{2}$ fertilization as well as which
crops and regions of the world would be more affected. Relying on
a supply and demand model also allows to compute welfare effects of
climate change and how they are distributed among consumers and producers
in various regions of the world.

However, these approaches present various limitations. Because these
models are not directly rooted on observational data, it is unclear
whether the assumptions about farmer behavior may be realistic in
real-world settings. These approaches are also deterministic, so they
don't directly provide a measure of uncertainty regarding the relationship
between changes in the distribution of weather conditions and agricultural
outcomes. In general, these model need to be extensively and carefully
calibrated to perform well within the sample and tend to perform poorly
when used out of sample. In fact, a key criticism of this approach
is that results tend to be highly dependent on the crop model used,
which has led some call for an overhaul of modeling approaches that
favor multi-model ensembles \citep{rotter_cropclimate_2011}. 

An important development in this literature the Agricultural Model
Intercomparison and Improvement Project (AgMIP, \url{https://agmip.org}),
which aims to improve biophysical crop modeling \citep{rosenzweig_assessing_2014}.
Similarly to CMIP models, intercomparison projects allow modelers
to compare model outputs based on identical scenarios and to learn
about the sources of discrepancies across individual models. The consensus
seems to be that the future of biophysical crop modeling resides with
multi-model ensembles. 

Moving to multi-model approaches not only facilitates model improvements,
but also allows researchers to sample from a wider range of crop models
when conducting climate change impact analyses. This helps better
characterize model-driven uncertainty of climate change impact projections.

Although the development to multi-model ensembles seems welcome, it
also means that research projects in this area involves relatively
large pre-established teams which can present a barrier to entry for
individual researchers, especially students. However, there seems
to be an important role for economists to play in the coupling of
these multi-model ensembles with supply-demand and trade models.

Finally, a present limitation of biophysical approaches is the major
emphasis on the major staple field crops such as wheat, corn and rice.
Cereal crops represent about a fifth of the total agricultural value
produced so these approaches have so far overlooked  the effects on
many other parts of the global agricultural sector. There is a limited
number of models focused on specialty crops or livestock production,
which seem like important directions of research. But then again,
models are likely to do well within the region of calibration and
pose limitations when trying to apply these models to other regional
contexts. See \citet{antle_climate_2017} for a review of the use
of process-based models along with economic models.

\subsection{The Ricardian approach\label{subsec:The-Ricardian-approach}}

The introduction of the Ricardian approach in \citeauthor{mendelsohn_impact_1994}
(1994) was a reaction to earlier studies based on biophysical approaches
that allowed for relatively little farmer adaptation to climate change
(e.g. \citealp{adams_global_1989,adams_global_1990,easterling_simulations_1992-1,kaiser_farm-level_1993,adams_reassessment_1995}).
In retrospect, these studies unsurprisingly tended to point to relatively
large damages primarily driven by losses in crop yields.

The idea behind the Ricardian approach is that one can assess future
climate change impacts capturing for the full range of farmer adaptations
without having to model these adaptive choices explicitly. The key
conceptual assumption is that farmers are already adapted to their
local climate. That is, they would have adopted production practices
and choices that are the most beneficial given the local climate,
prices and technology. Because land is a fixed factor of production,
the demand for land generates economic rents. The discounted stream
of these rents are capitalized in the value of land. So if these rents
originate from more beneficial climatic conditions, then climate would
be capitalized in the value of land.

Empirically, the Ricardian approach attempts to recover the marginal
value of climate by exploiting the cross-sectional spatial variation
in farmland values and climate across a large region. This constitutes
a hedonic analysis of the characteristics of land \citep{rosen_hedonic_1974,palmquist_land_1989}.
The regression can be expressed as:

\begin{equation}
y_{it}=\bar{Z}_{it}\beta+X_{it}\gamma+\alpha_{t}+\epsilon_{it}
\end{equation}

where $y_{it}$ is farmland value per acre in location $i$ (e.g.
county or district) and year $t$, $\bar{Z}_{it}$ is a vector of
climate variables defined over the previous 30 years ($\bar{Z}_{it}=\sum_{s=t-30}^{t-1}Z_{is}/30$),
$X_{it}$ is a vector of control variables, $\alpha_{t}$ is a year
fixed effect and $\epsilon_{it}$ is an error term. The hope in this
analysis is that the inclusion of control variables will reduce concerns
regarding omitted variable bias and lead to unbiased estimates of
$\beta$. 

The researcher can then couple these hedonic estimates of the marginal
value of climate on farmland values $\hat{\beta}$ with climate change
projections $\Delta\bar{Z}_{i}$ to derive climate change impacts
$\Delta\hat{y}_{i}=\Delta Z_{i}\hat{\beta}$. In principle, these
impacts account for the full range of farmer adaptations.

The first implementation of the approach (i.e. \citealp{mendelsohn_impact_1994})
was in the context of US agriculture and relied on county-level farmland
values from the 1978 and 1982 Census of Agriculture. The main specification
estimated these 2 cross-sections separately and regressed farmland
value per acre on linear and quadratic terms of seasonal (January,
April, July, October) temperature and precipitation along with a series
of physical (e.g. average soil characteristics) and economic controls
(e.g. population density and income per capita). 

The most striking finding at the time was that applying a uniform
warming of 5°F warming and a 8\% increase in precipitation, an approximation
of early IPCC projections, suggested slightly beneficial impacts for
US agriculture. However, the results were substantially different
depending on the regression weights used in the analysis. As indicated
by \citet{solon_what_2015}, when regression coefficients differ dramatically
with different regression weights, it may be a sign of misspecification
or un-modeled heterogeneity. The results were nonetheless in stark
contrast to previous work.

The approach and the implementation in \citet{mendelsohn_impact_1994}
generated considerable criticism (see \citealt{cline_impact_1996,kaufmann_impact_1998,darwin_impact_1999,quiggin_impact_1999}).
As summarized in \citet{schlenker_will_2005}, the main criticisms
included (1) that the Ricardian approach does not account for adjustments
costs, (2) that the regression results were not stable across regression
weighting schemes, and (3) the inappropriate treatment of irrigation.
This last point probably gained the most traction. Farmers in certain
regions can rely on a water supply for irrigation instead of directly
from precipitation. As a result, the shadow value of climate should
in principle differ across irrigated and non-irrigated regions. Econometrically,
that means that the researcher should estimate separate coefficients
for irrigated and non-irrigated areas. A simple dummy for irrigation
would not suffice as that simply alters the intercept.

This precise idea was proposed in \citet{schlenker_will_2005} which
showed that when the MNS model is restricted to the mostly non-irrigated
Eastern half of the US, the Ricardian model points to large negative
damages, rather than benefits. Results also become stable across regression
weights. In a related study, \citet{schlenker_impact_2006} proposed
a new set of climate variables including degree days variables commonly
used for predicting crop phenology and plant biomass growth. The main
results mirrored the findings of \citet{schlenker_will_2005}. Interestingly,
the results in \citet{schlenker_impact_2006} were robust to the inclusion
of state fixed effects. This is striking because it means that warmer
areas within states, tend to exhibit lower farmland values even after
controlling for land quality characteristics and other economic controls
such as population density and income per capita.

Various improvements to the Ricardian approach have been introduced
over time. \citet{timmins_endogenous_2005} notes that Ricardian estimates
may be biased when land is heterogeneous within locations (e.g. counties)
and land owners allocate land use optimally. This problem arises due
to spatial or administrative aggregation of data of parcels under
different land use and with differing shadow values of climate. \citet{fezzi_impact_2015}
also point out issues related to the common issue of spatial aggregation
in Ricardian models. Using a detailed database of land values for
Great Britain, they find that aggregation conceal important interactions
between temperature and precipitation. This results in substantial
biases in projection of climate change impacts. This is particularly
problematic for this literature given that farmland value data is
often only available at aggregate scales for privacy reasons.

Farmland values reflect expectations about future land rents. \citet{severen_forward-looking_2018}
point out that if land market players expect climate change to affect
future rents, then those expectations should be capitalized as well,
biasing Ricardian estimates. They test whether climate change projections
(based on two GCMs) appear to be already capitalized in US farmland
markets using a cross-sectional approach. The paper subsequently proposes
a corrected ``forward-looking'' Ricardian approach that addresses
this potential bias. One potential shortfall of this implementation
is that the test is based on cross-sectional evidence, so it is unclear
if climate change projections are correlated with unobservable determinants
of land values. An alternative approach would be to track changes
in farmland values over time as more information about climate projections
is made public. Another challenge here is that the US public views
regarding the existence and origin of recent climate change are highly
polarized between urban and rural areas across the US \citep{leiserowitz_climate_2013}.
Thus the regional divided in these views could be correlated with
the extent of non-farm pressures on farmland markets.

More recently, \citet{ortiz-bobea_role_2019} revisited the Ricardian
analysis of US farmland values and found that large damages found
in previous studies (e.g. \citealp{schlenker_will_2005,schlenker_impact_2006})
appear to be driven by factors outside the agricultural sector. Essentially,
large climate change damage estimates in recent farmland values cross-section
appear driven by non-farm omitted variables. Using a century of farmland
value data, the study finds that climate change impact estimates are
statistically insignificant when relying on older farmland value cross-sections.
The study finds that this result stems from major changes in the farmland
value cross-section over time. A convergence of evidence suggest such
changes in the cross-section are linked to the rise of non-farm pressures
which are correlated with climate within states (e.g. rise in recreational
demand for land in cooler areas of certain states). These changes
in farmland values appear unrelated to technological change or other
forces within the agricultural sector. To circumvent biases from the
capitalization of non-farm pressures, the study proposes a Ricardian
model based on farmland rental prices, rather than farmland asset
values. This approach was employed in \citet{hendricks_potential_2018}
that conducts a Ricardian analysis of cropland rental prices in the
central US to assess the potential gains from innovations that reduce
heat and water stress.

The Ricardian approach continues to be used across many contexts.
Perhaps its most attractive features are its simplicity and the conceptual
elegance of how it resolved previous thorny debates about capturing
adaptation. The approach has important drawbacks that seem difficult
to overcome. Despite efforts to make the approach more ``structural''
\citep{seo_measuring_2008}, it still remains a bit of a ``black
box'' regarding underlying mechanisms that are important for policy
or developing priorities for adaptation.

Perhaps more importantly, empirical economics is well engaged in a
``credibility revolution'' \citep{angrist_credibility_2010} where
much attention is given to the quality and credibility of research
designs. The Ricardian approach is based on an empirical strategy
that is fundamentally vulnerable to omitted variables. The identifying
variation in the approach comes from the spatial variation in land
values and climate. If unobserved drivers of land values, of which
there are many, are omitted and happen to be correlated with climate,
which is plausible, then estimates are biased. Previously proposed
fixes, like the inclusion of state fixed effects do not fundamentally
address this issue as shown in \citet{ortiz-bobea_role_2019}. The
inclusion of state dummies that makes the estimation be based on the
within-state variation in prices and climate, could actually amplify
omitted-variable bias if those operate more strongly within states
than across states. Current research standards clearly favor research
designs based on longitudinal data in which researchers can convincingly
control for time invariant confounders.

Still, a few attempts have been made to control for unknown omitted
variables in a cross-sectional setting. That includes the introduction
of the Spatial Durbin Error Model \citep{lesage_introduction_2009,elhorst_applied_2010}
in \citet{ortiz-bobea_economic_2016} or that of a ``Spatial First
Differences'' estimator in \citet{druckenmiller_accounting_2018}.
The underlying idea of these approaches is that unobservables may
be spatially dependent in a way that slightly differs from that of
climate. These approach harness this information to control for or
subtract out the influence of these confounders.

\subsection{Panel profit and productivity approaches\label{subsec:Panel-profit-and}}

The use of longitudinal data to analyze the effect of weather conditions
on agricultural outcomes has a very long tradition in agricultural
economics \citep{hodges_effect_1931,schickele_farm_1949,stallings_weather_1960,morgan_use_1961-1,stallings_measure_1961,shaw_effect_1964,oury_allowing_1965,black_evidence_1978-1}.
This early literature was not initially concerned with climate change,
but with forecasting crop production to anticipate price swings, and
understanding the nature of agricultural production risk. It was only
until the late 1980s (e.g. \citealp{adams_global_1989}) that articles
exploring empirically the implications of climate change on agriculture
started to appear in leading agricultural economics journals like
the \textit{American Journal of Agricultural Economics} (AJAE). However,
these initial studies were fundamentally based on biophysical crop
models.

The surge in interest in econometric panel approaches to analyze potential
climate change impacts on the agricultural sector can probably be
traced back to \citeauthor{deschenes_climate_2007} (2007), which
offered a counter narrative to the negative findings in \citet{schlenker_will_2005}
based on the Ricardian approach. See \citet{blanc_use_2017} for a
review on the use of panel models in assessing climate change impacts
on agriculture.

The approach in \citet{deschenes_climate_2007} was to rely on presumably
random year-to-year fluctuations in weather conditions to explain
variations in profits within US counties. This approach allows controlling
for time-invariant omitted variables that may be correlated with climate
and may plague the Ricardian approach. Their model can be expressed
as:

\begin{equation}
y_{it}=Z_{it}\beta+X_{it}\gamma+\alpha_{i}+\alpha_{st}+\epsilon_{it}
\end{equation}

where $y_{it}$ is net revenue or ``profit'' per acre in county
$i$ and year $t$, $Z_{it}$ is a vector of weather variables (e.g.
precipitation and temperature), $X_{it}$ is a vector of time-varying
control variables, $\alpha_{i}$ is a county fixed effect, $\alpha_{st}$
is a state-by-year fixed effect and $\epsilon_{it}$ is an error term.

Conceptually, the model estimates a short-run effect of climate change
on profits. The idea is that farmers cannot adjust many input decisions
to unanticipated changes in weather. As a result the effect of weather
on profits would be appear more detrimental than it would be if the
farmer were able to fully adjust inputs in the long run. Because the
study found very small effects of climate change on profits in their
implementation, their interpretation is that climate change would
have beneficial effects on US agriculture with the current technology.

As pointed out in \citet{fisher_economic_2012}, the implementation
in \citet{deschenes_climate_2007} suffered from a series of shortcomings.
Among the main issues were problems related to weather data quality
which tend to attenuate the magnitude of the $\beta$ coefficients.
Another concern is the inclusion of state-by-year fixed effects $\alpha_{st}$
which arguably ``wipe out'' a lot of the weather variation used
to identify effects on profits. The main issue is that with so little
variation within state-year, estimates might be imprecise and potentially
more vulnerable to biases associated with measurement error (see \citealp{griliches_errors_1986}).

A conceptual limitation of the \citet{deschenes_climate_2007} implementation
is the nature of the outcome variable. The study constructs a ``profit''
variable from the US Census of Agriculture by subtracting total county
expenses from total sales divided by the acres of land in farms. Technically
this is a net revenue variable that can be problematic in the analysis.
The potential issue relates to the role of inventories, which farmers
use in a countercyclical manner to smooth the effect of unusual economic
and weather conditions. That is, on a bad year, a farmer may sell
more than they produced, whereas she might sell less than what is
produced on a good year. Similarly, farmers may avoid large purchases
(e.g. tractor) in bad years whereas they might go forward with them
in good years. This behavior will tend to bias contemporaneous weather
effects toward zero. One potential fix to this issue is to estimate
a distributed lag model to account for weather shocks in previous
years (see \citealp{deschenes_economic_2012}). 

When some of the concerns above are addressed (see \citealp{fisher_economic_2012}
and \citealp{deschenes_economic_2012}), the projected impacts of
climate change on profits are negative. It would seem as if the degree
to which these effects are large or small lies on the eye of the beholder,
particularly on how these short-run effects translate to long-run
effects. Indeed, this panel approach only accounts for farmer adjustments
that could happen within the year (or a couple of years with a distributed
lag model). 

More recently, a few studies have relied on measures of Total Factor
Productivity (TFP) rather than on net revenue variables. A key advantage
is that TFP data, when well constructed, accounts for changes in inventory.
For instance, the Economic Research Service (ERS) within the US Department
of Agriculture (USDA) develops a high quality US state-level TFP panel
dataset that can be harnessed for climate-related research. While
the spatial resolution is coarser than US Census data (state versus
county), the greater temporal resolution (annual) is particularly
useful given that identification is based on the within-location variation
in weather. Examples of such studies include \citet{liang_determining_2017},
\citet{ortiz-bobea_growing_2018} and \citet{ortiz-bobea_anthropogenic_2021}.

Following the framework proposed in \citet{ortiz-bobea_anthropogenic_2021},
consider an aggregate production function of the form $Y_{it}=e^{f(Z_{it})}A_{it}X_{it}U_{it}$
where $Y_{it}$ is aggregate agricultural output in state $i$ and
year $t$, $e^{f(Z_{it})}$ is the effect of weather, $A_{it}$ is
a neutral productivity factor, $X_{it}$ is a measured aggregate input
and $U_{it}$ is an unmeasured aggregate input. Taking logs, first
differences and rearranging yields the following expression that approximate
the growth rate of TFP:

\begin{equation}
\Delta\ln TFP_{it}\equiv\Delta\ln Y_{it}-\Delta\ln X_{it}=\Delta\ln A_{it}+\Delta f(Z_{it})\beta+\Delta\ln U_{it}
\end{equation}

where $\Delta$ denotes change. By definition, TFP growth is the growth
in aggregate output that cannot be explained by growth in measured
input. In the proposed model, that TFP growth is the sum of three
factors, a technological improvement reflected in $\Delta\ln A_{it}$,
a weather effect embedded in $\Delta f(Z_{it})$ and changes in unobserved
inputs $\Delta\ln U_{it}$. The key point is that one can harness
fluctuations in TFP to capture and characterize the effect of weather
fluctuations on agricultural production, net of input responses.

Using this approach, \citet{ortiz-bobea_growing_2018} analyzes the
effect of weather conditions on US agricultural TFP. The analysis
is conducted separately by Climate Hub regions which are agro-climatically
coherent regions proposed by USDA for adaptation planning. The key
finding in the study is that Midwest agriculture is growing increasingly
sensitive to higher temperatures. This is found by conducting a Wald
test of stability of regression coefficients between the two halves
of the sample. The study also provides some ideas of the drivers.
The trend appears related to two compounding factors, the growing
sensitivity of the crop output to higher temperatures, and the increasing
specialization in crop production in Midwest. 

The rising sensitivity to high temperature in Midwest agriculture
could appear as a form of ``maladaptation'' but it remains unclear
whether this rising sensitivity results from a desirable tradeoff
with higher productivity. More research is needed to better understand
these trends, especially regarding the timing of these vulnerability
changes over time. 

\subsection{Statistical crop yield models\label{subsec:Statistical-crop-yield}}

A statistical crop yield model typically regresses a panel of crop
yields on various weather variables. These models had traditionally
been the focus of agricultural meteorologists. See \citet{decker_developments_1994}
for a history of that field. However, these models have grown increasingly
popular among agricultural economists interested in exploring a specific
sub-channel through which climate could affect agricultural production
(e.g. \citealp{tack_effect_2015,ortiz-bobea_is_2018,shew_yield_2020}).

The key strength of these models relative to biophysical approaches
is that they are grounded on observational data. This has several
advantages. First, these models account for actual farmer management
decisions when the data is based on crop yields collected from farmers.
Because farmers can sometimes respond to weather fluctuations within
the growing season, results can be interpreted as reflecting short
run adaptation to weather fluctuations. This interpretation is not
really correct when crop yield data are based on field trials in which
generally management decision are pre-determined by a researcher.

Statistical crop yield models have also been used to indirectly detect
whether farmer adaptation to climate change has already occurred.
Two approaches have been proposed. The first is to test whether crop
yields are growing less sensitive to extreme weather that are becoming
more common under climate change (e.g. high temperatures). This is
typically achieved by testing for stability of regression coefficients
over time, which naturally requires relatively long panels. But it
remains ultimately unclear whether changing coefficients over time
result from changes in crop cultivars, changing inputs or management
techniques, which constitute some form of adaptation, or from a statistical
artifact such as changing weather data quality over time. More emphasis
should be given to teasing out the sources of these changes, ideally
by coupling statistical models with other production information (e.g.
crop cultivars, data on management, etc.).

The second approach of indirectly detecting farmer adaptation is to
test for regional heterogeneity of weather coefficients. For instance,
one should expect that farmers in region with greater exposure to
high temperature to adapt to such environmental conditions over time.
Econometrically, this means that the effect of high temperature should
appear to be less detrimental in warmer places than in cooler places.
So this check consists on testing for an interaction between a weather
variable and its climatology. This is the approach adopted by \citet{butler_adaptation_2013}
to analyze adaptation to high temperature in US corn yields. However,
this climate is likely correlated with other factors that may affect
crop yield sensitivity (e.g. soil quality) so this approach remains
vulnerable to time-invariant omitted variables that interact with
weather fluctuations.

One important shortcoming of many studies in this area is that they
do not inform us about the types of adaptations or inputs adjustments
farmers may be engaging. Importantly, the fact that inputs are typically
unobserved makes the interpretation of short run effects more difficult.
For instance, when a farmer increases irrigation intensity in response
to higher temperatures or lack of precipitation means that the effect
of these undesirable weather conditions will appear attenuated. In
the other hand, if inputs are complementary to weather conditions,
farmers may cut labor and fertilization in response to undesirable
environmental conditions, which would further reduce crop yields.
Thus, depending on the nature of the input response, the estimate
yield effect may appear either exacerbated or attenuated. It is critical
that researchers acknowledge that farmer input decision are often
times correlated with weather fluctuations, and that results should
be interpreted accordingly.

Another important area of debate in this literature is the nature
of the weather variables and how they are modeled. A fundamental challenge
here is the mixed frequency of the data with high-frequency daily
weather predictors throughout the growing season affecting the ultimate
crop yield at harvest \citep{ghanem_what_2020}. Unlike for the biophysical
models, statistical models cannot directly incorporate daily weather
conditions. This would represent too many regressors which happen
to be highly correlated between neighboring days. As a result, researchers
undertake a variable selection more or less based on first principles
(e.g. what conditions are known to be important for crop yield determination)
and data-driven criteria (e.g. in-sample or out-of-sample measures
of model fit or other criteria).

Precipitation and temperature are naturally considered two fundamental
climatic variables affecting crop production. Crops need water to
grow so precipitation is critical in rain-fed systems. Too little
or too much precipitation is presumably detrimental to crop yields,
suggesting there is an optimal level of precipitation. Similarly,
very cold condition are detrimental to crop growth and very hot conditions
accelerate evapotranspiration and can cause heat stress, so extreme
temperature are presumed to be detrimental. This also suggest the
existence of an optimal temperature range. As a result, perhaps the
most basic crop yield statistical model takes the following form:

\[
y_{it}=\beta_{1}T_{it}+\beta_{2}T_{it}^{2}+\beta_{3}P_{it}+\beta_{4}P_{it}^{2}+\phi_{s}(t)+\alpha_{i}+\epsilon_{it}
\]

where $y_{it}$ is crop yield (or its logarithm) in location $i$
and year $t$, $T_{it}$ stands for growing-season average temperature,
$P_{it}$ represents growing-season precipitation, $\phi_{s}(t)$
represents a regional time trend (typically a $year$ variable interacted
with a regional dummy), $\alpha_{i}$ is a location fixed effect,
and $\epsilon_{it}$ is an error term. The introduction of the quadratic
terms seek to the existence of optimal levels for temperature and
precipitation. However, this imposes symmetry to the response function.

Note how this model reduces dramatically the dimensionality of the
problem by averaging and aggregating temperature and precipitation
over the growing season. This model boils down the entire growing
season to just four variables in an attempt to make the model tractable.

One drawback of averaging temperature over the growing season, is
that is conceals the distribution of temperature within the growing
season. This was a key contribution in \citet{schlenker_nonlinear_2009},
an influential study in this literature. The study estimated the potential
impacts of climate change on US crop yields later this century by
coupling crop yield models for corn, soybeans and cotton with GCM
projections. Its key contribution is the estimation of non-linear
effects using a generalization of the concept of degree-days that
harnesses the intra-daily distribution of temperature over the growing
season. Specifically, the study shows that exposure to temperature
above around 30°C are particularly detrimental for these crops. Because
anthropogenic climate change is projected to increase the frequency
of high temperature, then yields were projected to decline if growing
regions and the crop yield response function remain stable. Analogous
studies have found large damages from higher temperatures in other
regions of the world \citep{hsiang_climate_2013,gammans_negative_2017}. 

Note that the temporal aggregation of temperature conceals its underlying
high-frequency distribution. For instance, two days with the same
average temperature can exhibit different extrema, so averaging conceals
the amount of time exposed to very high or low temperatures. The approach
proposed in \citet{schlenker_nonlinear_2009} is to rely on temperature
variables that capture the amount of time spent at each temperature
interval. I cover in greater detail the technical aspects of how to
construct exposure variables as well how to estimate these models
in subsection \ref{subsec:Estimating-non-linear-effects}. 

Perhaps one of the most striking findings in this literature is the
dominating role that temperature plays in explaining historical variations
and future projections of crop yields relative to precipitation. Changes
in temperature typically explain about 80 to 90\% of the projected
climate change impacts. Given the fundamental role that water availability
plays in the functioning of plants, this is particularly surprising.
To explore this puzzle \citet{lobell_critical_2013} combines biophysical
models to replicate the results obtained using statistical approaches.
The study finds exposure to high temperature, as measured by degree
days above 30°C, are associated higher Vapor Pressure Deficit (VPD)
which contributes to water stress, which negatively affects yield.
In essence, the story is that high temperature affects water availability.
As result, other studies have relied on VPD as a predictor in statistical
crop yields models (e.g. \citealp{roberts_agronomic_2012,lobell_greater_2014}).

The interpretation that high temperature affects water availability
raises the question about the appropriateness of precipitation as
a measure of water supply. Indeed, season-long precipitation variables
are very crude measures of how much water is effectively available
to crops. In the same way that temporal aggregation in temperature
was potentially problematic, a similar shortcoming affects precipitation.
More concretely, it's not only the amount of precipitation that matters,
but also its timing throughout the growing season. When precipitation
is highly concentrated over short periods of time, a lot of that water
is lost via deep percolation or as runoff. Thus, two growing seasons
can have the exact same total precipitation, but one may be considerably
dryer than the other.

What this highlights is that precipitation, as a measure of water
flow, is not ideal for capturing the yield effect of water availability.
In fact, what matters to crops is how much water is effectively available
throughout the growing season. Water availability is thus a stock
rather than a flow. Unfortunately, there are no widespread network
of stations measuring soil water content, even in countries like the
US. However, there are model-generated raster datasets of soil water
availability obtained from Land Surface Models (LSM). These LSM aim
to describe the evolution of soil water content and other factors
given a set of exogenous drivers such as soil characteristics, land
cover, and fine scale atmospheric conditions (e.g. hourly or sub-daily
precipitation, temperature, solar radiation, etc).

To address this concern \citet{ortiz-bobea_unpacking_2019} rely on
model-generated measures of soil moisture to unpack the climatic drivers
of crop yields for six major US field crops. The study pointed to
three main findings. First, the adoption of soil moisture variables
substantially improves model fit relative to standard models. Second,
climate change impact projections from models based on soil moisture
variables remain similar to those from traditional models based on
precipitation. Finally, the study finds that accounting for soil moisture
reduces the relative role of temperature variables in climate change
impact projections. This means, as hypothesized by previous studies,
that temperature effects were partly reflecting the effects of dry
conditions in previous models. Note that the relationship between
high temperature and drought is not unidirectional. While higher temperature
do increase evapotranspiration and reduce soil water content, droughts
can also cause temperatures to rise \citep{seneviratne_investigating_2010}. 

The discussion about the importance of the timing of soil moisture
availability raises the question about how to properly account for
the timing of environmental conditions within the season. One basic
approach is to adopt weather variables for various seasons of the
year. A more recent approach introduced in \citet{ortiz-bobea_unpacking_2019}
relaxes the additive separability assumption in \citet{schlenker_nonlinear_2009}
and allow for the effects of temperature and soil moisture to have
non-linear and time-varying effects on crop yields. This implementation
is based on a tensor product spline or ``bi-dimensional spline''
where crop progress within the season and the level of the variable
are the two dimensions. I provide details about how to estimate this
model in subsection \ref{subsec:Estimating-within-season-varying}. 

Moving forward, work in this area could focus on ways to improve our
understanding of mechanisms through which environmental conditions
affect crop yields including farmer management decisions, crop genetics,
soil characteristics and their interactions.

\subsection{Mixed statistical and biophysical approaches\label{subsec:Mixed-statistical-and}}

Perhaps one of the main methodological divides in the literature assessing
climate change impacts on agriculture is whether the models are biophysical
or statistical in nature. This methodological dichotomy often coincides
with disciplinary boundaries and is thus reflected in publishing outlets.
One of the major issues with this divide is that it could slow down
scientific progress if conclusions from studies based on alternative
approaches are perceived as being biased due to the underlying methods
used.

To explore this question, \citet{lobell_comparing_2017} conducted
a systematic review of the existing literature to compare prediction
of climate change impacts from both biophysical and statistical approaches.
The main conclusion is that when studies follow best practices in
their respective literature, these methods point to impacts that are
largely similar. This means that the previously held perception that
statistical model tend to be too ``pessimistic'' is not really supported
by the existing literature. In a somewhat related study \citet{moore_economic_2017}
find that the differences between biophysical and empirical studies
is small when CO2 fertilization is controlled for.

Recent studies have also started to compare or even combine biophysical
and statistical crop yields models. For instance, \citet{roberts_comparing_2017}
conduct a comparison of a statistical model and a process-based model
and find that a combined ``hybrid'' model tends to outperform individual
models in terms of prediction. In addition, that hybrid model tends
to point to projected impacts of climate change that fall in between
those of the individual approaches.

One key advantage of conducting similar analyses based on alternative
approaches within the same study is to better characterize model uncertainty.
For instance, \citet{liu_similar_2016} show that both biophysical
and empirical models point to similar temperature responses on global
wheat yield. This helps address common criticisms from reviewers coming
from one specific ``camp'' of the methodological divide.

Perhaps a fruitful direction of research in this area is the integration
of statistical analyses within broader inter-comparison projects like
AgMIP. There is indeed a great deal of cross-fertilization that could
occur across these different methods. These efforts help bridge the
divide across disciplines by seeing these approaches as complementary
rather than substitutes.

\subsection{Joint estimation of short and long run responses\label{subsec:Joint-estimation-of}}

One of the central methodological dilemmas in the literature relates
to the estimation of either short run or long run estimates of climate
change impacts. While the Ricardian approach conceptually captures
long-run adjustments by assuming farmers are already adapted to their
local climate, the approach remains vulnerable to time-invariant omitted
variables. On the other hand, panel approaches control for time-invariant
unobservables, but are generally perceived as only being able to capture
within-season short-run responses to weather fluctuations, and thus
unable to capture longer-run adjustments to a changing climate. 

However, recent research efforts have sought to combine certain elements
of cross-sectional and panel approaches to jointly estimate short
and long run responses. For instance, \citet{moore_adaptation_2014}
introduce a ``hybrid'' model to jointly estimate short-run and long-run
adaptations to weather fluctuations and climate variations in European
agriculture. The model takes the form:

\[
y_{ijt}=\beta_{1}\bar{W}_{ijt}+\beta_{2}\bar{W}_{ijt}^{2}+\beta_{3}\left(W_{ijt}-\bar{W}_{ijt}\right)^{2}+\beta_{4}X_{ijt}+\phi_{j}(t)+\alpha_{j}+\epsilon_{ijt}
\]

where $y_{ijt}$ represents farm profits or yield in region $i$,
country $j$ and year $t$, $W$ represents weather variables (temperature
and precipitation), $\bar{W}$ represents climate (average of annual
weather over the preceding 30 years), $X$ are control variables,
$\phi_{j}(t)$ is a country-specific time trend and $\alpha_{j}$
is a country fixed effect. This specification primarily harnesses
the cross-sectional variation in climate within countries (given there
is a country fixed effect and there is relatively less temporal variation
in climate) and the temporal variation in weather anomalies. According
to the authors, combining these estimates allows depicting an outer
``envelope'' describing the long run response function, and a series
of short-run response functions that are tangent to the long-run response
function when $W=\bar{W}$. This approach combines both cross-sectional
and time-series variation to estimate different parameters. These
parameters are subsequently combined to assess short and long run
responses. However, note that the cross-sectional estimation is still
vulnerable to omitted variables operating within countries. 

A different approach consists in trying to harness slow changes in
the within dimension of a panel model to detect evidence of adaptation
to a changing climate. This is the strategy in \citet{burke_adaptation_2016}
which applies a ``long difference'' approach to a panel of crop
yields to assess the influence of recent climate trends on changes
in the sensitivity of crop yields. The study first posits a standard
panel data generating process of the form:

\[
y_{it}=\alpha+\beta_{1}z_{it}+\beta_{2}z_{it}^{2}+c_{i}+\epsilon_{it}
\]

where $y_{it}$ is crop yield in county $i$ and year $t$ and $z_{it}$
is a weather variable (e.g. temperature). The county fixed effect
$c_{i}$ captures all time-invariant unobserved factors. 

The study argues that this model only captures short run changes on
yield. The study then posits that if farmers are adjusting input choices
(e.g. crop cultivars) to adapt to recent changes in climate, then
this would be reflected in a change in the response function over
time that would renders extreme weather less damaging. That, is, the
response function would trace out an ``outer envelope'' that allows
for longer term adjustments.

To explore this hypothesis, the study first proposes averaging the
model above over a multi-year period $a$, to obtain a new model of
the form $\bar{y}_{ia}=\beta_{1}\bar{z}_{ia}+\beta_{2}\bar{z}_{ia}^{2}+c_{i}+\bar{\epsilon}_{ia}$
where $\bar{y}_{ia}$ is average yield over the multi-year period
of time $a$ and $\bar{z}_{ia}$ is the multi-year average weather
over the same period. Defining an analogous equation for a more recent
period $b$, and taking the differences between these equations yields
what the authors refer to as the ``long difference'' model:

\[
\bar{y}_{ib}-\bar{y}_{ia}=\beta_{1}\left(\bar{z}_{ib}-\bar{z}_{ia}\right)+\beta_{2}\left(\bar{z}_{ib}^{2}-\bar{z}_{ia}^{2}\right)+\left(c_{i}-c_{i}\right)+\left(\bar{\epsilon}_{ib}-\bar{\epsilon}_{ia}\right)
\]

which simplifies to:

\[
\Delta\bar{y}_{i}=\beta_{1}\Delta\bar{z}_{i}+\beta_{2}\Delta\bar{z}_{i}^{2}+\epsilon_{i}
\]

Note that the estimated parameters $\beta_{1}$ and $\beta_{2}$ are
the same as those specified in the original equation. In the absence
of any adaptation, the parameters in this new model would be identical.
However, if farmers are adapting to recent climate trends, then those
parameters would shift to make the response function flatter. Applying
this approach to US corn yields, the study finds that the parameters
of the ``long difference'' model as very similar to those of the
baseline panel model. The study thus concludes that little adaptation
has occurred over several decades.

One should highlight that it is unclear whether farmers perceive recent
changes in weather patterns as a permanent shift in climate. One should
not generally expect agents to make permanent shifts in production
practices in response to transient changes environmental changes.
Moreover, note that if farmers are adapting to a changing climate,
then the baseline panel model is misspecified as we should expect
adaptations to be reflected in the short run response function as
well.

It is important to clarify that the idea that panel estimates only
capture the short-run response is not entirely correct. In a panel
model with location fixed effects estimating a non-linear response
to weather fluctuations, which is common practice, the identifying
variation stems from within location deviations from the mean of each
variable. In locations with higher average values of the weather variables
(e.g. warmer or wetter climates), the deviations for the quadratic
terms are larger. Effectively, this means that group means, in this
case climate, plays a role in the identification. That is, cross-sectional
variation plays a role in the identification of weather effects. See
\citet{mcintosh_identifying_2006} for an explanation.

In a key recent study, \citet{merel_climate_2021} show that panel
estimates are a weighted average of the short-run and long-run response
functions to weather and climate. That study also derives the conditions
under which the short-run response approximates the long-run response.
Intuitively, the extent to which the panel estimate represents more
or less the short or the long-run response function depends on the
ratio of the within time-series variation to the cross-sectional variation
of the weather variable. If a weather variable exhibits greater degree
of variation in the cross-section than in the within dimension, then
the response function more closely approximates the long run response
function. This means that for small panels with units exhibiting similar
climates, the estimated response function represents primarily the
short run response function. In contrast, estimating global response
functions over a large panel with units with large climatic differences
(but relatively smaller within variations) approximates the long-run
response function. In a similar spirit, \citet{gammans_reckoning_2020}
propose an approach to recover the global long run response function
by harnessing variations across adjacent climates. A limitation of
these models is that they apply to a subclass of data generating processes.

\subsection{Retrospective climate change impacts \label{subsec:Retrospective-studies}}

The literature on climate change impacts on the sector has been mostly
concerned about future projected impacts. Most studies estimate or
calibrate a model based on historical data and make predictions about
potential future outcomes under alternative climate scenarios and
assumptions about farmer adaptation and agricultural markets. However,
anthropogenic forces have already altered the climate system as summarized
by the IPCC (\citealp{pachauri_climate_2014}). Our climate is about
1°C warmer than during pre-industrial period. Given that agriculture
is highly climate sensitive, it seems natural that these climate changes
may have already affected agricultural production. 

The first studies trying to analyze the observed impacts of climate
change have overwhelmingly focused on teasing out the effect of recent
climate trends on yields of major field crops. \citet{lobell_global_2007}
analyze the relationship between country-level cereal yields and growing
season weather. Specifically, the study computes first differences
of yields and weather variables to control for slowly changing unobservables
over time such as management practices. The study finds that there
is a clear negative response of maize, wheat and barley yields to
higher temperatures. The study then couples these estimates with observed
trends in climate variables (1961-2002) and finds the cumulative impact
on crop yield for these 3 crop amounts to 40Mt or \$5 billion per
year. 

In a related study, \citet{lobell_climate_2011} revisits this question
and focuses maize, wheat, rice, and soybeans, which represent about
75\% of calories that humans consume directly or indirectly. Relative
to previous work, the study does a better job at matching weather
conditions to the growing season of each crop based on the crop calendar
complied in \citet{sacks_crop_2010}. Other examples of more regional
studies include \citet{nicholls_increased_1997} for Australia, \citet{lobell_climate_2003}
for the US, \citet{tao_climate_2006,tao_climatecrop_2008} for China,
\citet{lobell_analysis_2005} for Mexico. 

As indicated in \citet{porter_food_2014}, these studies have largely
focused on attributing the effect of recent climate trends on crop
yields without unpacking the anthropogenic sources of these climate
trends. More recently, \citet{ortiz-bobea_anthropogenic_2021} coupled
an econometric model with counterfactual climate simulations to conclude
that anthropogenic climate change has slowed global agricultural productivity
growth by about 20\% over the 1961-2020 period. This loss is equivalent
to losing about 7 years of productivity growth over the same period.
The strategy in that study is to first estimate a global panel model
regression country-level agricultural TFP on annual weather variables.
The scope of the study differs from previous ones because the productivity
estimates encompass the entire agricultural sector, and not just crops.
The study then links the econometric estimates with weather trajectories
for 1961-2020 coming out of climate models from CMIP6 from both a
historical run with observed human emissions (the ``historical''
experiment) and a historical run without human emissions (the ``hist-nat''
experiment). This allows to compute the cumulative TFP growth over
the sample period under these two scenarios. Taking the difference
is thus interpreted as the impact of anthropogenic climate change.
One of the caveats in that study is that it does not recover how agricultural
TFP would have responded to weather in the counterfactual world.

In a related study focusing on global crop yields, \citet{moore_fingerprint_2020}
develops an approach that extends the literature on detection and
attribution in climate science to global crop yields. Rather than
relying on multiple climate models, the paper relies on multiple runs
from a single climate model to characterize the internal variability
of the model. The study concludes that the patterns of yield growth
observed on maize, wheat and rice production have less than a 10\%
chance of having arisen in the absence of anthropogenic climate change.
Specifically, the study finds that anthropogenic climate change has
reduced annual calorie production related to these 3 crops by about
5\% per year on average since 1961.

\subsection{Statistical crop quality models\label{subsec:Statistical-crop-quality}}

The focus on crop yield overlooks the fact that weather conditions
can affect crop quality \citep{soares_preserving_2019}. A small but
growing number of economic studies have analyzed the implication of
extreme weather conditions and rising carbon dioxide atmospheric concentrations
on crop quality and its potential implication under climate change.
These models also exploit longitudinal variation in panel model with
fixed effects. What typically differs is the nature of the dependent
variable, which can vary greatly across crop quality classification
systems and countries. For instance, wheat quality is typically captured
by its moisture and protein content as well as its milling and baking
qualities. Wheat is graded and classified so aggregate data is typically
obtained as the share of output falling into different grading categories.
For other grains like rice, the quality characteristics vary substantially
across countries reflecting very heterogeneous consumer preferences.

A major obstacle in this area is obtaining comprehensive panel data
on crop quality. Most previous research is either based on process-based
models (e.g. \citealp{erda_climate_2005,asseng_climate_2019}) or
on the statistical analysis of quality in relatively small samples
(e.g. \citealp{rao_cultivar_1993}). The economics literature in this
area is relatively thin but differs from agronomic research which
focuses on direct physical effects on quality. On the other hand,
economic studies tend to focus on tradeoffs between quality and quantity
or seek to quantify the relative contribution of yield and quality
changes on farmer revenues.

For instance, \citet{kawasaki_quality_2016} explores the effect of
extreme weather on rice yields and quality in Japan. The study finds
that while high temperature improves yield, it also reduces rice quality,
leading to an overall negative effect on farmer profits. While most
economic research has focused on cereal crops, there is some limited
work on fruits and vegetables. \citet{dalhaus_effects_2020} explores
the effect of temperature on apple yields and quality in Switzerland.
The study finds that the detrimental quality effects of spring frosts
can be substantial and can play a larger contribution to farmer revenue
than yield effects. Relatedly, some research has focused on wine quality
\citep{erda_climate_2005,ashenfelter_using_2010,ashenfelter_economics_2016}.

Moving forward, there is much to be learned regarding the impact of
anthropogenic climate change on crop quality. One of the main obstacles
seems the availability of reliable longitudinal datasets. This will
require new efforts of data collection or partnerships to track quality
to better understand the ongoing processes. There is also limited
understanding of how climate change may affect micro-nutrient availability,
which are critical for small scale farmers who primarily rely on their
own production for subsistence. Greater emphasis on how post-harvest
management and climatic conditions affect quality is also needed.

\subsection{Modeling planting and harvesting decisions\label{subsec:Modeling-planting-and}}

A growing season for annual crops is essentially determined by the
time ranging from planting to maturation or harvest. The decision
of which crop or cultivar to plant, and when, is critical for agricultural
production. For instance, in moisture-limited regions with highly
seasonal rainfall, planting prior to the arrival of the first rainfall
events could lead to plant death which would require new planting.
A similar constraint exists in temperature areas when planting too
early in the spring, when the increased risk of frost can jeopardize
crop emergence. 

In addition, the timing of planting and the choice of the crop cultivar
largely determines the timing of when specific stages of plant growth
occur during the calendar year. Planting a long-season cultivar in
a region with a short season could mean that the crop would not reach
maturity during the usual harvest period, which can be problematic.
For instance, fall frost can damage the crop in temperature regions.
Importantly, the flowering period which tends to occur around the
middle of the growing season for annual crop can be particularly vulnerable
to heat and moisture stress \citep{fageria_physiology_2006}.

Farmers have formed expectations about climate which guides the choices
about planting (crop and cultivar choice as well as timing) before
the weather conditions unfold throughout the growing season. This
is well known but there is relatively little research in economics
regarding how farmers form expectations that guide the nature and
timing of planting decisions. Changes in the length of the growing
season are likely important channels through which climate change
will affect farmers. For instance, the expansion of the frost-free
period in temperate regions may expand growing seasons allowing farmers
to grow longer season crops and cultivars of giving farmers more flexibility
with their planting dates \citep{ortiz-bobea_modeling_2013}. 

While the the current spatial distribution and drivers of planting
dates for major crops has been characterized \citep{sacks_crop_2010},
less is known about how farmers are adjusting their practices in response
to a changing climate. There is evidence of a multi-decadal trend
toward earlier planting in US Midwest that appears beneficial to crop
yields \citep{kucharik_multidecadal_2006,kucharik_contribution_2008}.
However, it is still unclear whether these trends are primarily driven
by new cold-resistant varieties or by climate trends.

In addition, climate change is also increasing the intensive and variability
of rainfall events, which could lead to excessive moisture during
parts of this could be detrimental. For instance, excessive rain in
the Spring reduces the ability of heavy farming equipment to enter
fields without causing severe soil compaction, which is detrimental
to root development and causes long term damage to crop yields \citep{hamza_soil_2005}.
These events can be extremely disruptive as the US floods in the Spring
of 1993 and 2019 exemplified.

More generally, the influence of weather shocks and climate on the
decision of what and when to plant remains under-explored. We still
need a more systematic understanding of the potential barriers precluding
farmers from making optimal decisions in a changing climate. This
is particularly critical in the context of perennial crops such as
fruits trees and vineyards where planting is extremely costly and
affects farming performance for many years or decades. For instance,
it is still unclear whether farmers have begun to shift crops or varieties
in specifically in response to climate change and what types of information
they rely upon for making these decisions. For instance, would farmers
presented with information about recent and projected trends in their
locale chose different cultivars of a perennial crop?

The discussion so far has primarily focused on planting. However,
the decision of whether to even harvest is also important. Crop abandonment
occurs when farmers decide not to harvest a crop they previously planted.
Conceptually, this occurs when the cost of harvesting exceeds its
expected benefits. Expected benefits depend on yield and output price
(as well as any form of production-based government subsidy). Harvesting
cost depends on input prices such as fuel and are not necessarily
proportional to yield in the case of field crops. One can also factor
in the cost of post-harvest management and storage as part of this
harvest cost. There are numerous reasons why expected benefits might
be lower than harvesting costs. It could be that either yield or output
prices are considerably lower at harvest time than the farmer anticipated
at planting. It is also possible that harvesting costs become unexpectedly
high, which might happen in the presence of an unexpected disruption
to labor necessary for harvest (e.g. sudden immigration policy or
shock that restricts access to the labor supply).

When these crop abandonment decisions are driven by unexpected drops
in yield, say from unusual weather, it is not uncommon to refer to
the situation as ``crop failure''. Note that crop failure is the
result of an economic decision and not a physiological or agronomic
condition. The decision not to harvest a crop is inherently economic
in nature. One could easily envision a farmer deciding to harvest
a crop with extremely low yields if there is a sufficiently high output
price.

The literature on climate-induced crop abandonment or failure is relatively
thin. The first economics study on this question seems to be \citet{mendelsohn_what_2007},
where a Ricardian-style cross-sectional model is estimated to predict
average crop failure rates in the US context. The study relies on
reported crop failure rates at the county level based on 5 years of
data from the US Census of Agriculture collected from 1978 to 1997.
The study finds that about 39\% of the cross-sectional variation in
crop failure can be explained by soils and climate.

More recently, and using longitudinal variation in a panel, \citet{cui_beyond_2020}
explores how weather shocks not only influences US crop yields but
also the fraction of planted acres farmers end up harvesting. As with
other panel studies, it is unclear to what extent these historical
relationships can be extrapolated in the long run under climate change.
More research is certainly needed to understand the heterogeneity
in crop abandonment decisions and how it is influenced by other sources
of risk (e.g. crop and storage prices) and risk management strategies
(e.g. crop insurance).

So far the discussion has focused on planting and harvesting decisions
surrounding a single season. However, many regions of the world have
one or more seasons for the same or different annual crops \citep{siebert_global_2010}.
The number of harvest per year is also referred to as cropping frequency
or intensity. Areas with more than one harvest per year tend to be
located in warmer regions with sufficient precipitation to sustain
multiple seasons. 

This is an area that has received considerably more attention by natural
scientists than economists. Crop intensity has received growing attention
because of its potential to increase global crop production without
expanding croplands \citep{ray_increasing_2013,wu_global_2018} although
recent work indicates limited room for increasing cropping intensity
\citep{waha_multiple_2020}. A particular focus is how climate change
could affect the cropping intensity. Using a biophysical modeling
approach, \citet{seifert_response_2015} find that the area suitable
for the most common form of double cropping in the US (winter wheat
followed by soybeans) rose by 28\% from 1988 to 2012. The study also
finds that the suitable area could double or triple depending on the
future climate scenario.

Naturally, a rise in suitable area doe snot imply that actual area
under double cropping will increase given that changes in cropping
intensity affects yields \citep{challinor_crop_2015}. Moreover, there
is a large discrepancy between the area suitable for double cropping
and the area currently under double cropping suggesting that there
are other constraints farmers faced that have not yet been well documented.
\citet{gammans_double_2019} develop an empirical model based on observed
double cropping area in the US to assess potential expansion of agricultural
production under a warming climate. Climate change is likely to play
a key role in driving changes in cropping intensity \citep{iizumi_how_2015,cohn_cropping_2016}.

\subsection{Irrigation and other input adjustments\label{subsec:Irrigation-and-other}}

Water is an essential input in crop production. Farmers can obtain
water via precipitation, but also from irrigation water coming surface
or underground sources such as rivers, lakes or underground aquifers.
Irrigation has played a fundamental role in the development of agriculture
of many nations, including the US \citep{edwards_role_2018}. The
main emphasis in the economics literature on climate change and agriculture
relate to 1- the role of irrigation in explaining how farmers cope
with environmental change, and 2- understanding the sources, consequences
and solutions to irrigation water misallocation.

Irrigation consists in 1- moving water from a source (e.g. river,
lake, aquifer) on or close to an agricultural field, and then 2- applying
that water throughout the field where it can reach the root system
of crops. The first point relates to the water source (surface or
groundwater). The second point relates to the irrigation technology
(flood, sprinkler, drip, etc.). Surface irrigation typically requires
major infrastructure to manage water flow from the source to the field.
That includes canals and water holding infrastructure like dams or
reservoirs. This infrastructure investments are substantial so they
can require collection action from an association or government. Groundwater
irrigation from an aquifer requires drilling and using a pump. It
does not require the type of common infrastructure to transport water
over potentially long distances like surface water irrigation.

A fundamental point is that having access to water via irrigation
fundamentally changes how farmers cope with changing climatic conditions.
For instance, \citet{ortiz-bobea_growing_2018} shows that agriculture
is much more sensitive to weather fluctuations in the Eastern parts
of the US than in the mostly irrigated Western regions of the country.
This points relates to an early debate regarding the role of irrigation
in the Ricardian literature. In \citet{mendelsohn_impact_1994}, the
hedonic model did not include an irrigation variable in the cross-sectional
regression. \citet{darwin_impact_1999} pointed this out and proposed
an alternative model with irrigation as an additional variable. However,
water from irrigation is not separable from other climatic inputs.
Empirically, this means that irrigation and climatic variables interact,
so that the marginal effect of weather on agricultural outcomes depends
on irrigation. In addition, and as pointed out in \citet{schlenker_will_2005},
irrigation water is often subsidized in the US and its long term availability
is uncertain, so performing a Ricardian model over irrigated areas
to make long term inferences about climate change impacts can be misleading.
As a result, \citet{schlenker_nonlinear_2006} propose a Ricardian
model focused on the mostly rainfed Eastern parts of the country.
Focusing on the Eastern US and non-irrigated areas has become a common
sample restriction in order to avoid the complex issues surrounding
irrigation.

There seems to be three major factors that complicate the long term
analysis of irrigation in a changing climate. First, irrigation infrastructure
(e.g. canals, dams, etc.) is often subsidized and water is typically
supplied to agricultural users below its cost of provision. Understanding
the long term implication of irrigation for the agricultural sector
requires understanding these true costs, but obtaining that data is
difficult. One of the implications is that farmers may treat water
as an abundant resource, and therefore not invest in irrigation technologies
that conserve water.

The second complicating factor are water rights. The rules surrounding
how water is allocated between agricultural users and between agricultural
and non-agricultural users can be highly complex and vary considerably
by region. In the US context, water in the Western US is governed
by a prior appropriation doctrine (first users to claim water have
priority) whereas water rights in the Eastern US follow riparian rights
(users close to the water source have priority). There are also important
issues regarding the common pool nature of groundwater resources and
the potentially perverse incentives that arise that may lead to more
rapid water depletion.

The third complicating factor is the the future availability of water,
especially in a changing climate \citep{elliott_constraints_2014}.
In the case of aquifers with slow natural recharge (e.g. Central Plains
aquifer in the US), the use of water is akin to mining a non-renewable
resource. But in other aquifers the dynamics of recharge and the interactions
with surface water systems make projections future water availability
highly uncertain. A warming climate will also increase evaporative
demand, placing new constraints on water availability for irrigation
\citep{fischer_climate_2007}. Also, surface water supply is linked
to precipitation in a watershed, so the future availability of water
for irrigation at a given location may depend on the water cycle in
potentially distant locations. This is particularly the case in agricultural
regions that depend on water from recurrent glacier melt such as in
the Indus, Ganges and Brahmaputra river basins. But these complicating
factors do not preclude to research documenting the nature of farmer
adjustments to changing environmental conditions over historical periods.

Indeed, an important direction of research documents how farmers adapt
to changing water scarcity. For instance, \citet{hornbeck_historically_2014}
explore the advent of groundwater irrigation in US Central Plains
following the Second World War. A key characteristic of this region
is the presence of one of the largest groundwater aquifers in the
world, which was previously inaccessible to farmers with pre-war technology.
The study analyzes how agricultural production evolved on either side
of the aquifer boundary. The study shows how counties with access
to the aquifer first became more resilient to droughts, but then progressively
specialized in more water-intensive crops which in turn increased
drought sensitivity. Counties without access to the aquifer maintained
drought-resilient agricultural systems. In a related study, \citet{hornbeck_enduring_2012}
explores the long term impact of the American Dust Bowl and analyzes,
among other things, the role of irrigation in helping farmers cope
with an unprecedented environmental disaster and drought.

A more recent example includes \citet{hagerty_adaptation_2020}, which
relies on high-resolution land cover data in California to derive
short and long run changes in farmer cropping choices over time in
response to changes in water availability. The paper exploits changes
in institutional settings that lead certain farmers to have more water
than others. This type of work requires detailed institutional knowledge
of water allocation arrangements and regulations. In a related study,
\citet{arellanogonzalez_intertemporal_2020} show that access to a
groundwater banking project decreased drought risk, which in turn
increased the farming transition from lower-value annual crops to
higher-value perennial nut crops in California.

Research has also focused on credibly teasing out the value of irrigation
water to farmers. These valuation models consist in hedonic models
where access to irrigation water is one of the features of the land.
The rising availability of repeated sales data now allows the estimation
of panel models with parcel or location fixed effects which allows
the more credible identification of these values relative to cross-sectional
designs \citep{buck_land_2014,mukherjee_irrigated_2015}.

There is also a growing literature exploring the role of irrigation
in helping stabilize agricultural production and thus prevent conflict.
For instance, \citet{gatti_can_2021} find that irrigation infrastructure
can help mitigate the effect of growing-season rainfall shocks on
conflict in Indonesia. 

The rise of smart technologies are also providing access to new datasets
to track farmer behavior in ways that were impossible before. For
instance, \citet{christian_monitoring_2021} track high frequency
data on water use in Mozambique and find what appear like inefficiencies
in farmer decisions. The study then introduces a randomized control
trial providing information aimed at improving farmer water management.
Field studies like appear promising in helping enhance farmers decisions
about water allocations in a changing climate.

\subsection{Market equilibrium and trade \label{subsec:Trade-and-general}}

Most of the discussion so far has focused on direct impacts of extreme
weather or climate change on agricultural production without any particular
consideration to market equilibrium and price formation. Climate change
is not a localized idiosyncratic shock, but a shift that affects virtually
every economic agent in the world. As a result, domestic and international
trade and markets are likely going to play a central role in modulating
how climate change impacts are distributed within and across nations.

One of the earliest studies to endogenize prices in a country-level
context is \citet{adams_global_1990}. In that study, the authors
linked biophysical process-based crop models with a partial equilibrium
market model. While the goal of the crop model is to translate changes
in climate into changes in agricultural productivity, the role of
the economic model is endogenously determine prices and quantities
produced in the agricultural sector. However, these are partial equilibrium
models where some features of the global economy are considered exogenous
(e.g. demand for food, incomes, etc). Other examples based on partial
equilibrium models include \citet{adams_reassessment_1995}, \citet{reilly_us_2003}
and \citet{janssens_global_2020}.

It was perhaps \citet{rosenzweig_potential_1994} that first introduced
a general equilibrium framework to the analysis of climate change
impacts on world food production. In that study, the authors linked
biophysical crop models with national agricultural sector models in
an overall framework representing all economic sectors with endogenous
supply and demand. More recent work in this area relies on the Global
Trade Analysis Project or GTAP (e.g. \citealp{randhir_trade_2000};
\citealp{hertel_poverty_2010};\citealp{baldos_global_2014} and \citealp{moore_economic_2017}).
For a guide to general equilibrium modeling in agriculture see \citet{hertel_chapter_2013}.

In a large comparative study, \citet{nelson_climate_2014} contrast
how nine global economic models influence estimates of climate change
impacts when subjected to standardized yield impacts. Among the economic
models, they consider both partial and general equilibrium models.
They find that the largest differences coming out across these economic
models are in terms of responses in agricultural production, cropland
area, trade, and prices. They find that these differences originate
from model structure and specification. In particular they find that
the ability to convert land to agriculture, to intensify agricultural
production and the propensity to trade are some of the most critical
factors in these models. It appears that there are still important
uncertainties related to the role of these market forces in modulating
climate change impacts.

In a recent study, \citet{costinot_evolving_2016} propose a general
equilibrium framework based on micro level data from 1.7 million agricultural
fields to explore the role of trade and within country reallocations
in modulating the effects of climate change on global agricultural
welfare. A key contribution is the ability to consider a large number
of fields within countries that allows analyzing adjustments both
within and across countries. The study finds that changing comparative
advantage will drive crop substitution within countries which will
greatly reduce climate change impacts. They find that climate change
would reduce global GDP by about 0.26 percent, which is about one
sixth of total crop value. Perhaps surprisingly, they find trade adjustments
would play a very little role in explaining the magnitude of this
result.

More recently, \citet{gouel_crucial_2021} revisit this work and find
drastically different results. They find that international trade
plays a comparable role to within-country crop reallocation. It appears
that the main critical factor explaining the differences is the choice
of the counterfactual. In \citet{costinot_evolving_2016}, the authors
appear to constrain the export shares to remain unchanged under climate
change. Instead, constraining bilateral import shares to remain the
same yields a much larger role for trade. Interestingly, this new
study finds large regional heterogeneities in impacts with large losses
for net-food-importing countries and benefits for agricultural-exporting
countries due to more favorable terms of trade.

I should highlight that the trade literature has most closely developed
in tandem with biophysical crop process-based models. The interactions
with the empirical literature based on statistical and econometric
models remains limited. It seems like greater collaboration and integration
accompanied by systematic model comparisons are needed to resolve
ongoing debates.

\subsection{Understudied problems and unsettled questions\label{subsec:Some-unsettled-questions}}

Here I discuss problems in the literature that seem understudied or
that remain unsettled. This list is by no means exhaustive and reflects
my own preferences as a researcher.

One of our roles as researchers is to identify mechanisms and policies
that could enhance adaptation to a changing climate. This includes
identifying institutional barriers to adaptation. For instance, governments
pay billions of dollars in subsidies every year to farmers. Some governments
have also implemented major policies that create sizable non-food
markets for agricultural products, such as the Renewable Fuel Standard
in the US. However, the effect of farm policies in enhancing or hampering
adaptation to climate change remains largely unexplored. \citet{ortiz-bobea_growing_2018}
finds that certain US agricultural regions are growing increasingly
sensitive to rising temperatures and this is partly due to regional
specialization. The role that policies and trade have played in this
remains unclear but seems plausible. There may also be unintended
consequences to large government programs. For instance, \citet{annan_federal_2015}
find some indications that crop insurance could provide a disincentive
to adapt to extreme temperatures.

Perhaps one of the most understudied issues pertaining to agriculture
and climate change is climate justice. A lot of the global work has
emphasized inequities in terms of cross-country impacts. But more
research is needed to understand inequities within countries and how
these are potentially exacerbated by existing socio-political systems
that perpetuate social exclusion.

Another area that has received little attention is the economics of
research and development (R\&D) and innovation in a changing climate.
The literatures on R\&D and climate change have evolved mostly separately
and the time is ripe for integration and collaborations between these
fields. This is critical, because agricultural innovations are for
the most part developed outside the farm and they are not easily transferable
across bio-climatic zones. In addition, the returns to R\&D take years
if not decades to materialize and climate is rapidly changing. Are
current global and regional investments and infrastructure adequate
under a changing climate? There is also no research on the potential
role of the private sector in the R\&D ecosystem under climate change.

An environmental challenge that is likely to become more prevalent
in a warming climate is soil salinity. Soil salinity arises from salt
water intrusion in low-lying agricultural coastal areas (e.g. Bengal
region) or from the evaporation of large quantities of irrigation
water. Soil salinity is toxic for many cultivated plants which negatively
affect yields. Breeding or planting salinity-tolerant crops likely
imposes a cost to farmers in these areas. For instance, \citet{finkelshtain_substitutability_2020}
recently analyzed the substitutability between freshwater and non-freshwater
sources in irrigation in Israel. More research on these matters under
climate change are needed and will likely require inter-disciplinary
collaboration with non-economists to better characterize the process
of water salinization in the future.

Researchers have allocated considerable effort to understanding a
few things very well. However, we still have a rudimentary understanding
on very important matters. For instance, we know relatively little
about the impacts of climate change on agricultural labor. There are
a few studies analyzing how weather shocks affect the labor supply.
For instance, \citet{branco_weather_2020} find that droughts tend
to increase the labor supply of rural households in non-agricultural
sector in Brazil. Some studies are also exploring how heat is affecting
human capital accumulation and the role that agriculture plays \citep{garg_temperature_2020}.
But more research is needed in this under-explored area. 

Overall, there is much more emphasis on agricultural production and
in agricultural fields, and less on what happens afterwards or beforehand.
For instance, there is less research on how weather can affect food
quality and post harvesting processing and associated crop losses.
Similarly, there is little understanding of the robustness of food
and agricultural supply chains.

We also have a limited understanding of how changing pest pressures
linked to changing biodiversity would affect agriculture. This is
particularly challenging because observational studies based on panel
data might not be necessarily well suited for such analysis. For instance,
a warm year does not lead to the same pest pressure than an equally
warmer climate would induce. There are not only farmer adaptations
to consider, but also ecological adjustments that remain largely unknown.
This invites a greater degree of collaboration and integration with
ecologists and other natural scientists.

Finally, economists have for the most part favored working on observational
empirical studies in isolation. However, there are fruitful collaborations
that can arise from closer integration with crop scientists, particularly
in the context or large inter-comparison projects such as AgMIP. These
collaboration can lead to major publications in interdisciplinary
outlets that could have greater impact on policy than what economists
can do on their own.

\section{Coding and other empirical matters\label{sec:Coding-and-other}}

The analysis of climate change impacts and adaptation in agriculture
involves specialized knowledge and training that is generally not
offered in graduate programs in economics, agricultural and resource
economics or other related fields. For instance, conducting empirical
research in this area requires manipulating large geospatial datasets
(e.g. weather data) which are more common currency in environmental
sciences than in economics. Unfortunately, this deficit in training
means that new researchers to this field often face substantial barriers
to entry.

This section provides a hands-on introduction to common tasks in this
area of research in an attempt to cover this ``hidden curriculum''.
Increasingly, these tasks are carried out in open-source programming
languages like R, which allows the flexible integration of workflow
from the manipulation of geospatial data to the regression analysis
with the ability to produce high-quality visualizations. This integration
also facilitates the reproducibility and replicability of research
projects. This is critical as a growing number of academic journals
require that papers be fully reproducible.

\textcolor{black}{The section provides a rationale behind how to code
these tasks, but also provides code and data to fully carry out these
tasks, including the ability to reproduce the figures in the chapter.
The reproduction code and data are hosted in a permanent repository
at the Cornell Institute for Social and Economic Research (CISER)
at Cornell University (\url{https://doi.org/10.6077/fb1a-c376}).}

\subsection{Aggregation of point weather data\label{subsec:Aggregation-of-point}}

As indicated in section \ref{sec:Basic-concepts}, the most basic
form of weather data comes from weather stations. These are point
datasets where weather observations are geo-referenced. Relying on
these direct observations may seem appealing but they present some
clear challenges for the non-specialist. The first inconvenience is
that the spatial distribution of weather stations may be sparse (see
Fig. \ref{fig:Spatial-distribution}). This means that the researcher
may not be able to accurately obtain weather information in the precise
locality under study. The second issue is that weather data are not
always quality-controlled. That means that the researcher needs to
spend considerable effort ``cleaning'' the data for suspicious or
implausible outliers. The third challenge is that weather observations
may be missing in some time periods. These issues of sparsity, data
quality and attribution, pose empirical challenges related to measurement
error and attrition bias which have important econometric consequences.

In cases when all units of observations are very close to weather
stations, say in the example of geo-referenced farm-level data, it
makes sense to rely on the closest weather station data. In cases
when the researcher is interested in weather conditions that fall
between weather stations, some form of interpolation may be necessary.
The most basic type of interpolation is an inverse-distance weighting
approach. 

Performing your own interpolations, however, is generally not ideal.
There are numerous high-quality datasets created by meteorologists
and climate scientists based on sophisticated interpolations or other
model based approaches (e.g. interpolation accounting for orography
or reanalysis). Those third-party products typically implement cutting-edge
approaches and undergo careful quality control procedures by dedicated
trained professionals. Trying to replicate similar work for an economic
research project is risky because these procedures are error-prone.
Readers and reviewers are rightfully skeptical of whether the researchers
were able to perform interpolation adequately. This requires researchers
to allocate non-negligible amount of effort describing and validating
their interpolation approach. 

The comparative advantage of economists does not lie on environmental
data interpolation. We are thus generally better off relying on datasets
developed by climate scientists unless there is a strong reason not
to do so. For instance, performing one's own interpolation may be
justified for study regions where existing gridded datasets have been
found to be unreliable.

\begin{figure}
\begin{centering}
\includegraphics[scale=0.15]{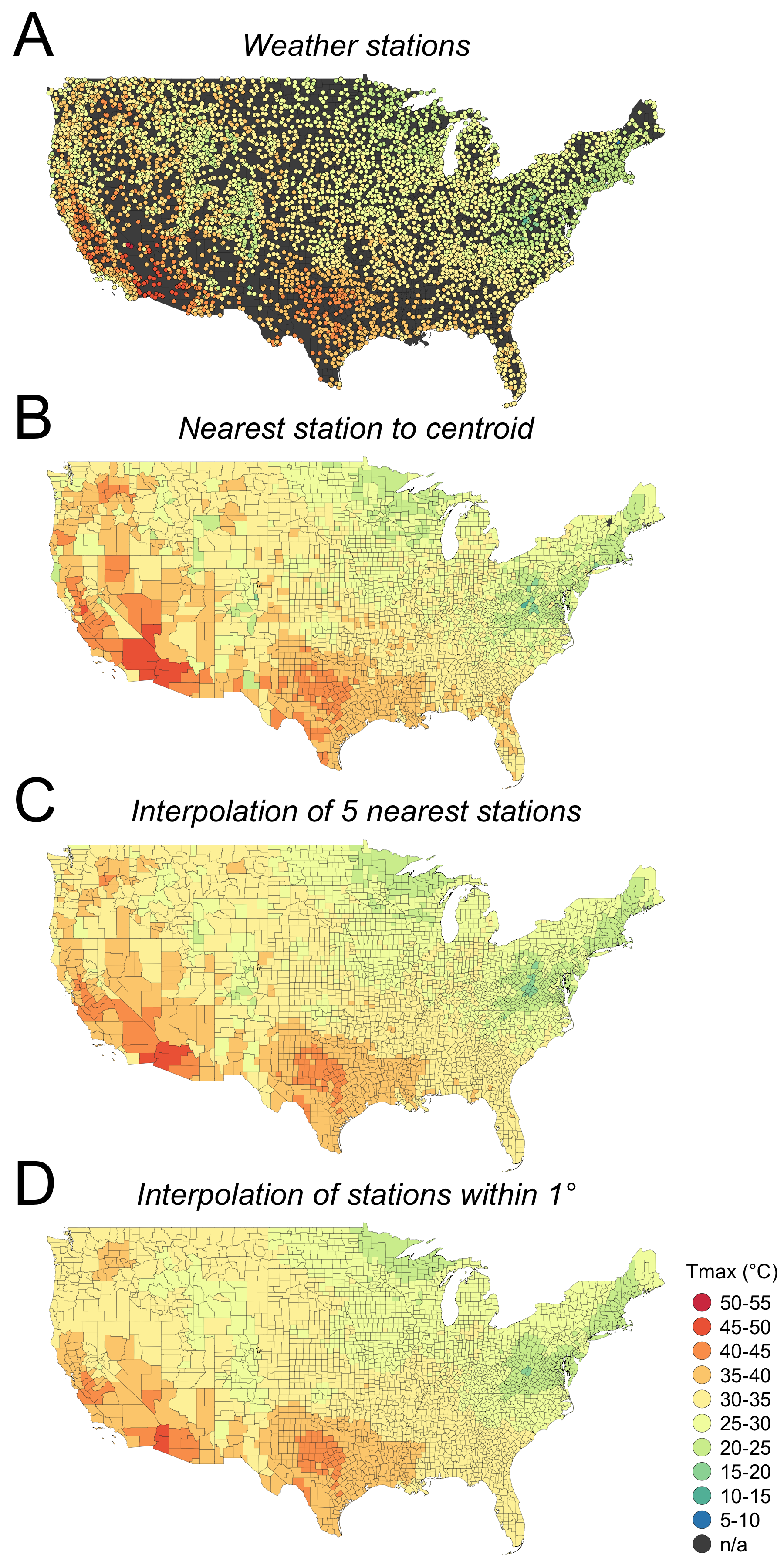}
\par\end{centering}
\noindent\begin{minipage}[t]{1\columnwidth}%
\textit{\footnotesize{}Notes:}{\footnotesize{} All maps show maximum
temperature (Tmax) on August 16 of 2020. A: Tmax over 6,785 weather
stations from the GHCN database falling over CONUS. B: Assigns value
of weather station closest to a county's centroid. C: Inverse-distance
spatial interpolation based on the 5 nearest stations to the county's
centroid. D: Inverse distance interpolation based on stations located
within 1° of the a county's centroid.}%
\end{minipage}

\caption{Spatial interpolation of weather station data.\label{fig:Spatial-interpol}}
\end{figure}

In the accompanying R code and data, I provide a simple introduction
to basic forms of spatial interpolation (see \texttt{1\_weather\_data.R}).
To illustrate these techniques I rely on weather station data from
the GHCN database falling over the lower 48 US states. I also select
maximum temperature (Tmax) on August 16, 2020 as the example. Figure
\ref{fig:Spatial-interpol}A shows Tmax on that day over more than
6,000 weather stations across the US. We can see some regions of the
country have a a relatively sparse network of weather stations.

Suppose we are interested in creating a county-level dataset based
on these weather stations. This means converting these 6,000 station-level
observations to a little more than 3,000 counties. Perhaps the most
basic type of mapping is to simply assign to a county the value of
the weather station that is nearest to the centroid of a county. This
is precisely what is shown in Fig. \ref{fig:Spatial-interpol}B. This
approach may work better in areas with homogeneous landscapes and
no mountain ranges. But this approach fully reflects any local noise
stemming from a particular weather station. 

An alternative approach is to perform a weighted average of nearby
weather stations. Figure \ref{fig:Spatial-interpol}C shows this procedure
for the $k$ nearest stations to a county's centroid (with $k=5$).
In computing these weighted averages, the weights are proportional
to the inverse of the distance to the county's centroid. This way
stations that are closer to the county are weighted more heavily.
Naturally, these weights must add to unity. This approach smooth out
local noise from any particular weather station. The downside, obviously,
is that it may smooth out temperature too much. especially over location
with major spatial climatic differences over short distances (e.g.
California). In addition, this approach may be problematic if weather
stations are extremely sparse. In such cases this approach could potentially
over-smooth weather conditions. One potential fix is to reduce the
number $k$ of neighboring stations used in the interpolation. In
general, this approach appears more suitable to situations with relatively
dense or homogeneous network of weather stations.

In other contexts, relying on all the weather stations within a certain
distance of a county could be preferable. That is the strategy used
in generating Fig.\ref{fig:Spatial-interpol}D. In that interpolation,
I chose all stations within 1° (\textasciitilde 111 km) and relied
on an inverse-distance weight. This procedure may be appropriate when
the station distribution is highly sparse and heterogeneous. If the
distance cutoff is too large, this approach may also over-smooth weather
conditions in regions with contrasting climates over short distances
(e.g. California). However, setting a distance cutoff that is too
small may lead to missing observations if station density is sparse
in certain parts of the study region.

I should note that interpolating temperature variables is prone to
smaller errors than precipitation. Temperature is a smoother field
and is more amenable to these procedures. Precipitation, on the other
hand, is spatially discontinuous, particularly at finer temporal scales
(e.g. daily), so interpolation can introduce substantial error at
fine spatial scales.

In some circumstances a researcher might be interested in harnessing
high spatial resolution from one dataset with the higher temporal
resolution of another. For instance, the PRISM data is available at
a daily scale starting in 1981. But the monthly PRISM data starts
in 1896. What if we were interested in deriving a daily PRISM-like
dataset prior to 1981 based on the PRISM grid? That is precisely what
\citet{schlenker_nonlinear_2009} did, where they harnessed the high
spatial resolution of the monthly PRISM data over 1950-2005, and combined
with with daily weather station data over the same period. Their goal
was to derive the daily distribution of temperature at the fine spatial
scale of the PRISM grid.

Here I illustrate how to do this for daily maximum temperature (Tmax)
and precipitation (ppt) using the monthly 4-km dataset from PRISM
and daily but relatively sparse weather station data from the GHCN
database. I focus our attention on weather conditions over CONUS for
August 16, 2020 and present the procedure step by step. 

The first step is to interpolate daily weather station to the PRISM
grid. The goal of this step is to ``infuse'' the PRISM grid cells
with the temporal variation from the weather station data. Note that
the PRISM grid has 872,505 cells of which 481,631 fall over land,
whereas the GHCN database has maximum temperature data for 6,028 stations
and precipitation data for 14,492 stations.\footnote{The weather stations counts concern balanced weather stations for
the month of August of 2020.} This means that this interpolation has a larger ``target'' dataset
(PRISM grid) than the ``source'' dataset (stations). This will expectedly
lead to fairly smoothed out patterns at a very fine scale and might
not respect fine-scale spatial differences arising due to orography
or proximity to the coasts. Specifically, I select the 5 nearest weather
stations for each PRISM grid cell and compute an inverse-distance
weighted average of each variable. This is repeated for each PRISM
grid cell, noting that the weather station population varies by variable.
This step leads to 31 daily interpolated layers for each variable
for the month of August of 2020. 

The second step consists in ensuring that the daily interpolated PRISM
data are consistent with the monthly PRISM data. That is, we constrain
the daily data so that when aggregated it matches exactly the monthly
PRISM data. To do this for temperature, I start by subtracting monthly
average from the daily interpolated data layers. Essentially, this
only leaves the local anomalies in the interpolated data. I then proceed
to add the monthly average from the reference PRISM grid to these
daily interpolated anomalies. This reference monthly file has a finer
and more accurate rendering of spatial climatic differences. For precipitation,
I scale the daily precipitation variables based on the ratio of the
total precipitation from the daily interpolated files and the reference
monthly PRISM data layer. This procedure ensures that I do not create
``ghost'' rainfall in places and days where it did not rain.

\begin{sidewaysfigure}
\includegraphics[scale=0.27]{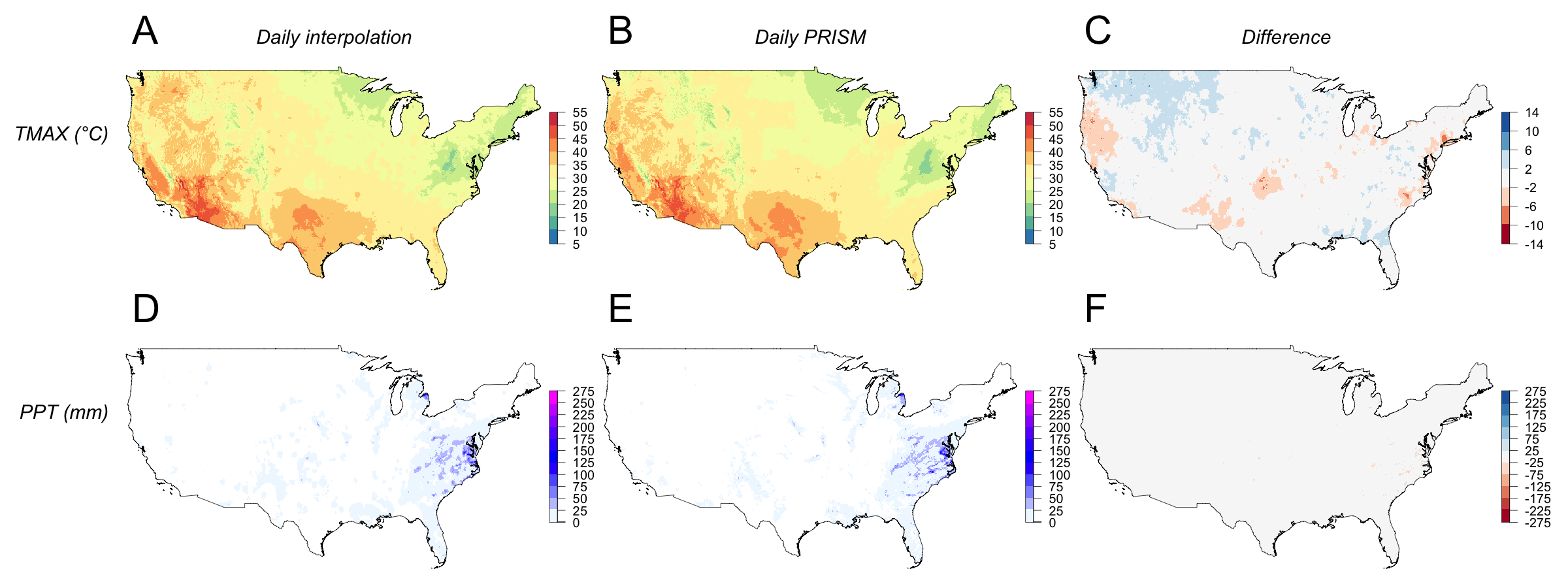}

\noindent\begin{minipage}[t]{1\columnwidth}%
\textit{\footnotesize{}Notes:}{\footnotesize{} All maps show maximum
temperature (Tmax) on August 16 of 2020. A: Tmax over 3,431 weather
stations from the GHCN database. B-C: Inverse-distance spatial interpolation
based on the stations located within 0.5 and 1° of the each county's
centroid. D: Inverse-distance interpolation based on the 5 nearest
stations to each county's centroid.}%
\end{minipage}

\caption{Spatial interpolation of weather station data.\label{fig:Spatial-interpol-1}}
\end{sidewaysfigure}

I summarize the results of this interpolation in figure \ref{fig:Spatial-interpol-1}.
Fig. \ref{fig:Spatial-interpol-1}A shows the resulting interpolated
maximum temperature for August 16, 2020. Notice that this interpolation
technique preserves very fine-scale spatial variation close to mountains
and coasts. As previously mentioned, PRISM provides daily data since
1981 so we can contrast our interpolation with theirs, shown in panel
B. The spatial patterns are very similar, but not identical as indicated
in panel C showing the difference between panels A and B. The PRISM
team likely operates a more sophisticated technique than what I employ
here.

Figure \ref{fig:Spatial-interpol-1}D shows the interpolation for
daily precipitation for the same day. The spatial patterns of precipitation
appear slightly less realistic than the daily PRISM data shown in
panel E, but the differences are surprisingly small as indicated in
panel F. However, note that differences can be very large (e.g. shows
rain in panel D and no rain in panel E) in certain places due to differences
in interpolation techniques.

While simple interpolation techniques like this appear to work, I
invite the reader to exhibit extreme caution when performing these
calculations. The output from these procedures is not validated which
means that we have a very limited understanding of the type of the
underlying measurement error. Advanced users are invited to seek additional
resources in kriging and more advanced geospatial statistic tools.

\subsection{Aggregation of gridded weather data\label{subsec:Aggregation-of-gridded}}

Here we turn our attention on how to efficiently aggregate gridded
weather data by administrative units (e.g. counties). As discussed
in section \ref{subsec:Weather-data}, gridded or raster data are
matrices with data on a particular variable (e.g. precipitation) linked
to information on how the cells of the matrix map into into space
and time. These are essentially geographically and temporally referenced
matrices. This characterization is important because it means that
manipulating such data can be greatly improved by relying on matrix
algebra which are performed efficiently by programming software like
R. Weather raster files can come in a variety of formats but perhaps
the most common is NetCDF. These are typically multi-layer raster
files in which each layer corresponds to a time period. For instance,
it is common to see daily weather datasets provided as multi-layer
annual files (1 file with 365 layers). 

The way that each grid cell in a raster is mapped into a particular
location on Earth is determined by a few parameters, including the
extent, spatial resolution, and projection. The extent defines the
``boundaries'' of the raster. Because rasters are rectangular, the
extent is simply a matrix providing information on the southern, northern,
western and eastern limits of the raster. These limits are typically
provided in degrees (but not always). A global dataset can thus range
from -180° W to +180° E, for rasters centered in Europe, and from
-180° S to 180°N, although it is rare to see rasters extend to the
poles. The spatial resolution of the raster simply indicates the size
of a grid cell in degrees, although sometime also in meters. Finally,
the projection is the rule by which the surface of the Earth is flatten
for the purpose of visualization. Gridded weather datasets are generally
``unprojected'' using a simple latitude-longitude (or lat-lon) projection.
This portrays the gridded surface of the earth as a flat rectangle
on a computer screen. Importantly, changing the projection of a raster
can be time consuming (at least in R) because the boundaries of each
grid cell need to be reprojected. As a result, it is often preferable
to change the re-project polygon data (e.g. administrative boundaries)
to match that of the raster data, and not the other way around.

The key point to efficiently map gridded data to administrative units
is to recognize gridded data as a matrix. A multi-layer raster containing
daily information for $N$ grid cells and $T$ time periods can be
through of as an $N\times T$ matrix, which I denote $G$ (for \uuline{g}ridded).
The target aggregated dataset would be a $n\times T$ matrix, which
I denote $A$ (for \uuline{a}ggregated or \uuline{a}dministrative).
We usually have $n\ll N$, but not always (more on this later). The
``trick'' is to consider the transformation of $G$ into $A$ as
a matrix multiplication of the form:

\[
\underset{n\times T}{A}=\underset{n\times N}{P}\times\underset{N\times T}{G}
\]

Matrix $P$ is a transformation or projection matrix that converts
the gridded data $G$ into aggregate data matrix $A$. That transformation
is simply a weighted average of rows in $G$ to obtain aggregate matrix
$A$. To clarify this further, the first row of $P$ corresponds to
the first administrative unit in the aggregated dataset (e.g. a country,
state, county, etc.). The first column of $P$ corresponds to the
first grid cell in the gridded dataset. Thus the first row corresponds
to a vector of weights for computing a weighted average of grid cells
in $G$ in order to obtain the data for the first administrative unit,
which is located on the first row of $A$.Values in each row in $P$
must add to 1, and only grid cells falling within the administrative
unit should have non-zero positive values. 

That means that $P$ is mostly full of zeros and could be adequately
represented as a sparse matrix on a computer. For a sparse matrix,
a computer only holds the values and location of non-zero values in
the memory. This is a memory efficient way of storing large matrices
and for computing fast matrix multiplications.

The projection matrix $P$ can be created in various ways. Perhaps
the most basic approach is to give equal weight to every grid cell
falling within each administrative boundary. However, as we saw in
Fig. \ref{fig:raster-aggregation}, giving the same weight to every
location within a county or state may be misleading because it does
not convey information about the locations within the administrative
units where the economic activity of interest operates or relies upon.
This is particularly critical for large administrative areas like
countries, or large geographically diverse states, like in the Western
parts of the US.

A more common approach of constructing the projection $P$ matrix
is for it to reflect weights that correspond to areas where relevant
economic activities take place. For instance, in the case of economic
studies exploring the effect on overall economic activity, it is common
to see gridded weather data aggregated over areas where population
are located within the administrative units. In the case of agriculture,
it is common to rely on cropland, or cropland and pastures, as the
underlying land cover information to construct such weights. For global
studies one can rely on global datasets describing the global distribution
of agricultural lands (e.g. \citealp{ramankutty_farming_2008}). For
more regional studies, one can rely on finer scale land cover datasets.
In the US context, it is common to rely on either the USDA NASS Cropland
Data Layer (CDL) or USGS National Land Cover DataBase (NLCD). The
CDL is an annual product with 30m resolution and is differentiated
by crop. The NLCD is also 30m but has broad land cover categories
as it pertains to agriculture (e.g. cropland or pasture).

The advantage of performing spatial aggregation in this manner is
that the researcher only needs to compute $P$ once. Matrix $P$ is
essentially a matrix of aggregation weights. The spatial aggregation
can be done in a fraction of a second by performing the matrix multiplication
above because these matrix operations can be done very fast with current
computers. The main bottleneck in the spatial aggregation is reading
matrix $G$ into the memory and perhaps writing matrix $A$ to the
disk. This workflow presents a substantial gain in speed when aggregating
very large datasets with either high spatial or temporal resolution.
The alternative typically consists of using canned functions that
perform overlays of raster and polygons for each raster layer. Such
strategies have a substantial overhead that make the rapid aggregation
of large datasets infeasible.

\begin{figure}
\includegraphics[scale=0.25]{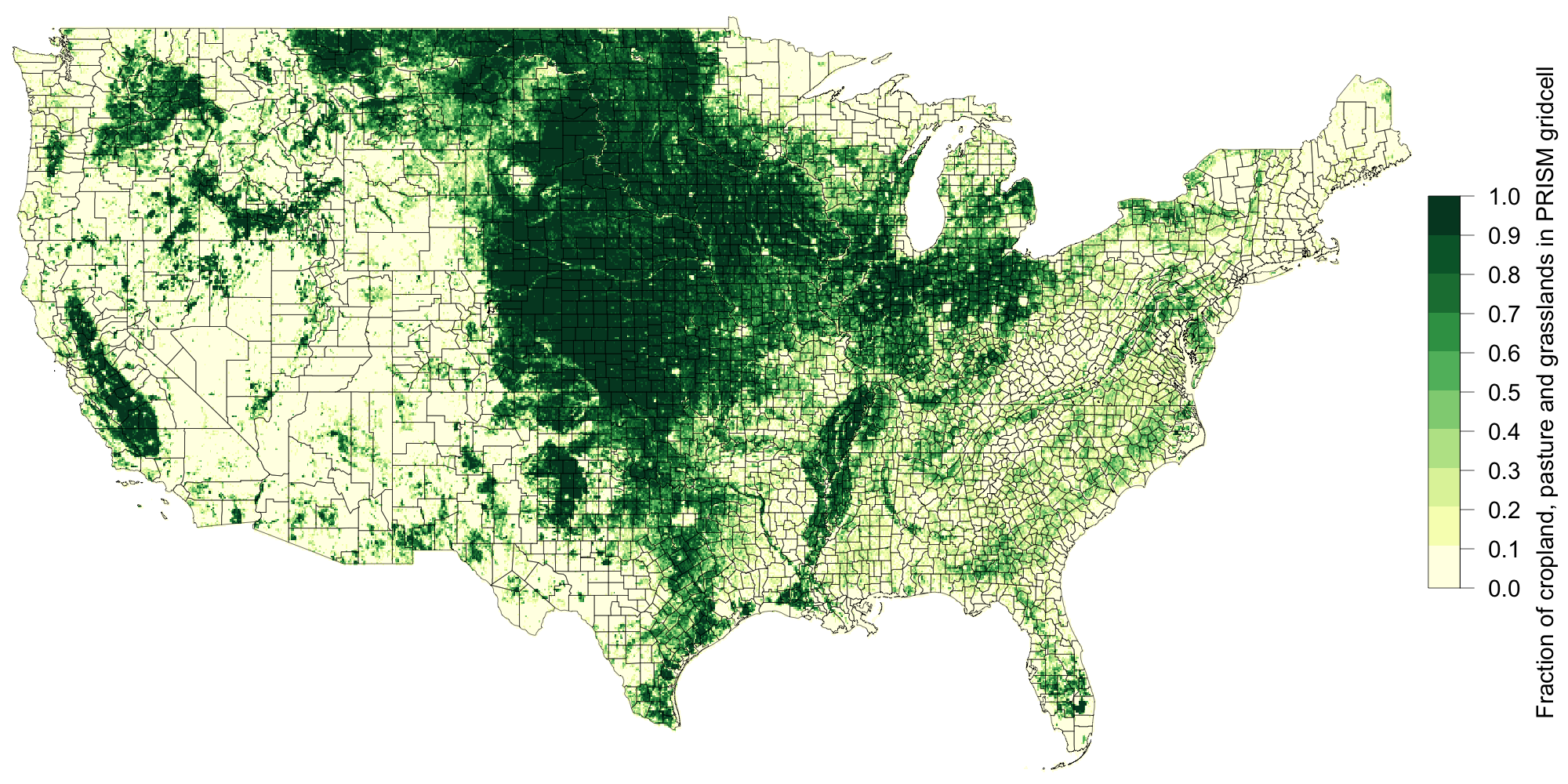}

\noindent\begin{minipage}[t]{1\columnwidth}%
\textit{\footnotesize{}Notes:}{\footnotesize{} The cropland, pasture
and grassland shares are computed based on 30m land cover data from
the National Land Cover Database (NLCD) for 2016.}%
\end{minipage}

\caption{Fraction of cropland, pasture and grassland in each PRISM gridcell.
\ref{fig:cropland cover}\label{fig:cropland cover}}
\end{figure}

The R code included with this chapter provides an illustration of
how to compute the matrix $P$ based specific types land cover falling
within each US county to aggregate PRISM data (see \texttt{1\_weather\_data.R}).
The first step is to create a new land cover raster that matches the
target PRISM raster grid. This aggregation can be done by computing
``zonal statistics'' which is simply computing the frequency of
small 30m land cover pixels falling within each PRISM grid cell. Fig.
\ref{fig:cropland cover} shows the fraction of each PRISM grid cell
covered by cropland, pasture and grassland. As expected, areas in
the Midwest and the Central Plains are mostly allocated to these classes.
The second step is to derive a vector of weights that sum to 1 within
each county, based on the land cover fractions of the grid cell falling
within each county. The third step is to store these aggregation weights
in a sparse matrix $P$. The final step is to perform the matrix multiplication
to perform the aggregation.

Importantly, note that the interpolation of weather station data described
in subsection \ref{subsec:Aggregation-of-point} can also be conducted
using sparse matrices. In fact, the implementation shown in the R
code provided does precisely that. In other words, performing averages
of observations over nearby weather stations does not require a loop.
Doing a loop would be tremendously inefficient and possibly make the
project intractable for large datasets. The aggregations described
here are simply linear transformations of matrices, whether the source
matrix represents raster or point data, and where the transformation
matrix $P$ is a sparse matrix with aggregation weights.

\subsection{Estimating non-linear effects\label{subsec:Estimating-non-linear-effects}}

Now that we have covered the basics of how to aggregate weather data
to administrative units, I focus our attention to the estimation of
nonlinear effects of temperature. Specifically, I illustrate how to
estimate the semi-parametric model introduced in \citet{schlenker_nonlinear_2009}.
Rather than estimating the effect of temperature on crop yields, that
study examined the effect of \textit{exposure} to different levels
of temperature on crop yields. I emphasize the term \textit{exposure},
because the underlying variables in the analysis are actually measures
of the amount of \textit{time }spent at various temperature intervals
or ``bins''. 

The motivation behind this approach is that temporal averaging of
temperature conceals the exposure to extreme temperatures. Two temperature
sequences may have the exact same average, but one may exhibit considerably
more exposure to high temperature. The underlying hypothesis is that
exposure to various levels of temperature affect agricultural outcomes
(e.g. crop yield) very differently, so capturing the varying effects
of exposure to the entire temperature distribution may yield deeper
insights. 

Capturing such linearities requires information on how much time is
spent at each temperature level or ``bin''. These bins are typically
1°C wide, ranging from say 0° to 40°C. Intra-daily (e.g. hourly) data
is often impossible to obtain in a reliable fashion so deriving these
intra-daily distributions often requires making assumptions about
the temperature-time path. A common assumption is that temperature
follows a sine curve passing between the minimum and the maximum temperature
of each day. This naturally requires data on daily minimum and maximum
temperature.

\begin{figure}
\includegraphics[scale=0.35]{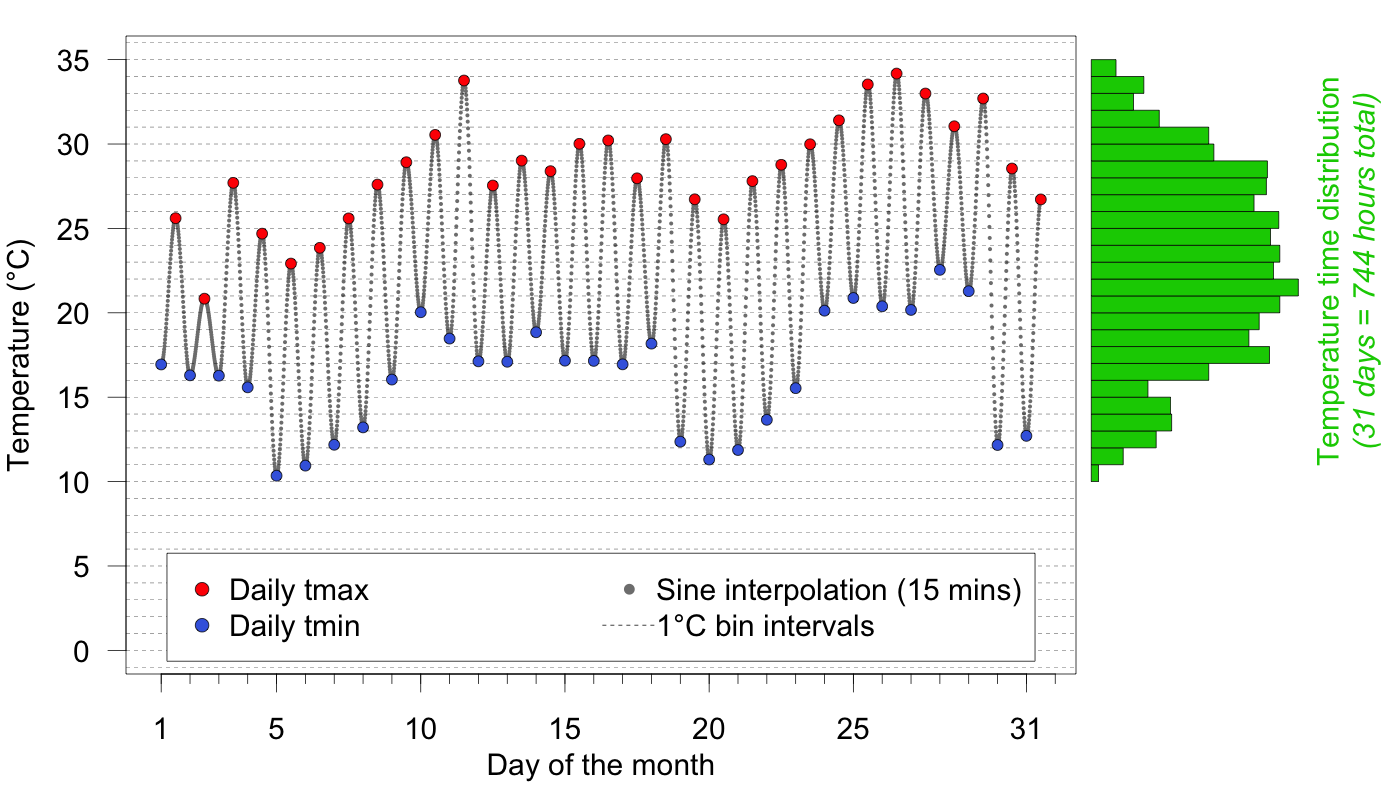}

\noindent\begin{minipage}[t]{1\columnwidth}%
\textit{\footnotesize{}Notes:}{\footnotesize{} The figure represent
minimum (Tmin) and maximum (Tmax) temperature data from a random location
in the lower 48 states from the PRISM dataset during the month of
August of 2020. The grey dots describing a sine curve between consecutive
Tmax and Tmin are obtained via interpolation every 15 minutes. Temperature
intervals of 1°C are highlighted in dashed lines over the entire month.
The green distribution on the right shows the underlying temporal
distribution of temperature throughout the month. It is essentially
a histogram of the 15-min points that I just described. The distribution
adds to the total number of hours (or days) in the month.}%
\end{minipage}

\caption{Illustration of the construction of temperature bin exposures from
daily minimum and maximum temperature. \label{fig:bin distribution}}
\end{figure}

The R code provided illustrates how to build exposure data from a
sequence of daily minimum and maximum temperature. The results for
a random location in the US is shown in Fig. \ref{fig:bin distribution}.
The figure shows the daily sequence of Tmax and Tmin with in red and
blue points, respectively. To this I add an interpolated intra-day
temperature trajectory. The code generates a series of points at 15-minute
intervals based on a double sine curve that passes through Tmax and
Tmin of consecutive days. Computing the exposure bins simply consists
in determining the frequency of these 15-minute interval points throughout
the month. This distribution is shown in green on the right side of
the figure. Note that the support of this distribution is time and
is measured in hours or days, not °C. The distribution describes how
much time is spent in each temperature interval. By construction,
summing over all the bins over a month adds to the number of hours
or days in that month. Importantly, note that the temperature ``bins''
described here are not counts of days in which the average temperature
falls in a particular interval. 

It is important to emphasize that these bins we are discussing are
exposure bins. They are not dummy variables that count whether the
average temperature fell within a particular bins. That approach would
not capture within intra-day variation in temperature. This is a common
point of confusion. In addition, note how averaging temperature over
time, say over the entire month, would conceal exposure to extreme
temperatures within the month.

\begin{figure}
\includegraphics[scale=0.25]{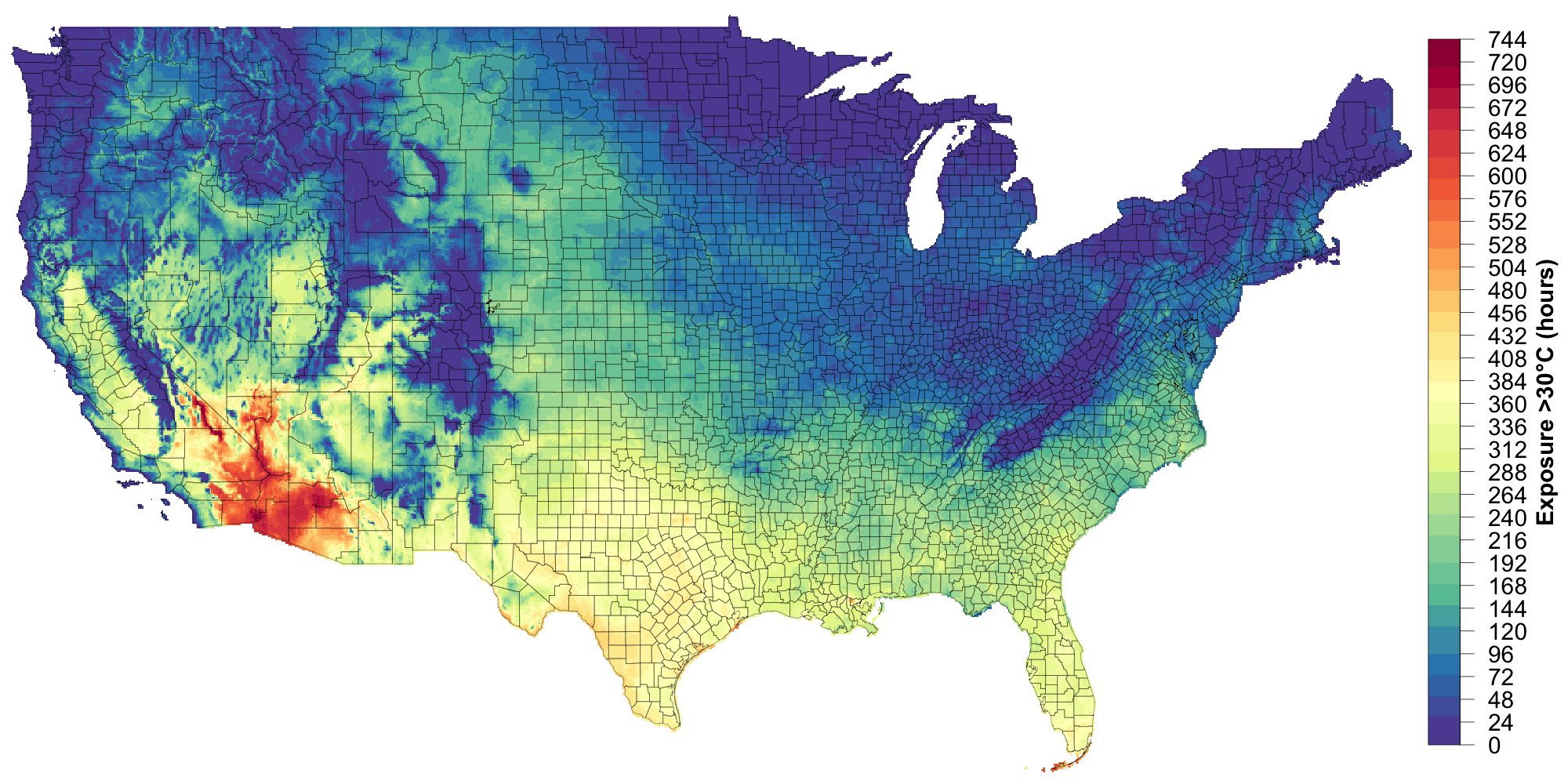}

\noindent\begin{minipage}[t]{1\columnwidth}%
\textit{\footnotesize{}Notes:}{\footnotesize{} The derivation of this
exposure relied on daily data from PRISM.}%
\end{minipage}

\caption{Map of exposure above 30°C in August of 2020. \label{fig:map EDD}}
\end{figure}

We can perform the ``binning'' exercise for all grid cells located
over the US and for each month of the year. In the code provided I
do this for every month over the 1981-2020 period for bins ranging
from $-10$ to $50$°C in 1°C intervals. For each month, the code
creates a file with 61 layers (one per bin) corresponding to the amount
of time spent in each bin. I illustrate the exposure above 30°C for
August of 2020 in Fig. \ref{fig:map EDD}. This is essentially showing
the sum of all the exposure bins above 30°C in August (depicted for
one grid cell in Fig. \ref{fig:bin distribution}) over all PRISM
grid cells. In the map you can appreciate the importance of latitude
and orography in explaining cross-sectional differences in temperature
exposure. For instance, certain parts of the Southwest experienced
more than 720 hours above $30$°C. August has 31 days or 744 hours,
meaning that some of these regions experienced temperatures above
$30$°C over virtually every single moment of the month. In contrast,
other regions of the country experienced little to no exposure beyond
$30$°C (dark blue).

Note that these exposure bins for temperature are related to the degree-day
concept discussed in subsection \ref{subsec:Degree-days-and-agriculture}.
In fact, degree days can be computed directly from these exposure
bins. Specifically, computing degree-days between a low threshold
$\underline{h}$ and a high threshold $\bar{h}$ can be written as:

\[
DD_{\underline{h}\textrm{ to }\bar{h}}=\sum_{k=\underline{h}}^{\bar{h}-1}z^{k}\times\left(\underline{h}-30+1\right)
\]

where $z^{k}$ is the exposure (e.g. in days) to the $k$-th temperature
bin. For instance, when computing growing degree-days between 8 and
32°C, we would set set $\underline{h}=8$ and $\bar{h}=32$. When
computing ``extreme'' degree-days, say above 30°C, we would set
$\underline{h}=30$ and $\bar{h}$ at a very high level that is never
reached (e.g. 60°C). In essence, the expression above approximates
the area under the temperature curve and between a lower and an upper
threshold.

Now let's move to how to represent the effect of temperature exposure
on crop yield following the approach laid out in \citet{schlenker_nonlinear_2009}.
The underlying data generating process presented in that study is
of the form:

\[
y_{it}=\intop g(h)\phi_{it}(h)d(h)+p_{it}+p_{it}^{2}+\psi(t)+\alpha_{i}+\epsilon_{it}
\]
where $y_{it}$ is the log of yield in location or county $i$ and
year $t$, $p_{it}$ is growing-season precipitation, $\psi(t)$ is
a time trend, $\alpha_{i}$ is a county fixed effect and $\epsilon_{it}$
is the error term. The first term represents the effect of temperature
on crop yield, where $g(h)$ is the marginal effect of temperature
$h$ on yield, and $\phi_{it}(h)$ is the growing-season density at
$h$ in that location and year. The integral simply means that the
product of marginal effect and exposure is summed over the entire
temperature range, given that the integrating variable is $h$. This
continuous representation is theoretical and is not tractable for
estimation. However, that integral can be approximated in various
ways empirically. In the original study, the authors provided 3 ways
of approximating this function, using a piece-wise linear function,
a step function and a Chebyshev polynomial of degree 8.

In this chapter and in the included R code I illustrate how to estimate
these models using step functions, natural cubic splines, and Chebyshev
polynomials (see \texttt{2\_nonlinear\_effects.R}). I focus on US
corn yields for the 1981-2020 period east of the 100th meridian West
in order to focus on mostly rainfed counties. I set a growing season
ranging from August to September. I also bottom and top code the exposure
data so that bins range from 0 to 38°C. This avoid having too little
exposure at the tails of the temperature distribution, which makes
the estimation noisier.

Perhaps the most basic and intuitive approximation is to use a step
function. If the temperature range during the growing season ranges
from say 0 to 40°C, then the approximation with eight 5°C steps can
be represented as:

\[
y_{it}=\sum_{k=1}^{8}\beta_{k}z_{it}^{k}+p_{it}+p_{it}^{2}+\psi(t)+\alpha_{i}+\epsilon_{it}
\]

where $z_{it}^{k}$ is the amount of time spent in the $k$-th step
or temperature interval. This means that we essentially estimate separate
coefficients for each temperature interval. The regression coefficients
are to be interpreted as the effect on yield of spending an additional
hour (or day) in that particular bin.

\begin{figure}
\begin{centering}
\includegraphics[scale=0.22]{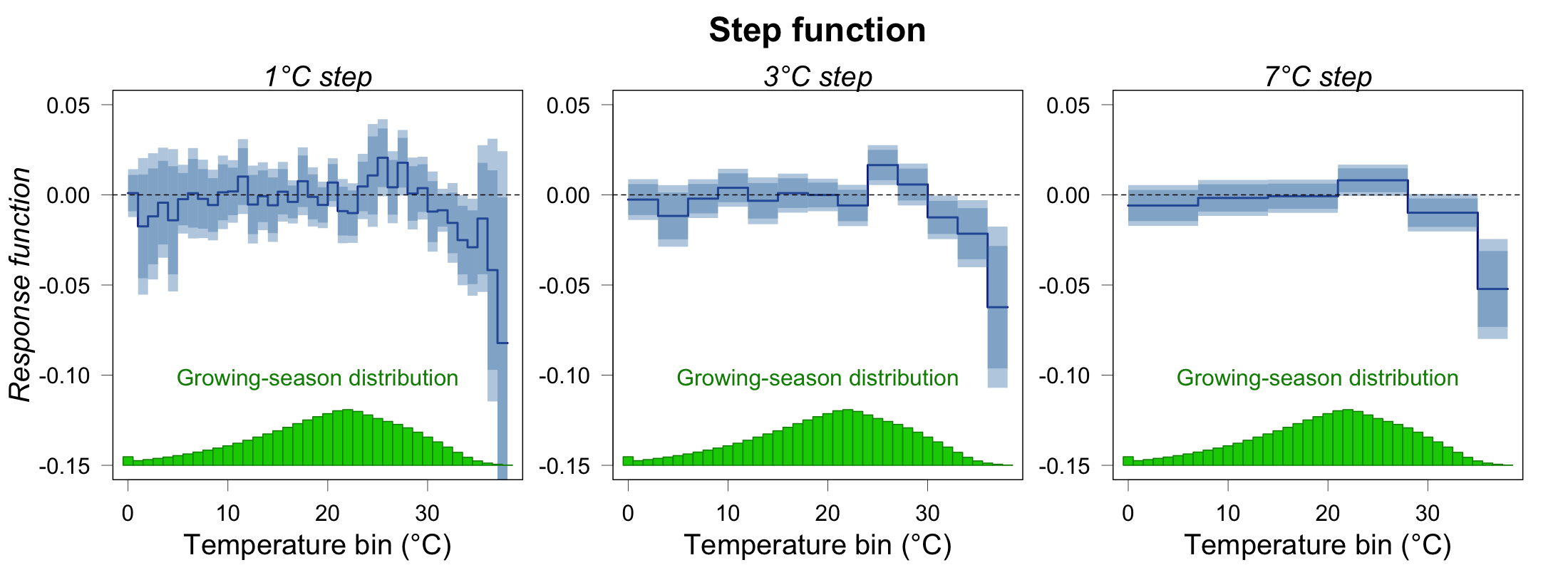}
\par\end{centering}
\noindent\begin{minipage}[t]{1\columnwidth}%
\textit{\footnotesize{}Notes:}{\footnotesize{} Standard errors are
clustered at the state and year level. The data covers US corn yields
over the 1981-2020 period east of the 100th meridian West. Precipitation
variables were included in the regression but coefficients are not
shown. The blue colored bands around the response function corresponds
to 95 and 99 percent confidence intervals. The green histogram represents
the growing-season of exposure to all temperature bins.}%
\end{minipage}

\caption{Effects of temperature exposure on corn yields based on step functions
of varying widths. \label{fig:step fun}}
\end{figure}

Figure \ref{fig:step fun} illustrates the marginal effects of models
with step functions of different widths. In each panel the blue line
represents the estimates for each step together with 95 and 99 percent
confidence intervals in blue. The green distribution underneath the
response function describes the growing-season temporal distribution
of temperature exposure in the sample. By construction, that distribution
sums to 183 days between April and September.

The first panel of Fig. \ref{fig:step fun} shows the response function
with 1°C steps. This leads to a fairly noisy response function, particularly
toward high levels of temperature. As a result, although the point
estimates suggest exposure to temperature above 30°C appear detrimental,
these effects are not statistically different from zero. This imprecision
likely results from the high degree of collinearity between neighboring
bins. Indeed, the time spent on any given year between 31 and 32°C
is always very correlated to the time spent between 32 and 33°C, and
so on for other bins. This is problematic and thus model like this
are better avoided. The role of bin size has also been discussed in
\citet{carter_identifying_2018}.

One natural way to avoid collinearity is to aggregate bins into wider
steps or intervals. The second and third panels in Fig. \ref{fig:step fun}
show the response functions for models based on steps that are 3°
and 7°C wide, respectively. As expected, the resulting response functions
are more precisely estimated. Both model also how that temperature
exceeding 30°C are detrimental to corn yield. For instance, the last
model indicates that an additional day of exposure to the last bin
over the growing season (183 days) reduces crop yield by 0.05 log
points, or about 5\%. That is a substantial reduction.

Note that the step function approach assumes that the marginal effects
between neighboring steps are unrelated. That is, the model does not
impose any structure on how smooth the response function could be,
which could render the estimation unnecessarily noisy. 

One way to address this is to allow marginal effects to vary smoothly
across neighboring temperature bins. This can be implemented with
a natural cubic spline or a Chebyshev polynomial. Both of these approaches
involve a basis matrix $B$ which is used to project the temperature
bins (say $J$ bins) into a smaller space (say of size $K$) and thus
reduce dimensionality prior to estimation ($K<J$). That is, we are
able to estimate only $K$ parameters to represent the marginal effects
of $J$ individual bins.

To illustrate, the basis matrix $B$ of a natural cubic spline with
$J$ degrees of freedom evaluated over $K$ temperature bins has $K$
rows and $J$ columns. The basis matrix maps a $nT$-by-$K$ matrix
$Z$ of data, with columns representing temperature bin exposures,
into a $nT$-by-$J$ matrix $X$ of transformed variables used in
the regression analysis. In matrix form, this mapping can be represented
as:

\[
\underset{nT\times J}{X}=\underset{nT\times K}{Z}\times\underset{K\times J}{B}
\]

This essentially reduces the underlying binned data with $K$ bins
to $J$ regressors. Naturally, we select $J\ll K$ in order to substantially
reduce the dimensionality of the temperature space. We can now write
the regression model in the following form:

\begin{align*}
y_{it} & =\sum_{j=1}^{J}\sum_{k=1}^{K}\left(\gamma_{j}B_{j}^{k}z_{it}^{k}\right)+p_{it}+p_{it}^{2}+\psi(t)+\alpha_{i}+\epsilon_{it}\\
 & =\sum_{j=1}^{J}\gamma_{j}\underbrace{\sum_{k=1}^{K}B_{j}^{k}z_{it}^{k}}_{x_{it}^{j}}+p_{it}+p_{it}^{2}+\psi(t)+\alpha_{i}+\epsilon_{it}
\end{align*}

where $B_{j}^{k}$ is the element in the $k$-th row and $j$-th column
of the basis matrix $B$. The term $z_{it}^{k}$ corresponds to one
row (observation $it$) and the $k$-th column of $Z$, and $x_{it}^{j}$
corresponds to one row (observation $it$) and the $j$-th column
of $X$. That is, rather than estimating $K$ separate coefficients
for each individual bin $z^{k}$, we end up with only $J$ regressors
$x^{j}$. 

After the estimation, one can recover the marginal effects evaluated
at each of the $K$ bins by pre-multiplying the vector $\hat{\Gamma}$
(containing the $J$ estimated coefficients) by the basis matrix: 

\[
\underset{K\times1}{\hat{\beta}}=\underset{K\times J}{B}\times\underset{J\times1}{\hat{\Gamma}}
\]

This operation returns a vector $\hat{\beta}$ of temperature effects
evaluated at each one of the $K$ original temperature bins in matrix
$Z$. One can also easily derive an estimate of the variance of these
temperature effects as follows: 
\[
\underset{K\times K}{Var(\hat{\beta})}=\underset{K\times J}{B}\times\underset{J\times J}{Var(\hat{\Gamma})}\times\underset{J\times K}{B^{'}}
\]

Obtaining standard errors for the marginal effects consists in selecting
the squared root of the diagonal elements of $Var(\hat{\beta})$.

\begin{figure}
\begin{centering}
\includegraphics[scale=0.23]{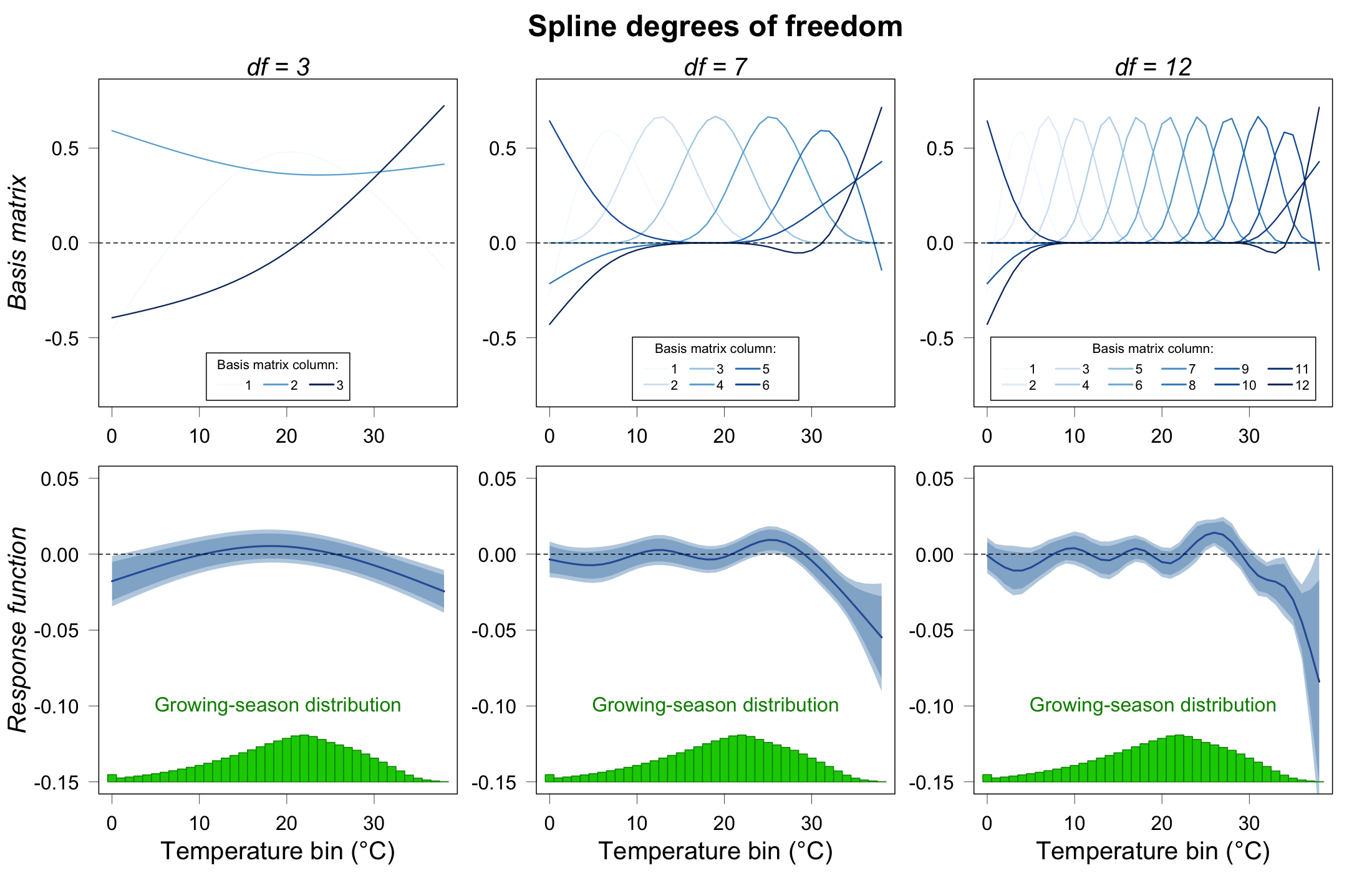}
\par\end{centering}
\noindent\begin{minipage}[t]{1\columnwidth}%
\textit{\footnotesize{}Notes:}{\footnotesize{} Standard errors are
clustered at the state level. The data covers corn yields over 1981-2020
east of the 100th meridian West. Precipitation variables were included
in the regression but coefficients are not shown. The blue colored
bands around the response function corresponds to 95 and 99 percent
confidence intervals. The green histogram represents the growing-season
of exposure to all temperature bins.}%
\end{minipage}

\caption{Effects of temperature on corn yields based on natural cubic splines.
\label{fig:spline}}
\end{figure}

To fix ideas, I illustrate this estimation technique using cubic natural
splines with different degrees of freedom in Fig. \ref{fig:spline}.
Each panel in the top row shows the the columns of the basis matrix
$B$ for splines with 3, 7 and 12 degrees of freedom. With temperature
bins defined over the 0 to 38°C, the basis matrices have sizes of
$39\times3$ , $39\times7$ and $39\times12$, respectively. The top
row shows these basis matrices performs linear transformation of neighboring
bins that are mapped into a reduced number of regressors. This means
that each regressor roughly reflects the effects of fluctuations in
temperature exposure occurring in neighboring bins. Importantly, these
regressors can be transformed ``back'' to the original support,
as previously mentioned and as illustrated in the second row of the
figure.

Each panel in the bottom row of Fig. \ref{fig:spline} shows the response
functions (the vector $\hat{\beta}$ derived above) along with 95
and 99 percent confidence bands (based on $Var(\hat{\beta})$) corresponding
to these splines with varying degrees of freedom. I also depict the
underlying distribution of temperature exposure over the growing season
in green.

The first panel of Fig. \ref{fig:spline} corresponds to a spline
with 3 degrees of freedom. This response function clearly exhibits
limited flexibility. This is evident by the symmetric response function
shown on the lower left panel. Note, however, how the spline with
7 degrees of freedom allows for a more flexible estimation of the
response function. This response function shows that exposure above
30°C are clearly detrimental to crop yields. 

Moving to the most flexible specification with 12 degrees of freedom,
shows a response function that is a bit more unstable but still shows
a distinct pattern where exposure to temperature exceeding 30°C are
clearly detrimental. Note that the response function because more
imprecise at high levels of temperature. This is likely driven by
the high flexibility of the model over a temperature range with relatively
little variation.

\begin{figure}
\begin{centering}
\includegraphics[scale=0.23]{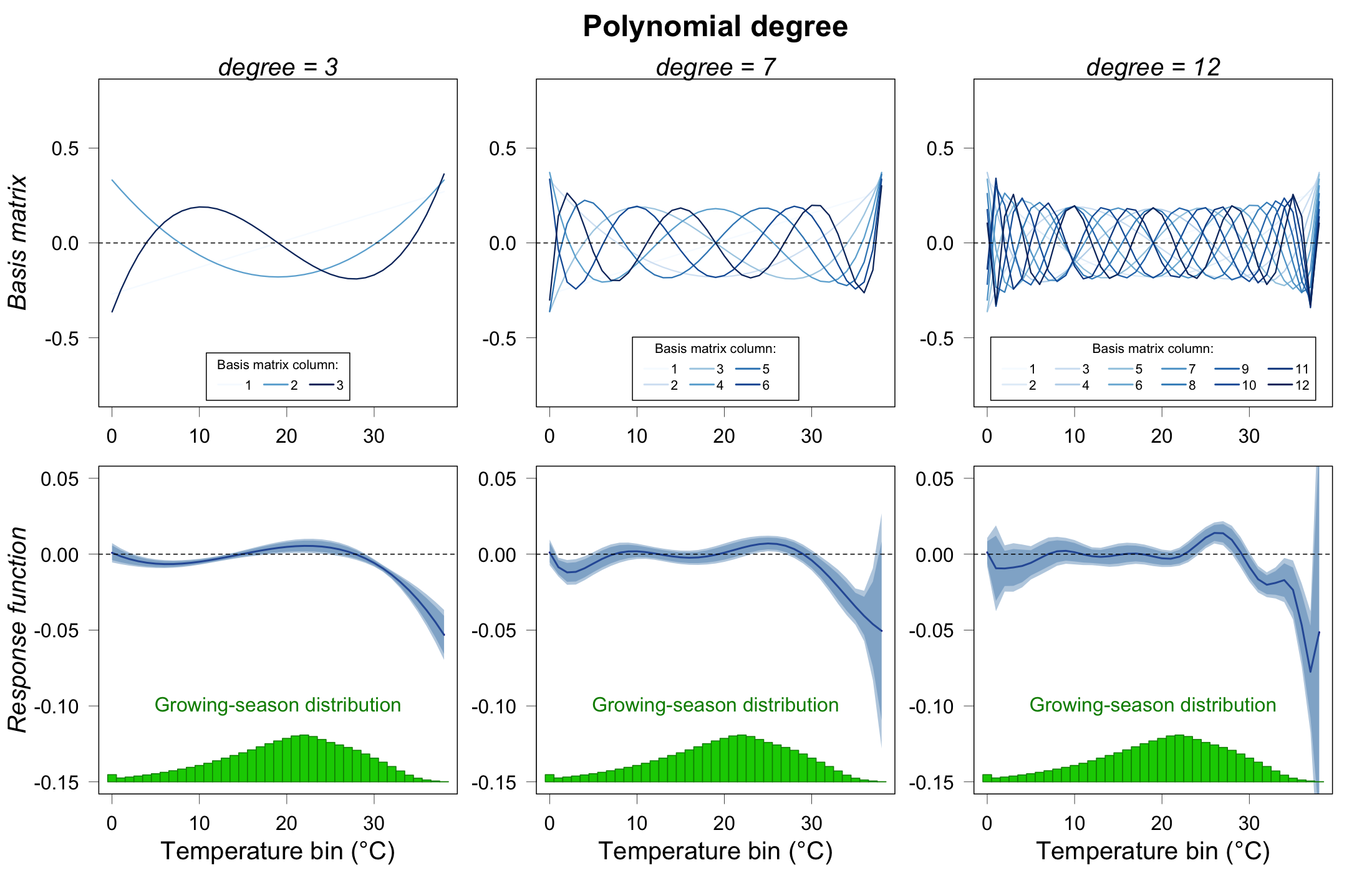}
\par\end{centering}
\noindent\begin{minipage}[t]{1\columnwidth}%
\textit{\footnotesize{}Notes:}{\footnotesize{} Standard errors are
clustered at the state level. The data covers corn yields over 1981-2020
east of the 100th meridian West. Precipitation variables were included
in the regression but coefficients are not shown. The blue colored
bands around the response function corresponds to 95 and 99 percent
confidence intervals. The green histogram represents the growing-season
of exposure to all temperature bins.}%
\end{minipage}

\caption{Effects of temperature on corn yields based on Chebyshev polynomials.
\label{fig:poly}}
\end{figure}

It is also possible to estimate these temperature effects based on
a Chebyschev polynomial. The procedure is very similar to a spline
given this strategy also involves a basis matrix. The first row of
Fig. \ref{fig:poly} shows the columns of the associated basis matrix
for Chebyshev polynomials of degree 3, 7 and 12. Unlike for the spline,
note that the values are not locally defined. What that means is that
each regressor carries information regarding exposure to all bins
of the temperature distribution. As a result polynomials tend to be
a bit less table around the extremes than splines. The bottom row
of the figure shows the response functions and the underlying distribution
of temperature exposure. Similar to the spline, we find that exposure
above 30°C appears detrimental to corn yields. Note how allowing too
much flexibility in the polynomial lead to much noisier effects around
the right tail of the distribution.

Overall, this section shows that allowing too much flexibility in
the semi-parametric response function leads to relatively noisy effects
at the extreme end of the temperature distribution. We also see that
not allowing for enough flexibility tends to understate the effects
of extreme temperatures.

I should highlight that there has not been a formal exploration of
the advantages of these semi-parametric approaches. These models tend
to fit the data better than alternative models based on average temperature.
But it is unclear to what extent these approaches could reduce issues
related to spatio-temporal aggregation bias. 

To conclude, note that an important assumption of the approach presented
here is that the effects of temperature exposure are additive throughout
the growing season. In other words, the timing of temperature exposure
is irrelevant. This obviously contradicts agronomic conventional wisdom
stipulating that the timing of weather conditions is particularly
important for crop yield determination. I now turn to a generalization
of this model to allow for time-varying effects within the growing
season.

\subsection{Estimating within-season varying effects\label{subsec:Estimating-within-season-varying}}

The timing of environmental conditions plays a critical role in agriculture
particularly in crop production \citep{fageria_physiology_2006}.
In the case of many field crops like cereals and leguminous crops,
the ability of plants to store biomass in useful parts of the plant
(e.g. grain) is in large part determined by the success of their flowering
process. If the flowering process falters, then the plant loses its
ability to store biomass in seeds (grain fill). Importantly, flowering
is a delicate stage of plant development that is fairly vulnerable
to environmental stresses. Flowering also occurs over a relatively
short period of time around the middle of the crop cycle in annual
crops, meaning that environmental conditions can have drastically
different effects throughout the growing season.

So how does this affect the estimation of statistical crop yield models?
Most standard statistical yield models do not consider the timing
of environmental conditions in great detail. For instance, the models
we explored in the previous subsection assume additivity of weather
effects on yield. That means that the timing of environmental conditions
within the season are irrelevant. 

There have been various efforts over the years to account for these
within season time-varying effects. The most basic approach is to
simply include monthly weather variables for various critical period
of the growing season. However, those approaches have been found not
to substantially improve model fit or lead to significantly different
conclusions than models that assume additivity (e.g. see the appendix
in \citealp{schlenker_reply_2009}). However, these efforts commonly
rely on calendar periods of the year rather than actual stages of
crop development (e.g. \citealp{gammans_negative_2017}).

Perhaps the earliest economic study exploring climate change impacts
on crop yields that accounts for biophysical features underlying the
non-additivity of weather in the growing season is \citet{kaufmann_biophysical_1997}.
In that study, the authors define weather variables over periods corresponding
to crop development stages. Many more studies have adopted this approach.
For instance, \citet{ortiz-bobea_modeling_2013} estimates a corn
yield model with 3 sub-seasons matching crop stages to analyze the
effectiveness of changing planting as an adaptation to rising damages
from a warming climate. Another example is \citet{welch_rice_2010}
that matches weather conditions to vegetative and ripening phases
of rice. In a recent study, \citet{shew_yield_2020} link weather
conditions to wheat development stages to analyze variations in sensitivity
to extreme heat across cultivars in South Africa.

One potential limitation of previous work is that it assumes that
the effects between neighboring portions of the growing season are
independent. That is, nothing in the modeling approach allows for
marginal effects of weather conditions to vary smoothly within the
growing season itself.

To address this limitation \citet{ortiz-bobea_unpacking_2019} introduce
a bi-dimensional spline that allows the effects of soil moisture and
temperature to vary smoothly in levels and throughout the growing
season. This model is akin to a bi-dimensional generalization of the
crop yield model presented in the previous subsection. We now not
only have exposure bins to various levels of environmental variables,
but time within the season also becomes a ``bin''. So rather than
fitting a spline over a vector of binned exposures, this approach
applies a tensor spline to a 2-dimensional set of bins.

That study introduces a conceptual model similar to the following,
where crop yield may be differently affected by the distribution of
environmental conditions throughout the growing season:

\[
y_{it}=\intop\intop g(h,p)\phi_{it}(h,p)d(h)d(p)+p_{it}+p_{it}^{2}+\psi(t)+\alpha_{i}+\epsilon_{it}
\]

where $\phi_{it}(h,p)$ describes the distribution of the environmental
variable $h$ at each level of progress $p$ in the growing season,
and $g(h,p)$ describes the marginal effect of the environmental variable
throughout the season. Although he I only show one environmental variable,
the study considers soil moisture and air temperature. The rest of
the specification is analogous to the specification in the previous
sub-section.

Similarly, one cannot estimate this model with a double integral,
but one can approximate the $\phi_{it}(h,p)$ with 2-dimensional bins
and then employ a semi-parametric technique to estimate $g(h,p)$.
Again the 2 dimensions are progress in the season and the level of
the environmental variable. For example, in the case of temperature
throughout the growing season, we can define exposure bins over 1°C
temperature intervals every week of the growing season. In this example
we thus have 1°C bins in the ``temperature'' dimension, and weekly
bins in the ``season progress'' dimension.

Let me illustrate how to construct the tensor spline necessary to
reduce the dimensionality of these 2D bins. Let's start by considering
the basis matrix for a natural cubic spline for the environmental
variable, which constitutes the first ``dimension''. Let's denote
this matrix $B_{1}$. With $J_{1}$ degrees of freedom evaluated over
$K_{1}$ bins, this matrix has a dimension of $K_{1}$ rows and $J_{1}$
columns. These bins can be 1°C interval if we are dealing with temperature.
Similarly, let's define a basis matrix $B_{2}$ for the second ``dimension''
which is the time or progress within the growing season. With $J_{2}$
degrees of freedom evaluated over $K_{2}$ bins, this matrix has a
dimension of $K_{2}$ rows and $J_{2}$ columns. The associated can
be weeks or similarly short time intervals within the growing season. 

The tensor basis matrix is constructed based on the kronecker product
of these two basis matrices such that $B=B_{1}\varotimes B_{2}$ is
a matrix with $K_{1}K_{2}$ rows and $J_{1}J_{2}$ columns. Essentially,
our 2D bin space has $K_{1}K_{2}$ bins ($K_{1}$ in the first ``variable''
dimension and $K_{2}$ in the second ``season progress'' dimension)
and the goal of this tensor basis matrix is to reduce the dimensionality
of this space prior to estimation. This means that our binned data
can be stored in a matrix $Z$ with $nT$ and $K_{1}K_{2}$ columns.
Each row of that matrix corresponds to one observation (e.g. a county-year
$it$). Thus the tensor basis matrix $B$ maps a $nT$ by $K_{1}K_{2}$
matrix $Z$ of 2-dimensional binned data, with columns representing
variable and progress bin exposures, into a $nT$ by $J_{1}J_{2}$
matrix $X$ of transformed variables used in the regression analysis.
In matrix form, this mapping can be represented as:

\[
\underset{nT\times J_{1}J_{2}}{X}=\underset{nT\times K_{1}K_{2}}{Z}\times\underset{K_{1}K_{2}\times J_{1}J_{2}}{B}
\]

One way to visualize how this dimensionality reduction works is to
extract one row of $Z$, let's call it $z_{it}$ with dimension $1$
by $K_{1}K_{2}$, and organize it in ``two dimensions'' in a matrix
$z_{it}^{2D}$ of dimension $K_{1}\times K_{2}$. This essentially
re-arranges the bins so that we have $K_{1}$ rows of environmental
variable bins, and $K_{2}$ columns of season progress bins. What
the tensor spline is essentially doing is the following transformation:

\[
\underset{J_{1}\times J_{2}}{x_{it}^{2D}}=\underset{J_{2}\times K_{2}}{B_{1}^{'}}\times\underset{K_{1}\times K_{2}}{z_{it}^{2D}}\times\underset{K_{1}\times J_{1}}{B_{2}}
\]

where $x_{it}^{2D}$ is the transformed variable for observation $it$
rearranged in two dimensions. 

The regression analysis is performed with $J_{1}J_{2}$ regressors,
rather than on all the $K_{1}K_{2}$ individual bins. After the estimation,
one can recover the marginal effects evaluated at each of the $K_{1}K_{2}$
bins by pre-multiplying the vector $\hat{\Gamma}$ (containing a vector
with the $J_{1}J_{2}$ estimated coefficients) by the tensor basis
matrix: 

\[
\underset{1\times K_{1}K_{2}}{\hat{\beta}}=\underset{K_{1}K_{2}\times J_{1}J_{2}}{B}\times\underset{J_{1}J_{2}\times1}{\hat{\Gamma}}
\]

This operation returns a vector $\hat{\beta}$ of temperature effects
evaluated at each one of the $K_{1}K_{2}$ original temperature bins
in matrix $Z$. To visualize marginal effects on a 2-dimensional space,
one would have to rearrange $\hat{\beta}$ in 2 dimensions to obtain
a matrix $\hat{\beta}^{2D}$ of dimension $K_{2}\times K_{1}$. This
would allow showing marginal effects with season progress bins on
the horizontal axis and the variable bins in the vertical axis. One
can also easily derive an estimate of the variance of these temperature
effects as follows: 
\[
\underset{(K_{1}K_{2})\times(K_{1}K_{2})}{Var(\hat{\beta})}=\underset{(K_{1}K_{2})\times(J_{1}J_{2})}{B}\times\underset{(J_{1}J_{2})\times(J_{1}J_{2})}{Var(\hat{\Gamma})}\times\underset{(J_{1}J_{2})\times(K_{1}K_{2})}{B^{'}}
\]

Obtaining standard errors for the marginal effects consists in selecting
the squared root of the diagonal elements of $Var(\hat{\beta})$.
For more details, I invite the reader to consult the supplementary
data in \citet{ortiz-bobea_unpacking_2019}.

This may seem a bit too theoretical, so the attached code and data
provides an implementation for US corn yields (see \texttt{3\_time-varying\_effects.R}).
In the example, I consider 1°C temperature bins between 0 and 35°C
($K_{1}=36$) and monthly temporal bins between April and October
($K_{2}=7$). Ideally, we would probably want to have finer scale
temporal bins (e.g. weeks or pentads) but the use of monthly bins
makes this illustration more tractable as it makes use of data already
generated for other parts of the chapter. I also select degrees of
freedom $J_{1}=6$ and $J_{2}=3$. As a result, our binned matrix
$Z$ has $36\times7=252$ bins. Our tensor basis matrix $B$ has $6\times3=18$
columns. All these steps are clearly annotated in the R script file.

\begin{figure}
\begin{centering}
\includegraphics[scale=0.33]{figures/fig11_timevarying}
\par\end{centering}
\noindent\begin{minipage}[t]{1\columnwidth}%
\textit{\footnotesize{}Notes:}{\footnotesize{} Standard errors are
clustered at the state and year level. The data covers corn yields
over 1981-2020 east of the 100th meridian West. Panel A shows marginal
effects of additional exposure to temperature bins in different parts
of the growing season. Marginal effects that are statistically different
from zero at a 95\% level are indicated with a star ({*}). Panel B
shows the density (in days) in each one of the $36\times7=252$ bins.
These bins vertically add to the amount of time in each month. Summing
across all months and temperature levels adds up to the total number
of days in the growing season (183 days).}%
\end{minipage}

\caption{Time-varying effects of temperature on US corn yields throughout the
growing season. \label{fig:timevarying}}
\end{figure}

Figure \ref{fig:timevarying}A shows the marginal effects in their
2-dimensional form ($\hat{\beta}^{2D}$). The panel shows that high
temperature above 30°C appear detrimental especially in the months
between June and August. Marginal effects that are statistically different
from zero at a 95\% level are highlighted with a star. In contrast,
high temperatures appear beneficial early in the season in April,
although I later show that result is barely significant.

Figure \ref{fig:timevarying}B shows the underlying density over the
$36\times7=252$ bins. As expected, the distribution of temperature
in April and and October is toward lower temperature, whereas the
distribution is toward higher temperatures in the summer months. This
seasonality will naturally affect the precision of estimates for temperature
bins that exhibit little to no exposure over certain parts of the
year. For instance, there is little exposure above 30°C in the months
of April and October. This should render the estimation of marginal
effects around that time particularly noisy.

\begin{sidewaysfigure}
\begin{centering}
\includegraphics[scale=0.23]{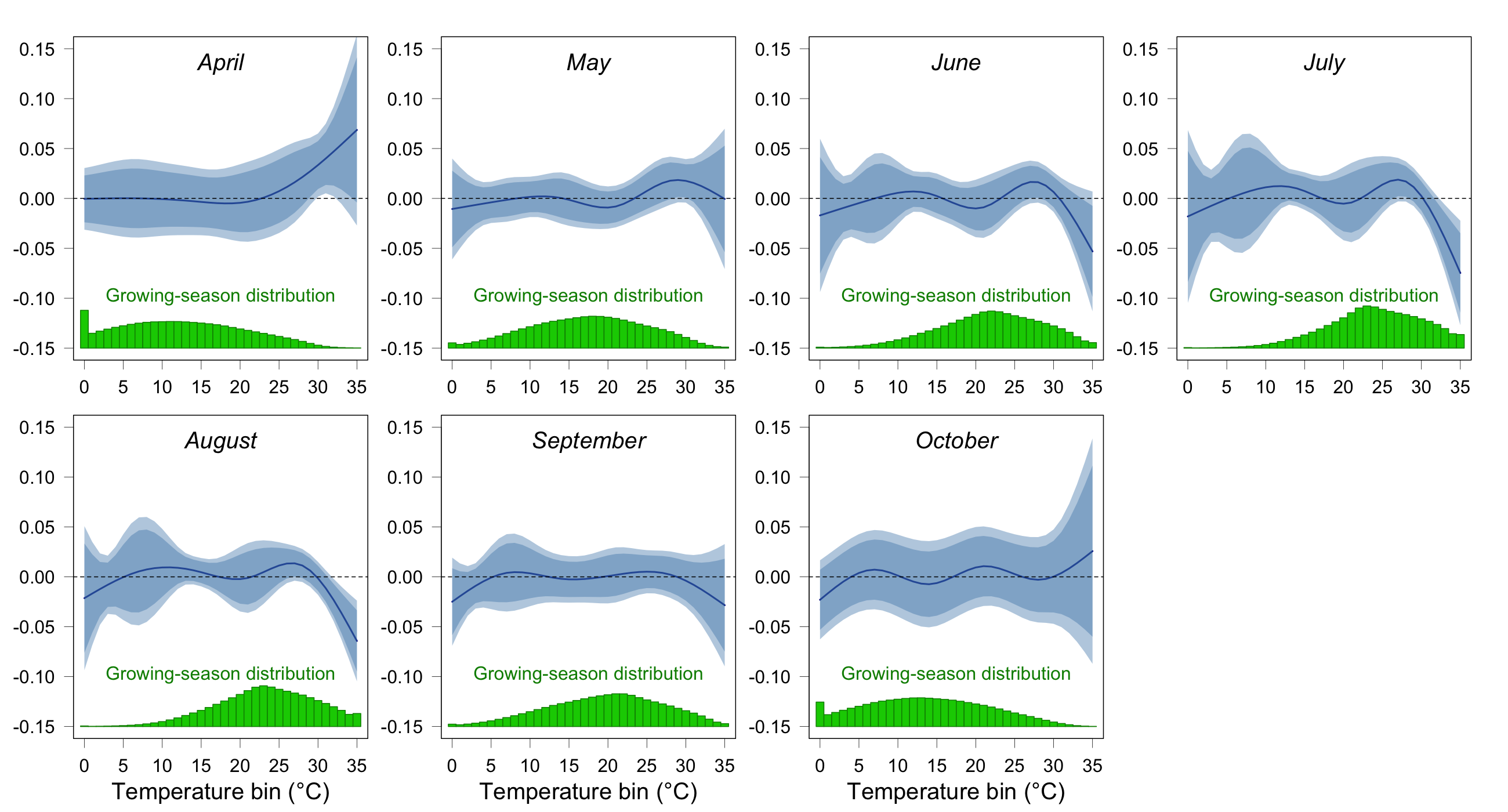}
\par\end{centering}
\noindent\begin{minipage}[t]{1\columnwidth}%
\textit{\footnotesize{}Notes:}{\footnotesize{} Standard errors are
clustered at the state level. The data covers corn yields over 1981-2020
east of the 100th meridian West. Precipitation variables were not
included. The blue colored bands around the response function corresponds
to 95 and 99 percent confidence intervals. The green histogram represents
the monthly exposure to all temperature bins.}%
\end{minipage}

\caption{Time-varying effects of temperature on US corn yields evaluated at
each month of the growing season. \label{fig:timevarying-2}}
\end{sidewaysfigure}

To explore this, Fig. \ref{fig:timevarying-2} shows a cross-section
of marginal effects evaluated at each temporal bin, that is, at each
month. The marginal effects in April and October at high levels of
temperature are rather noisy and are not significant at a 99\% confidence
level. Notice there is little to no density (green histogram) at high
levels of temperature in those months. In contrast, the marginal effects
in July and August are clearly significant at a 99\% confidence level
and those effects appear more precisely estimated.

This subsection illustrates how one can tractably harness variation
in environmental variables within the growing season to estimate within-season
time-varying effects.

This type of model seems particularly useful when researchers seek
to harness important within-season differences in crop yield sensitivity.
For instance, this might be particularly useful for modeling changes
in planting dates or growing seasons, or for estimating high-frequency
within-season yield forecasts. \citet{ortiz-bobea_unpacking_2019}
shows that this approach performs better out of sample than their
traditional counterparts that do not allow for time-varying effects
(presented in the previous subsection). However, the differences in
model fit are not enormous and the magnitude of projected impacts
of climate change (without changes in growing seasons) remain similar.
One potential reason for this is that very high temperature typically
occur in the summer when crops like corn are flowering. This means
that the detrimental effect of season-long exposure to extreme temperature
in traditional additive models could be primarily reflecting the effect
of high temperature (and drought) during the sensitive stages of crop
development that happen to coincide with the warmer summer months.

This model also has some limitations. First, the process of constructing
bins is somewhat more involved especially if constructing temporal
bins finer than a month. However, the code provided with this chapter
should alleviate many of these concerns. Another issue is that marginal
effects can be fairly imprecise around areas with little to no exposure
(see \ref{fig:timevarying}B). So perhaps this approach may perform
better with standardized variables that exhibit similar distribution
across different points of the growing season. That would ensure enough
density throughout the entire rectangular support of the 2-dimensional
spline.

Potential future research directions may include improvements to this
approach including the implementation of penalized B-splines, also
know as P-splines, to improve model fit. For instance, the approach
implemented in \citet{ortiz-bobea_unpacking_2019} to select the flexibility
of the tensor spline in both direction was based on a relatively onerous
grid search. There are more advanced approaches to implementing these
techniques. It might also be possible to explore multi-dimensional
splines that allow smooth interactions between environmental indicators
throughout the growing season. However, this is likely to require
large datasets and possibly experimental data to obtain independent
draws of environmental indicators which tend to be highly correlated
in observational settings (e.g. high temperature and drought).

For additional reader regarding the use of high-frequency weather
data for analyzing climate change impacts I invite the reader to consult
\citet{ghanem_what_2020}. This also relates to a broader literature
on model selection (see \citealp{cui_model_2018}).

\subsection{Spatial dependence\label{subsec:Spatial-dependence}}

A common characteristic of agricultural data is that neighboring locations
exhibit similar values. For instance, a map of crop yields, farm profits
or farmland values all appear spatially correlated even at large spatial
scales. There are multiple reasons for this. One of them has to do
with the spatial dependence of land characteristics such as soil texture,
slope or climate. These spatially-dependent factors influence the
performance of agriculture which in turns also ends up exhibit spatial
dependence. In addition, socio-economic factors can also play a role
including local and regional regulations, the presence of irrigation
or transportation infrastructure, or the proximity to certain markets
or population centers. 

In practice, econometric models do not incorporate all these spatially-dependent
drivers of our agricultural outcomes of interest. As a result, these
omitted variables end up in the error term which ends up also exhibiting
spatial dependence. Moreover, weather and climate variables, which
are used as predictors, are also spatially dependent. The combination
of spatially-dependent error terms and regressors leads particular
challenges in regression analysis. For starters, the assumption that
errors are independent no longer holds. So adopting standard errors
that are simply robust to heteroscedasticity is insufficient. Ignoring
positive spatial dependence in the error term in the presence of positively
spatially-dependent regressors leads to overconfident inference. Standard
errors in those regressions are simply wider than they appear. 

The challenge of spatial dependence is analogous to issues that arise
in the presence of clustered error terms and regressors \citep{cameron_practitioners_2015,moulton_random_1986,moulton_illustration_1990}.
A clustered variable is one that exhibits within-group (cluster) correlation.
In observational settings, this correlation is typically positive.
Ignoring the positive within-cluster correlation also leads to the
estimation of standard errors that are narrower than they truly are. 

One common approach to ``correct'' for spatial dependence is to
cluster standard errors at regional scales larger than the unit of
analysis. For instance, in a model based on US county data that would
be to cluster at the district or state level. What this approach assumes
is that there is a common ``shock'' to all locations within that
cluster but no correlation between locations across cluster boundaries.
In the case of US counties, that would mean that neighboring counties
exhibit correlated errors or regressors within states, but that there
is no correlation across state lines. Given that the drivers of many
agricultural outcomes are natural in nature and do not follow administrative
borders, this assumption is unlikely to be valid, at least \textit{a
priori}.

A more conceptually appropriate way to correct for spatial dependence
in a linear model is to correct directly for spatial dependence. This
would allow to account for smoother patterns of correlation between
neighboring locations irrespective of whether locations fall within
the same state or district. One approach to achieve this is the adopting
a spatial heteroscedasticity and autocorrelation consistent (HAC)
estimator of the variance covariance matrix introduced in \citet{conley_gmm_1999}.
The idea is close to a spatial analogue to the heteroscedasticity
and autocorrelation consistent estimator of the covariance matrix
introduced in \citet{newey_simple_1986} to address serial correlation.
The key idea behind the spatial HAC estimator is the use of a kernel
that weighs the cross-products in the computation of the covariance
matrix based on the spatial distance between observations. The implementation
requires a distance threshold (say 500 miles) beyond which one assumes
there is no correlation between observations. This is typically implemented
via a Bartlett window which assigns the value of 1 for the observation
at hand (distance equal 0) and decreases linearly down to 0 when one
reaches the threshold. Beyond that point the assigned weight is 0.

The R code that accompanies this chapter provides an implementation
of the spatial HAC standard errors (see \texttt{4\_spatial\_dependence.R}).
The implementation allows the specification of multiple distance thresholds
as well as that of various weighting kernels in addition to the more
common Bartlett window.

Another approach to account for spatial dependence is to harness the
spatial dependence in the estimation to obtain a more efficient estimator.
This can be achieved with a Spatial Error Model (SEM) which can be
estimated via GMM or MLE \citep{anselin_spatial_1988}. This approach
is relatively uncommon and can be found in a few studies in the literature
(e.g. \citealp{schlenker_impact_2006}). The difference between the
spatial HAC correction above and the SEM is analogous to the difference
between correcting for heteroscedasticity and estimating a Generalized
or Weighted Least Squares. In the former case one seeks to correct
the estimation of standard errors of an otherwise inefficient estimator,
whereas in the latter one tries to harness more information about
the distribution of the error term to derive a more efficient estimator.

One perceived disadvantage of the SEM is that it requires specifying
a weight matrix, imposing some structure on the nature of the spatial
dependence between neighboring observations. However, these concerns
are likely overblown \citep{lesage_biggest_2014}. Moreover, doesn't
the imposition of a kernel and a cutoff distance in the estimation
of the spatial HAC estimator also impose some form of structure to
the nature of spatial dependence? One approach to avoid unnecessary
criticism by reviewers unfamiliar with the SEM is to present the SEM
results along with results based on OLS (corrected for spatial dependence).
Note however, that the SEM and OLS estimates are consistent in the
absence of an omitted variable, but the presence of an omitted variable
induces different types of biases in these two estimators. This situation
permits the implementation of a spatial Hausman test \citep{pace_spatial_2008}.
The R code that accompanies this chapter provides an implementation
of the SEM model for balanced panel models. The implementation allows
the specification of various types of weight matrices. 

\begin{sidewaysfigure}
\begin{centering}
\includegraphics[scale=0.23]{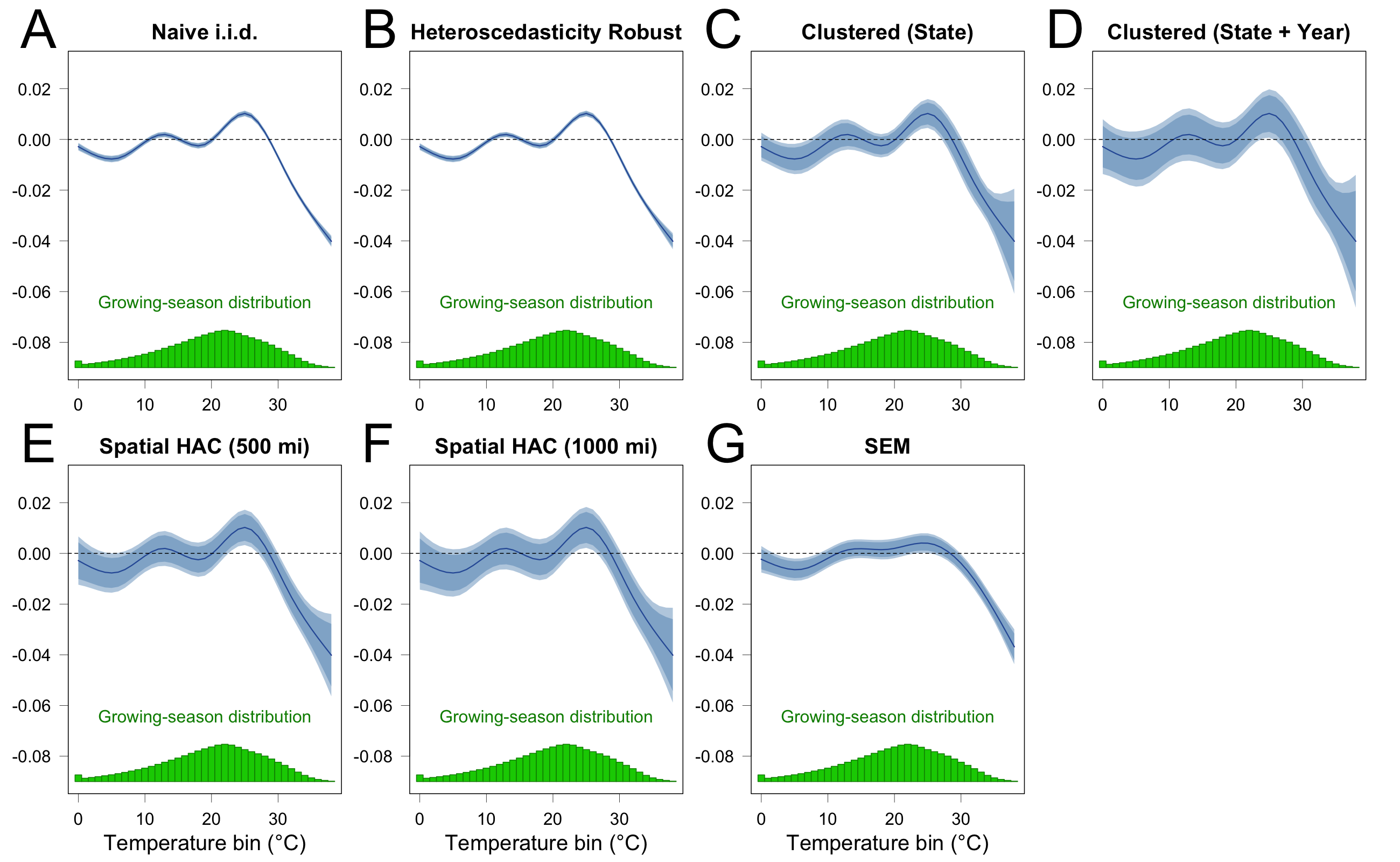}
\par\end{centering}
\noindent\begin{minipage}[t]{1\columnwidth}%
\textit{\footnotesize{}Notes:}{\footnotesize{} The data covers corn
yields over 1981-2020 east of the 100th meridian West. Precipitation
variables were included in the regression but coefficients are not
shown. The blue colored bands around the response function corresponds
to 95 and 99 percent confidence intervals for each type of standard
error. The green histogram represents the growing-season of exposure
to all temperature bins. The response function is based on a natural
cubic spline with 7 degrees of freedom and 1°C bins between 0 and
38°C over the April-September growing season. Note the Spatial Error
Model (SEM) is estimated based on a smaller balanced panel.}%
\end{minipage}

\caption{Different standard errors and estimators for the effects of temperature
exposure on corn yields. \label{fig:spatial}}
\end{sidewaysfigure}

I illustrate the marginal effects of exposure to various temperature
levels with various standard errors for OLS and for the SEM in Fig.
\ref{fig:spatial}. Panels A through F are based on OLS and have the
same point estimates. Panel G is based on the SEM, which is a different
estimator, so the response function is different.

Panel A of Fig. \ref{fig:spatial} shows a confidence band that assumes
errors are independent and identically distributed. Even beforehand,
we know this assumption is incorrect because the error is spatially
dependent so these confidence band is deceptively narrow. The R code
shows that the residuals appear highly correlated and that a Moran's
I test rejects the hypothesis of no spatial correlation. Note that
correcting simply for heteroscedasticity in B does not change things
much.

Interestingly, Fig. \ref{fig:spatial}C and D show that clustering
the standard error by state or by state and year leads to much more
conservative standard errors which leads to a much wider confidence
band around our response function. While many of the effects of temperature
exposure in the 0-25°C largely indistinguishable from zero at conventional
levels, the effects of exposure beyond 30°C appear clearly detrimental.

Figure \ref{fig:spatial}E is based on the spatial HAC standard errors
proposed in \citet{conley_gmm_1999} using a cutoff distance of 500
miles. Panel F shows the confidence band when the cutoff is increased
to 1000 miles. Note these standard errors are of a similar magnitude
to those clustered by state or by state and year. This is an interesting
finding because it suggested that in some cases, clustering may yield
similarly conservative estimates despite not accounting for cross-cluster
correlations. 

Finally, Fig. \ref{fig:spatial}G shows estimates based on the SEM
estimated via Maximum Likelihood. Note that the implementation in
R only allows balanced panels. So while the OLS models where estimated
with an unbalanced panel of 2,248 counties over 1981-2020 and 69,190
observations, the SEM was estimated with a balanced panel of 599 counties
over the same period with only 23,960 observations. And yet, the standard
errors --which are conceptually correct-- are much narrower. As
explained above, this estimator is more efficient than OLS because
it harnesses information about spatial dependence in the estimation.

\subsection{Common robustness and sensitivity checks\label{subsec:Common-robustness-and}}

It is often common for reviewers in the peer review process to ask
authors to perform additional robustness or sensitivity checks. While
these requests can often feel like an unnecessary burden to authors,
these checks can help build more confidence on the results and clarify
to what degree certain modeling assumptions have an outsized influence.
This relates to a large degree to the classic call in \citet{leamer_lets_1983}
for the systematic adoption of sensitivity checks in empirical research. 

Perhaps the most basic type of check is about heterogeneity, which
can generally be manifested either in space or time. When estimating
weather effects on economic outcomes in a panel setting, it is generally
a good idea to test whether coefficients are stable across major regional
or temporal subsets of the data (e.g. with a Wald test). Understanding
heterogeneity is critical and can even become the primary focus of
a study. For instance, \citet{butler_adaptation_2013} find that corn
yields in hotter regions in the US appear less sensitive to extreme
heat relative to colder regions, which the authors suggest is indicative
of adaptation to climate change (see \citealp{schlenker_us_2013}
for a response). Analogously, \citet{ortiz-bobea_growing_2018} explores
temporal heterogeneity in the response function of US agricultural
TFP to document the rising sensitivity of Midwestern agriculture to
higher temperature.

Another aspect that often receives considerable scrutiny in empirical
work are nonlinearities in how weather or climate variables affect
agricultural outcomes of interest. For instance, \citet{mendelsohn_impact_1994}
included linear and quadratic coefficients for monthly temperature
and precipitation variables in the specification. This captures the
idea that while certain weather conditions are optimal for agricultural
production, extreme weather tends to be detrimental. The work of \citet{schlenker_nonlinear_2009}
also highlighted the importance of nonlinearities for temperature
by introducing a semi-parametric approach that flexibly harnesses
the influence of daily temperature exposure on crop yields. 

Sometimes, however, regression coefficients are difficult to interpret
directly, particularly when dealing with higher term polynomials or
splines. As a result, a useful approach is to compare the impact of
a uniform warming (e.g. +2°C) or a percentage change in precipitation
across alternative specifications. There is a broad recognition that
temperature effects may not only capture heat stress but also moisture
stress (see \citealp{lobell_critical_2013} or \citealp{ortiz-bobea_unpacking_2019})
so the emphasis tends to be justifiably more focused on projected
impacts than on the values of the estimated parameters themselves.
This is perhaps slightly different from other research areas where
the value of estimated parameters are quantities with a clear theoretical
grounding in economics (e.g. a price elasticity). In general, it is
advisable to compare models, not only in terms of their fit (ideally
out of sample) but also in terms of the associated impacts of some
specific change in the distribution of weather or climatic conditions.
This facilities comparison across models.

Another modeling choice that receives some scrutiny is the choice
of the weather dataset and how it is aggregated in space or time.
It is not uncommon to see researchers reporting results based on alternative
weather datasets, or alternative ways of spatially aggregating weather
to the unit of analysis. As indicated earlier in this chapter, one
can spatially aggregate gridded weather data based on a variety of
different of weighting schemes (e.g. cropland or cropland and pasture
weights, etc). The same goes for spatially interpolated weather station
data. The choice of seasons and how weather conditions are aggregated
over time, is also a common point of scrutiny. In general, it is difficult
to know \textit{a priori} the role of these modeling assumptions in
our results. It is thus generally the case that researchers present
results based on alternative definitions of the growing season.

An important aspect of conducting empirical research is appropriately
characterizing the uncertainty around estimated parameters. The assumption
that errors are independent and identically distributed is virtually
always violated in observational settings. When analyzing climate-economy
linkages, weather conditions tend to be strongly spatially correlated
across great distances (see previous subsection). And just like economic
outcomes, their unobservable drivers are also spatially dependent.
Ignoring this positive spatial dependence typically leads to overconfidence
by underestimating standard errors. In such cases, a common solution
is to correct for spatial dependence in a linear model using spatial
HAC standard errors (e.g. \citealp{conley_gmm_1999}). It is also
common to see researchers cluster standard errors at relatively large
regional levels (e.g. at the state level in a county-level panel setting)
in order to capture the contemporaneous dependence. However, such
clustering approaches would not account for correlation across clusters
that may be occurring. But as shown in the previous subsection, these
approaches can point to similar standard errors.

It is also not uncommon to see the adoption of weighted regressions,
where observations are weighted based on some metric indicative of
each observation's variance, like acreage. The idea is that the variance
of observations in locations with small acreage is higher, so these
would be given a smaller weight in a weighted least squares estimation.
However, this approach poses some challenges regarding the interpretation
of the results if the introduction of regression weight drastically
alter the estimated coefficients. This may signal a misspecification
or an unaccounted heterogeneity (see \citealp{solon_what_2015} for
guidance on regression weights). The adoption of spatial HAC errors
seems more appropriate, at least conceptually, because it jointly
accounts for heteroscedasticity and spatial dependence.

An increasingly common exercise in empirical studies is the implementation
of ``placebo'' checks or tests \citep{eggers_placebo_2021}. The
idea behind a placebo check is that a treatment should not appear
to have an effect on an untreated unit of analysis. In our context,
for instance, it means that a weather shock in location $i$ at time
$t$ should have no effect on the outcome of interest in a different
location or time period. This is a bit tricky because weather is not
independently ``assigned'', meaning that similar weather shocks
occur in neighboring locations, so a weather shock in one location
may well appear to have an effect in neighboring locations simply
because of such contemporaneous correlations. We could also make an
analogous point in the time dimension in the presence of serially
correlated weather conditions (e.g. in places where droughts tend
to be multi-year phenomena).

A common placebo check in panel settings is to estimate the model
with lead weather. It is reasonable to assume that future weather
fluctuations, which are arguably impossible to predict accurately
in advance, should have no bearing on a current outcome. An analogous
spatial version of this check is to simply randomly reassign weather
of other locations to the outcome variable. In a large enough panel,
this ``reshuffling'' should result in results that are, on average,
insignificantly different from zero. However, note that it is still
possible to get a significant effect by chance. This is a key weakness
of such simple placebo checks, which are not actual statistical tests.

A preferable approach is to conduct a placebo test in the form of
a permutation test (also called randomization test). These are non-parametric
tests for which the researcher constructs the distribution of a test
statistic under the null hypothesis of ``no effect''. In a setting
estimating the effect of weather on an economic outcome, the idea
is to reshuffle the weather data across units, re-estimate the model
with the permuted weather predictors, store the coefficients and repeat
this process, say, ten thousand times. The resulting distribution
of this spurious coefficients should be centered around zero. The
researcher can then contrast the sample estimate, i.e. the estimate
with the correctly matched data, to this distribution to obtain a
p-value. This procedure seems straightforward when assessing a single
coefficient, but the analysis can be done when estimating a quadratic
relationship (e.g. see Figure 2 in \citealp{ortiz-bobea_anthropogenic_2021}).
Another approach that might help conducting multiple simultaneous
tests is to derive the distribution of the test statistic as the linear
combination of coefficients (e.g. computing the impact of a 2°C warming
at each iteration).

\begin{sidewaysfigure}
\begin{centering}
\includegraphics[scale=0.4]{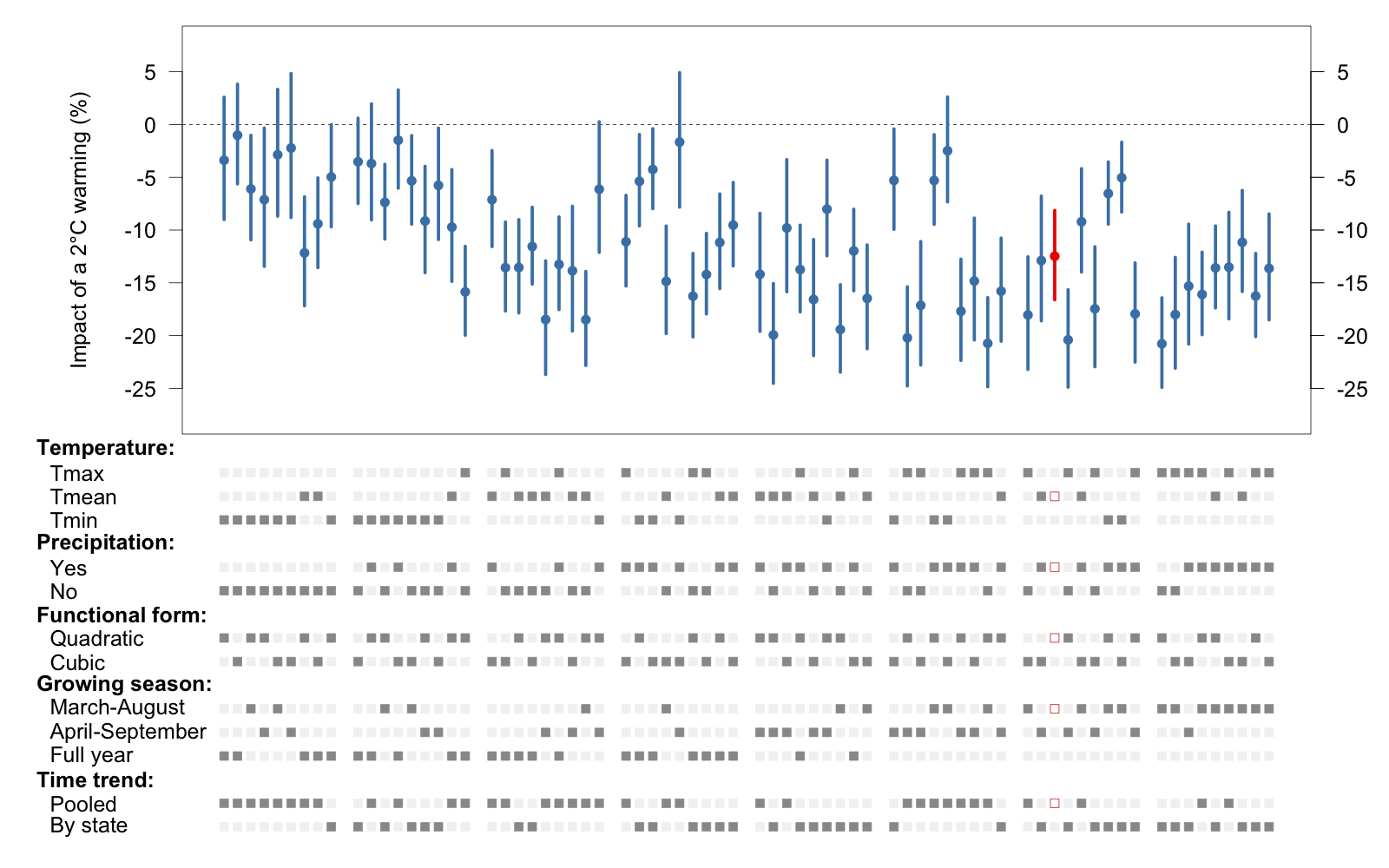}
\par\end{centering}
\noindent\begin{minipage}[t]{1\columnwidth}%
\textit{\footnotesize{}Notes:}{\footnotesize{} Standard errors are
clustered at the state level. The data covers corn yields over 1981-2020
east of the 100th meridian West. The models are sorted by adjusted
R2 with the best-fitting model shown on the right.}%
\end{minipage}

\caption{Estimated impact of a 2°C for 72 alternative specifications. \label{fig:specchart}}
\end{sidewaysfigure}

This section highlights that conducting empirical research in this
area requires making a relatively large number of modeling assumptions
along the way. In order to build confidence in our results, it is
critical to justify these assumptions and, ideally, to show to what
extend results are sensitive to these. However, presenting the results
of the paper under many alternative sets of modeling assumptions can
be tedious and make any paper feel overburdened. For instance, the
online appendix in \citet{ortiz-bobea_unpacking_2019} is 74 pages
long and includes 6 pages of single-spaced text, 57 supplementary
figures and 7 tables.

One strategy to show robustness checks in a more parsimonious way
is to present the main estimate in a study under alternative modeling
assumptions in a specification chart or curve \citep{simonsohn_specification_2019}.
This type of chart can summarize a large number of estimates in a
single visualization, which helps authors and readers better understand
the important of specific modeling assumptions. Rather than thinking
about these charts as a way to show ``robustness'', it is preferable
to think about this as an exercise in transparency and as a way to
characterize the potential uncertainty regarding model selection. 

The R code provided with this chapters provides an implementation
of a specification chart with a reproducible example (see \texttt{5\_robustness
\_checks.R}). In this example I estimate a statistical crop yield
model based on a panel of corn yields east of the 100th meridian West
over the 1981-2020 period. The dependent variable is log yield and
I consider a wide range of variation of the model specification. Specifically,
I explore the role of the temperature variable (Tmax, Tmean or Tmin),
the inclusion of a precipitation variable, the adoption of a quadratic
or cubic functional form for the weather regressors, the definition
of the growing season (March-August, April-September or the entire
year) and wether a quadratic time trend in the regression model is
estimate by state or pooled for all counties. Obviously, we could
add more checks.

It is very common for researchers to conduct analysis and show results
based on a particular ``baseline'' model. Let's say that our baseline
model is based on Tmean, includes precipitation, adopts a quadratic
response function, adopts a growing season defined over March-August,
and includes a common time trend for the sample. One way to show how
this model compares to alternative models is to conduct the analysis
on an exhaustive combination of these modeling assumptions. However,
comparing coefficients across models with different temperature variables
and functional forms can be intractable. A better approach is to compare
the implied effect of a, say, 2°C warming.

Figure \ref{fig:specchart} shows the effect of a 2°C warming for
72 different specifications. In this example, I have sorted the models
from the lowest to the highest fitting specification based on the
adjusted R2. Models could also be sorted based on any other criteria
including out-of-sample Means Squared Error, their point estimate
or simply preserve the original order in which the models were stored
in the input table. What is interesting here is that the best fitting
models (on the right) tend to point to larger negative impacts. The
baseline model we selected, highlighted in red, in not much different
than many other well-fitting models. 

A few things stand out while exploring this specification chart. First,
models based on Tmin tend to point to small impacts and perform relatively
poorly in terms of fit. Second, models based on Tmax tend to point
to larger impacts and fit better, especially when coupled with precipitation
variables. Third, there is no discernible difference between models
estimated with a quadratic or a cubic functional form. Fourth, the
shorter March-August season tends to fit better whereas the full annual
season performs more poorly. Finally, models estimated based on a
state-level time trend tend to perform better in terms of fit, and
also put to larger impacts relative to models with pooled time trends.

This exercise shows that modeling assumption can drastically alter
reported estimates. Being upfront and transparent about how the main
baseline results shown in the paper can change with modeling assumptions
should build confidence in the reader that results were not cherry
picked. Figure \ref{fig:specchart} is fully reproducible with the
included R code. The code also includes a function that allows the
creation of such charts with relative ease. I have also included a
reproducible example in \texttt{spec\_chart\_reproducible\_example.R}
that showcases all the capabilities of this function.

\section{Conclusion\label{sec:Conclusion}}

This chapter provides an overview of the economic literature analyzing
climate change impacts and adaptation in agriculture. The first subsection
cover some basic concepts and knowledge necessary to understand the
discussions and debate in the literature. This includes overviews
of common sources and the nature of weather and climate change data.

The following subsection cover a range of topics with an emphasis
on methods for analyzing various aspects of the impacts of climate
change on agriculture. I discuss early studies based on biophysical
approaches and their evolution and subsequent integration with trade
models. I then describe the emergence and motivation behind econometric
techniques and the explosion of empirical studies analyzing impacts
of weather fluctuations on agricultural outcomes based on longitudinal
data. I also discuss emerging efforts to link statistical and biophysical
techniques, as well as new methods to combine the advantages of cross-sectional
and longitudinal methods in the empirical literature. I also spend
some time discussing work exploring various mechanism of adaptation,
including irrigation, crop choices, planting and cropping frequency
decisions. I finally provide a very short overview of recent work
analyzing the role of trade in determining climate change impacts.
I conclude with some possible future directions of research and collaborations
with non-economists.

The last subsection is a unique effort to provide a hands on introduction
to new researchers in this literature. The subsection discusses very
practical empirical challenges as well as common tasks necessary to
estimate more sophisticated empirical models in the field. That part
also discusses the estimation of standard errors in the presence of
spatial dependence. I also spend a bit of time discussing common sensitivity
checks in the literature and introduce a parsimonious way of presenting
numerous robustness checks without overwhelming the reader. 

All of the figures in this chapter are fully reproducible. This code
represents the entire workflow including downloading some of the weather
datasets, transforming the weather data, spatially aggregating and
interpolating weather data, estimating various types of semi-parametric
models and the construction of various types of standard errors that
are relevant in this literature. It is my hope that this chapter removes
some of the barriers to entry to this field of research. Thank you
for reading.

\bibliographystyle{aer}
\addcontentsline{toc}{section}{\refname}\bibliography{references}

\end{document}